\documentclass[letterpaper,twocolumn,10pt]{article}
\usepackage{usenix}

\usepackage{tikz}
\usepackage{filecontents}
\usepackage{xcolor}
\usepackage[T1]{fontenc}    % use 8-bit T1 fonts
\usepackage{colortbl}

\usepackage{url}            % simple URL typesetting
\usepackage{booktabs}       % professional-quality tables
\usepackage{amsfonts}       % blackboard math symbols
\usepackage{nicefrac}       % compact symbols for 1/2, etc.
\usepackage{microtype}      % microtypography
\usepackage{xcolor}         % colors
\usepackage[textsize=tiny]{todonotes}
\usepackage{amsmath}
\usepackage{amssymb}
\usepackage{mathtools}
\usepackage{amsthm}
\usepackage{graphicx}
\usepackage{subfigure}
\usepackage{multirow}
\usepackage{tablefootnote}
\usepackage{makecell}
\usepackage{algorithmic}
\usepackage[linesnumbered,ruled,vlined]{algorithm2e}
\usepackage{wrapfig}
\usepackage{lipsum,caption}
\usepackage{pifont}
\usepackage{lipsum}
\usepackage{adjustbox}
\usepackage{tabularx}
\usepackage{enumerate}
\usepackage{etoolbox}
\usepackage{diagbox}
\usepackage{threeparttable} 
\usepackage{siunitx}
\usepackage{soul}
\usepackage{tikz}
\usepackage{tcolorbox}
\usepackage{mathrsfs}
\usepackage{cancel}
\usepackage{hyperref}

\usepackage[available]{usenixbadges}

\soulregister\ref7 % 为 ref 添加删除线
\soulregister\cite7 % 为 cite 添加删除线
\soulregister\underline7

\newtheorem{theorem}{Theorem}

\newcommand{\revise}[1]{{\color{black}#1}}
\newcommand{\newtext}[1]{{\color{black}#1}}
\newcommand{\delete}[1] {{}}
\newcommand{\ashr}[1] {[\![#1]\!]}

\newcommand{\method}{BLB}
\newcommand{\mbm}{\bold}
\newcommand{\HomAdd}{\boxplus}
\newcommand{\HomMul}{\boxtimes}
\newcommand{\Dec}{\operatorname{Dec}}
\newcommand{\Enc}{\operatorname{Enc}}
\newcommand{\Encode}{\operatorname{Encode}}
\newcommand{\Decode}{\operatorname{Decode}}

\newcommand{\pring}{\mathbb{A}_{N,q}}
\newcommand{\poly}[1]{\hat{#1}}

\newcommand{\cmp}{\operatorname{cmp}}

\newcommand{\fusion}{FineGrainFusion}
\newcommand{\mmcp}{\operatorname{matmul_{cp}}}
\newcommand{\mmcc}{\operatorname{matmul_{cc}}}

\usepackage{stmaryrd} % llbracket
\newcommand{\ashare}[1]{\llbracket #1 \rrbracket}

\definecolor{mygreen}{RGB}{226, 240, 217}
\definecolor{myorange}{RGB}{254, 231, 198}
\definecolor{myred}{RGB}{247, 206, 204}

\begin{document}

%don't want date printed
\date{}

\title{\underline{B}reaking the \underline{L}ayer \underline{B}arrier: 
Remodeling Private Transformer Inference\\with Hybrid CKKS and MPC}
% \ml{Layer Decomposition and Fine-Grained Fusion?}}
% %for single author (just remove % characters)
\author{
{\rm Tianshi Xu$^\dagger$}\\
Peking University
\and
{\rm Wen-jie Lu$^\dagger$}\\
TikTok
\and
{\rm Jiangrui Yu$^\dagger$}\\
Peking University
\and
{\rm Yi Chen}\\
Peking University
\and
{\rm Chenqi Lin}\\
Peking University
\and
{\rm Runsheng Wang}\\
Peking University
\and
{\rm Meng Li$^*$}\\
Peking University
% copy the following lines to add more authors
% \and
% {\rm Name}\\
%Name Institution
} % end author

\maketitle
\newcommand{\cofirstsymbol}{\textsuperscript{\textdagger}} % 使用 \textdagger 作为符号
\newcommand{\cofirsttext}{Corresponding author: meng.li@pku.edu.cn}

% 定义通讯作者的脚注标记
\newcommand{\corrauthorsymbol}{\textsuperscript{\textasteriskcentered}} % 使用 \textasteriskcentered 作为符号
\newcommand{\corrauthortext}{These authors contributed equally to this work.}

% 重置脚注符号计数器，确保它们从1开始
\let\oldthefootnote\thefootnote % 保存旧的脚注符号定义
\renewcommand{\thefootnote}{\fnsymbol{footnote}} % 将脚注符号改为符号类型

% 手动定义第一个脚注（共同一作）
\footnotetext[1]{\cofirsttext}

% 手动定义第二个脚注（通讯作者）
\footnotetext[2]{\corrauthortext}

% 这个脚注不会显示任何上标符号，直接在页面底部显示文本
\footnotetext{Our artifact's DOI is \href{https://doi.org/10.5281/zenodo.15590214}{10.5281/zenodo.15590214}.}
% 建议使用 hyperref 包来让 DOI 成为可点击的链接
\hypersetup{colorlinks=true, urlcolor=blue} % 可选：设置链接颜色

% 恢复脚注符号为默认（通常是数字），以防影响后续的脚注
\renewcommand{\thefootnote}{\oldthefootnote}
\begin{abstract}
    % \xts{all numbers in this paper need final check!!}
    % \xts{all captions and figures need double check!!!}
    %-------------------------------------------------------------------------------
    % Your abstract text goes here. Just a few facts. Whet our appetites.
    % Not more than 200 words, if possible, and preferably closer to 150.

    % As large transformer-based models are applied to various real-world applications, there is a growing privacy concern, especially in scenarios where user inputs and model weights must remain confidential. 

    This paper presents an efficient framework for private Transformer inference that combines Homomorphic Encryption (HE) and Secure Multi-party Computation (MPC) to protect data privacy. Existing methods often leverage HE for linear layers (e.g., matrix multiplications) and MPC for non-linear layers (e.g., Softmax activation functions), but the conversion between HE and MPC introduces significant communication costs. The proposed framework, dubbed \method, overcomes this by breaking down layers into fine-grained operators and further fusing adjacent linear operators, reducing the need for HE/MPC conversions. To manage the increased ciphertext bit width from the fused linear operators, \method~proposes the first secure conversion protocol between CKKS and MPC and enables CKKS-based computation of the fused operators.
    Additionally, \method~proposes an efficient matrix multiplication protocol for fused computation in Transformers. Extensive evaluations on BERT-base, BERT-large, and GPT2-base show that~\method~achieves a $21\times$ reduction in communication overhead compared to BOLT (S\&P'24) and a $2\times$ reduction compared to Bumblebee (NDSS'25), along with latency reductions of $13\times$ and $1.8\times$, respectively, when leveraging GPU acceleration.
\end{abstract}
\section{Introduction}
\label{sec:introduction}
% \ml{
% Problems with existing frameworks:
% \begin{itemize}
%     \item Limited flexibility: only consider coarse-grained fusion across neighboaring layers with limited fusion patterns;
%     \item High ciphertext bit width: consecutive linear operations significantly increase the bit widths for BFV, and, thus, the computation and communication cost;
%     \item Suboptimal fusion protocol: for complex operator fusion, e.g., self-attention, existing protocols requirea large number of rotations.
% \end{itemize}
% Faced with the three problems, BLB makes the following contributions:
% \begin{itemize}
%     \item Fine-grained fusion across operators w/ more flexible fusion patterns;
%     \item Hybrid CKKS and MPC framework with secure conversion;
%     \item Compact packing algorithm for self-attention layers.
% \end{itemize}
% }

Transformer architectures, e.g., BERT~\cite{kenton2019bert}, GPT~\cite{radford2019languagegpt2}, and LLAMA~\cite{touvron2023llama}, have achieved state-of-the-art (SOTA) performance in extensive real-world applications, including person re-identification~\cite{sarker2024transformer}, medical diagnosis~\cite{shamshad2023Transformers}, and voice assistant~\cite{cheng2022personal}. However, as these applications involve processing sensitive personal data, privacy concerns have emerged as a critical issue when deploying the Transformer models.

% Private Transformer inference has recently been studied to protect sensitive user data as well as 

% \ml{change ``Transformer'' to ``Transformer''.}

\begin{figure}[!tb]
    \centering
    \includegraphics[width=1.0\linewidth]{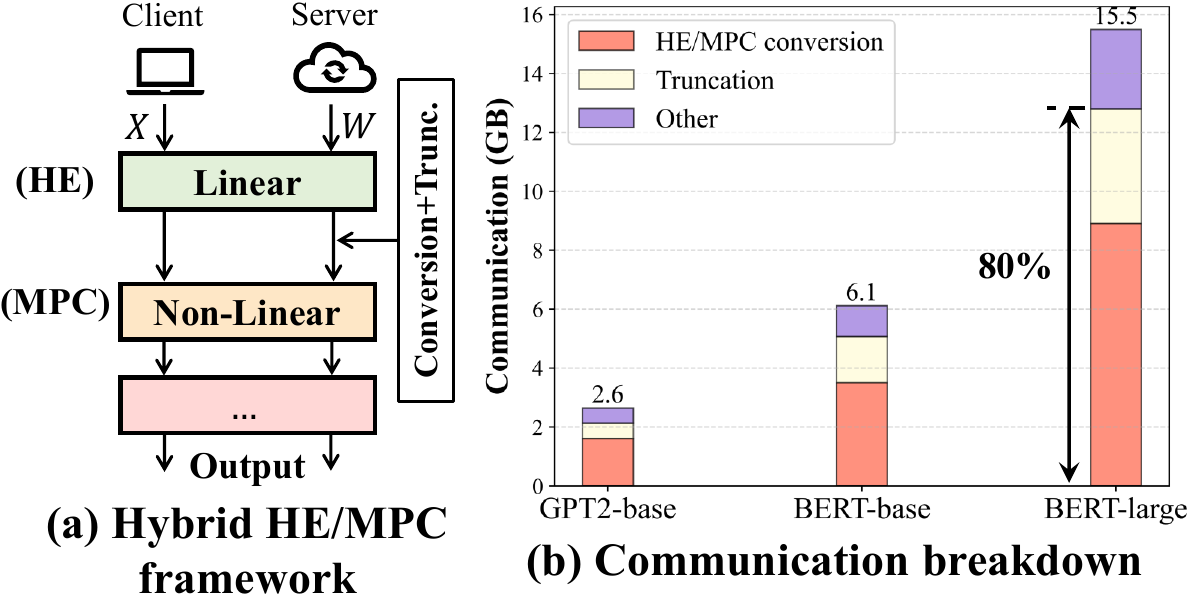}
    \caption{(a) Illustration of hybrid HE/MPC-based private inference; (b) Communication breakdown for GPT2-base, BERT-base, and BERT-large models based on the Bumblebee protocol~\cite{lu2023bumblebee}.}
    \label{fig:intro_profile}
    \vspace{-3pt}
\end{figure}

Private inference seeks to protect users' data privacy while ensuring users learn nothing about the model parameters except for the final results~\cite{Juvekar_Vaikuntanathan_gazelle_2018,rathee2020cryptflow2,huang2022cheetah}.
% Private inference seeks to protect sensitive model weights from users while ensuring that the server learns no information about the users' private inputs.
Recent advancements have proposed cryptographic frameworks based on Secure Multi-party Computation (MPC) for private Transformer inference~\cite{hao2022iron,wang2022characterization,zeng2022mpcvit,huang2024secbert,luo2024secformer,zeng2024securegpt,pang2023bolt,lu2023bumblebee}. 
% \ml{Add the references for OT-based methods, e.g., COINN, CoPriv, etc.}\xts{done.}
Among them, SOTA frameworks such as BOLT~\cite{pang2023bolt} and Bumblebee~\cite{lu2023bumblebee} employ a hybrid protocol combining Homomorphic Encryption (HE) and MPC. 
As shown in Figure~\ref{fig:intro_profile} (a), in the hybrid protocol, HE is applied to compute linear layers (e.g., matrix multiplications), and MPC is used for evaluating nonlinear layers (e.g., Softmax, GeLU, etc). Compared to other alternatives \cite{cryptoeprint:2024/136NEXUS,hussain2021coinn}, the hybrid protocol effectively combines the strengths of both cryptographic primitives: HE reduces communication cost for linear layers, while MPC ensures high computation accuracy for nonlinear layers \cite{huang2022cheetah,lu2023bumblebee,pang2023bolt}. Therefore, in this paper, \textbf{we focus on the hybrid HE/MPC protocol.}

% These hybrid HE/MPC protocols achieve comparable accuracy to plaintext inference while reducing communication and latency significantly~\cite{hao2022iron,pang2023bolt,lu2023bumblebee}.
% \ml{These references are weird.}\xts{done}

% As illustrated in Figure~\ref{fig:intro_profile}(a), the hybrid HE/MPC protocol separates the evaluation of linear and nonlinear layers: HE is applied to compute linear layers (e.g., matrix multiplications), while MPC

\begin{figure*}[!tb]
    \centering
    \includegraphics[width=1.0\linewidth]{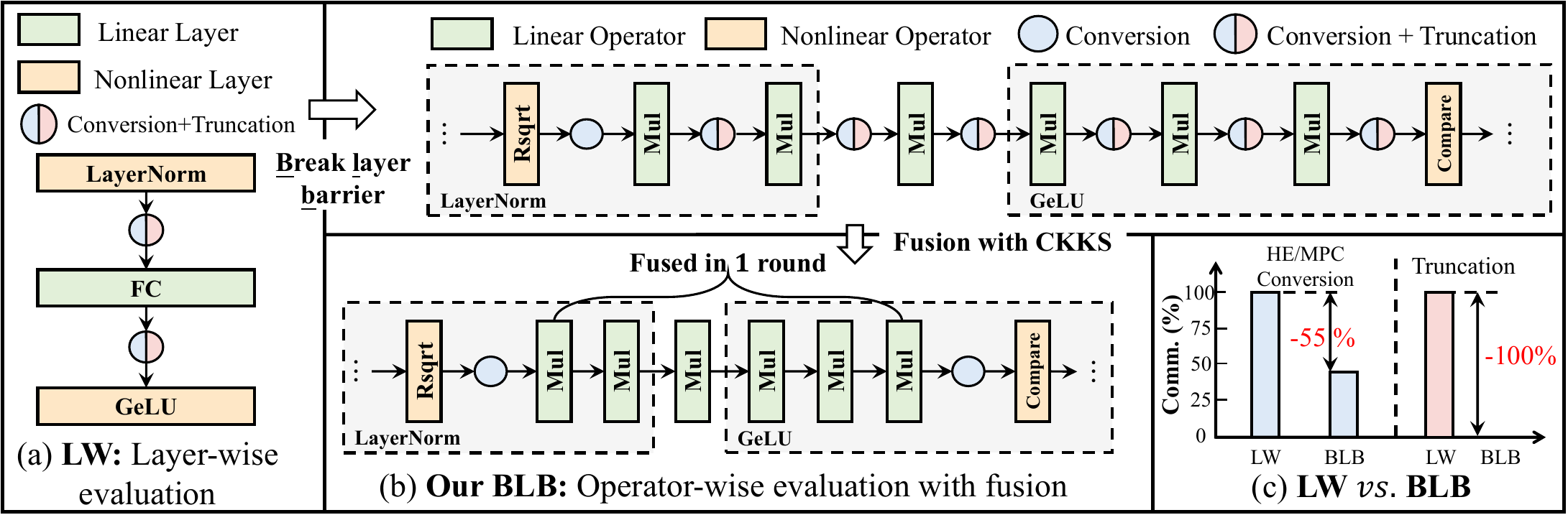}
    \caption{(a) Previous layer-wise (LW) paradigm; (b) Our proposed BLB paradigm; (c) BLB reduces communication costs by 55\% for HE/MPC conversion and eliminates all of the truncation communication, compared to~\cite{pang2023bolt,lu2023bumblebee}.}
    \label{fig:intro1}
    \vspace{-2pt}
\end{figure*}
% These frameworks generally fall into two categories: (1) Fully Homomorphic Encryption (FHE) schemes~\cite{} and (2) hybrid HE/Secure Multi-party Computation (MPC) schemes~\cite{}. The FHE scheme evaluates the entire neural network (NN) on encrypted data. However, since HE only supports linear operations, the FHE scheme requires high-order polynomial approximations for nonlinear functions, resulting in significant accuracy degradation. FHE schemes require a fine-tuning process to mitigate this, which is impractical for large language models (LLMs) due to their immense computational and data demands~\cite{}.
% In contrast, hybrid HE/MPC schemes leverage HE and MPC protocols to evaluate linear and nonlinear layers separately, achieving substantial improvements in both performance and accuracy. This approach combines the strengths of HE and MPC: HE provides low communication costs for linear layers, while MPC offers high accuracy for nonlinear layers. 

However, existing hybrid HE/MPC protocols still face significant communication overhead when applied to Transformer models. For example, BOLT~\cite{pang2023bolt} requires 59.61 GB of communication to perform a single inference on a BERT-base model. Bumblebee~\cite{lu2023bumblebee} introduces HE into nonlinear layer computation to reduce communication, but still generates \delete{16.37}\revise{15.5} GB communication for a BERT-large model. This high communication overhead presents a critical bottleneck, limiting their applicability to larger models. We profile the communication breakdown of Bumblebee in Figure~\ref{fig:intro_profile} (b) and identify two primary factors contributing to the substantial communication costs:

% \begin{enumerate}[9]
    % \item[\ding{182}]
\ding{182} \textbf{Excessive Truncations.} Hybrid HE/MPC protocols often rely on fixed-point arithmetic. A real number $x$ is represented as an integer $\lfloor x \cdot 2^s\rfloor$ with a \textit{scale} $s$, where $\lfloor \rfloor$ denotes round down. Multiplication of two fixed-point numbers $x$ and $y$ produces $\lfloor xy \cdot 2^{2s}\rfloor$ with scale $2s$. To restore the scale to $s$, the result must be right-shifted (truncated) by $s$ bits through an MPC-based truncation protocol~\cite{rathee2021sirnn}. As truncations are needed after each linear multiplication and require communication between the user and server, the overall communication cost becomes excessive.
% requiring an MPC truncation protocol~\cite{rathee2021sirnn}. Each truncation involves communication, and the high frequency of truncations significantly increases overall communication complexity. \ml{Need to explain what $\lfloor \rfloor$ means.}\xts{done}
    
% \item[\ding{183}]
% \xts{beaver triple, consecutive multiplication needs multiple rounds, communication is huge, introduce hybrid HE/MPC scheme in detail, what is a hybrid? How it works?}
\ding{183} \textbf{Frequent Conversions Between HE and MPC.}
As truncations and nonlinear layers are evaluated using MPC protocols while linear layers are computed with HE, conversions between HE and MPC are frequently required. The HE/MPC conversion involves transferring HE ciphertexts and invoking MPC protocols for bit width, scale, and field adjustment \cite{pang2023bolt}, leading to high communication overhead. As in Figure~\ref{fig:intro_profile} (b), truncations and HE/MPC conversions together account for over 80\% of Bumblebee's total communication.

% Truncations and other nonlinear functions are evaluated using MPC protocols, leading to frequent conversions between HE and MPC. The HE/MPC conversion protocol requires transferring multiple HE ciphertexts and executing MPC protocols to adjust bit width and scale, resulting in significant communication overhead. As a result, truncations and conversions together account for over 80\% % of the total communication costs in Bumblebee~\cite{lu2023bumblebee}.
% \ml{Better define the HE and MPC conversion process here.}\xts{done}
% \end{enumerate}

To address these problems, previous hybrid frameworks have attempted to fuse adjacent linear layers to reduce the communication overhead~\cite{xu2024privcirnet,pang2023bolt,zhang2024individual,chen2024accelerating}. \textbf{Here, linear layer fusion refers to using HE to evaluate consecutive linear layers within a single communication round,} thereby eliminating the conversion and truncation protocols required between the adjacent linear layers. For example, in convolutional neural networks (CNNs), PrivCirNet~\cite{xu2024privcirnet} fuses the convolution and batch normalization layers as well as the two following convolution layers in MobileNetV2. For Transformers, BOLT~\cite{pang2023bolt} fuses two consecutive matrix multiplications (MatMuls) in the self-attention layer. However, these methods often show insufficient communication reduction due to the following three limitations:
\begin{figure}[!tb]
    \centering
    \includegraphics[width=1.0\linewidth]{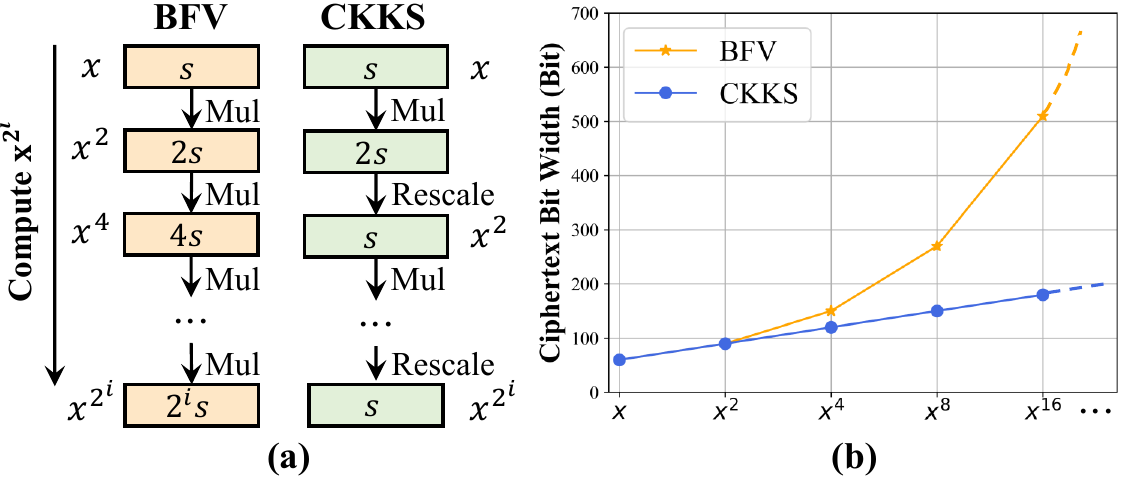}
    \caption{(a) CKKS can maintain the operand scale at $s$ with rescale, while BFV needs exponential scale growth when computing $x^{2^i}$; (b) The ciphertext bit width change for the BFV and CKKS schemes during computation.}
    \label{fig:intro2}
    \vspace{-3pt}
\end{figure}
% \ml{No need to use ``enumerate'' here?}
% \begin{enumerate}[9]

\ding{182}\textbf{ Limited Flexibility.} As shown in Figure~\ref{fig:intro1} (a), current frameworks follow a layer-wise (LW) evaluation paradigm, where they design protocols for each layer and evaluate a model layer by layer. This coarse-grained fusion across neighboring layers limits the fusion patterns and their applicability to Transformer models. For instance, only two consecutive MatMuls in a self-attention layer can be fused in one Transformer block~\cite{pang2023bolt}, resulting in limited communication reduction.
% missing many additional optimization opportunities.

% This leads to a coarse-grained fusion across neighboring layers with limited fusion patterns. For instance, only two consecutive matrix multiplications in the self-attention layer can be fused in the whole Transformer model~\cite{pang2023bolt}, missing many additional optimization opportunities. 

% \ml{Bit width or bit width? Unify.}\xts{done, ``bit width''}

\ding{183}\textbf{ High Ciphertext Bit Width.} SOTA hybrid HE/MPC protocols often use the Brakerski-Fan-Vercauteren (BFV) HE scheme for linear layers \cite{huang2022cheetah,hao2022iron,pang2023bolt,lu2023bumblebee}. However, the BFV scheme suffers from a significant expansion in operand scale when computing fused multiplications.
% and plaintext bit width with the multiplicative depth when computing fused linear layers,
% \ml{Still a bit con}
As shown in Figure~\ref{fig:intro2}, when repeatedly squaring an input \( x \) with an initial scale of \( s \), i.e., computing \( x^{{2^i}} \), the BFV scheme exhibits an exponential growth in scale, as the scale doubles after each square computation. Consequently, the bit widths of plaintexts and ciphertexts have to increase proportionally to maintain computation correctness~\cite{cheon2017homomorphic}. The high ciphertext bit width increases both the computation and communication costs, limiting the benefits of linear layer fusion.

% Previous frameworks utilize the Brakerski-Fan-Vercauteren (BFV) HE scheme to compute linear layers. However, the BFV scheme is not good at consecutive multiplications, as the operand scale expands exponentially with multiplication depth. As illustrated in Figure~\ref{fig:intro2}, in the BFV scheme, the lack of truncation causes the scale of operands to double after each multiplication. Consequently, the plaintext bit width grows exponentially, requiring ciphertext bit width to increase proportionally to maintain computation correctness~\cite{cheon2017homomorphic}. This large ciphertext bit width significantly raises both communication and computation costs, negating the benefits of fusion. \ml{Why does the ciphertext bit width need to increase exponentially? It's not very clear.}\xts{done}

% \ml{Define MatMul abbreviation when matrix multiplication appears the first time.}

\ding{184}\textbf{ Suboptimal MatMul Protocol.} As Transformer models process high-dimensional tensors while HE computes over one-dimensional vectors, the mapping from tensors to vectors, denoted as packing, directly impacts the HE computation complexity. Linear layer fusion imposes more restrictions on the packing algorithm as adjusting the packing of intermediate ciphertexts in fused linear layers involves complex homomorphic operations and incurs high computation complexity \cite{pang2023bolt}. In contrast, without layer fusion, it is possible to optimize the packing for each linear layer separately. As a result, BOLT~\cite{pang2023bolt} requires approximately $20\times$ more computationally-intensive HE rotations for fused MatMul compared to the unfused protocol~\cite{lu2023bumblebee}. While PowerFormer~\cite{park2024powerformer} proposes further packing optimization, it still incurs high costs for non-square matrices. % These suboptimal packing strategies significantly increase computational overhead, offsetting the advantages of layer fusion.

% \ding{184}\textbf{ Suboptimal Matrix Multiplication (MatMul) Protocol.} Layer fusion imposes stricter requirements on HE packing algorithms, where two-dimensional matrices in Transformer models must be encoded into one-dimensional vectors. Adjusting the packing of intermediate ciphertexts during consecutive linear layer computations is challenging due to the restriction to homomorphic operations. Consequently, the existing solution BOLT~\cite{pang2023bolt} requires approximately $10\times$ more HE rotations for $QK^T$ compared to non-fused approaches~\cite{hao2022iron}. \ml{compared to [clarify comparison, e.g., non-fused approaches or baseline methods]}\xts{done}. While subsequent work such as PowerFormer~\cite{park2024powerformer} introduces optimizations, it still incurs high rotation costs for non-square matrices. These suboptimal packing strategies significantly increase computational overhead, offsetting the advantages of layer fusion.
% \end{enumerate}
\subsection{Technical Details}
To overcome the limitations of existing solutions, we propose the Breaking the Layer Barrier framework, dubbed BLB, which features three key techniques designed to resolve the above three limitations:

% \ml{Nonlinear layers and nonlinear operators, still confusing}

% \begin{enumerate}[9]
\ding{182} \textbf{Breaking the Layer Barrier.} Nonlinear layers in Transformer models, e.g., GeLU, Layer Normalization (LayerNorm), Softmax, etc, are very complex and incur high communication costs when directly evaluating with MPC protocols \cite{rathee2021sirnn,hao2022iron}. SOTA hybrid HE/MPC frameworks \cite{pang2023bolt,lu2023bumblebee} propose piecewise linear approximations to improve communication efficiency\footnote{A detailed review of SOTA protocols for nonlinear layers is provided in Appendix~\ref{app:nonlinear_protocols}.}. As a result, each nonlinear layer involves computing a series of linear operators, e.g., element-wise multiplications, providing new fusion opportunities.
% that can be fused.
This insight motivates us to break the layer barrier and refine fusion granularity to neighboring operators.
% from entire layers to individual operators.
For example, as in Figure~\ref{fig:intro1} (b), the last two operators of the LayerNorm layer, the subsequent fully connected (FC) layer, and the first three operators in the following GeLU layer are all linear operators. Once fused, all HE/MPC conversions and truncations between adjacent linear operators can be eliminated, significantly reducing communication costs.
% The BLB framework is based on an important observation that the nonlinear layers in Transformers are much more complex than those in CNNs. SOTA hybrid HE/MPC frameworks~\cite{pang2023bolt,lu2023bumblebee,dong2023puma,zeng2024securegpt} introduce partial linear approximations for these nonlinear layers, such as exponentiation and GeLU\footnote{A detailed review of SOTA protocols for nonlinear layers is provided in Appendix~\ref{app:nonlinear_protocols}}. Therefore, these nonlinear layers contain many linear operators that can be fused.

% \ml{Can be improved.}\xts{improved}
However, realizing such fusion is challenging due to the diverse packing algorithms employed for different linear operators~\cite{pang2023bolt,cryptoeprint:2024/136NEXUS,park2024powerformer}. For example, MatMuls and element-wise multiplications often have different preferences for packing algorithms. Ensuring the computation correctness and efficiency of fused linear operators remains an open question. We propose the \fusion~paradigm to tackle the challenge systematically. We classify the operators into different categories so that operators of the same category can share a common packing algorithm. This enables us to define fusion patterns directly for different categories and design optimized packing algorithms for each fusion pattern. Through \fusion, neighboring linear operators following the defined fusion patterns can be correctly fused, significantly reducing the communication overhead.

% Ensuring the correctness of fused computations across these linear operators remains an unsolved problem. We propose the FineGrainFusion paradigm, which systematically tackles this challenge.
% Ensuring the correctness of fused computations requires addressing the conversion of different ciphertext packing methods during computation, which remains an unsolved problem.
% \ml{what are packing structures?}\xts{improved}
% \ml{While XXX, systematically analyzing XXX remains challenging.}
% First, we categorize operators, allowing operators of the same type to share a common packing method. This enables us to develop fusion patterns for various type combinations and design specialized packing algorithms for each pattern, ensuring correctness.
% With FineGrainFusion, we can fuse adjacent linear operators that match these patterns across the entire neural network (NN), thereby significantly reducing communication overhead.
% This approach enables the fusion of all adjacent linear operators that meet these patterns across the entire neural network (NN). Then, the HE-based evaluation of the fused operators can be performed in a single communication round, significantly reducing communication overhead.

\ding{183} \textbf{Hybrid CKKS and MPC Framework with Secure Conversion.} For fused linear operators, we argue that the Cheon-Kim-Kim-Song (CKKS) HE scheme is a superior choice as it can mitigate the high ciphertext bit width limitation. This is because CKKS supports an important \textbf{rescale} operation that enables to reduce the operand scale back to $s$.
% To address the high ciphertext bit width limitation, we argue that the Cheon-Kim-Kim-Song (CKKS) HE scheme is a superior choice for consecutive linear operator computations.
% functions similarly to truncation, maintaining the operand scale at $s$.
As illustrated in Figure~\ref{fig:intro2} (a), the rescale operation can restore the operand scale from $2s$ to $s$ after multiplication, enabling a controlled, linear increase of ciphertext bit width (Figure~\ref{fig:intro2} (b)). However, the existing CKKS and MPC conversion protocol proposed in MP2ML~\cite{boemer2020mp2ml} has severe security problems and may leak the information of computation results. Specifically, MP2ML directly adapts the BFV-MPC conversion protocol to CKKS and MPC but neglects that the CKKS encoding relies on a fast Fourier transform (FFT), which produces a narrowly distributed output. As a result, the sampled noise in MP2ML cannot effectively obscure the computation results, leading to high privacy risks. We carefully analyze the CKKS mechanics and develop the first secure and efficient conversion protocol for CKKS and MPC. Through the fine-grained operator fusion and CKKS-based HE computation, we can reduce the communication costs for HE/MPC conversion by 55\% and eliminate all truncations, compared to BOLT and Bumblebee~\cite{pang2023bolt,lu2023bumblebee}, as shown in Figure~\ref{fig:intro1} (c).

\ding{184} \textbf{Rotation-Efficient Fused MatMul Protocol.} 
Linear operator fusion poses extra constraints on the packing of intermediate ciphertexts. For fused MatMuls, it results in more homomorphic operations, e.g., HE rotations~\cite{pang2023bolt,park2024powerformer}. To reduce the computation costs, we design a rotation-efficient ciphertext-ciphertext MatMul protocol. Besides, we observe that the multi-head attention computation in Transformers naturally leads to batched MatMuls, which enables packing multiple batches into a ciphertext to reduce HE rotations. We also apply the baby-step-giant-step (BSGS) optimization \cite{pang2023bolt,ju2023neujeans} to further reduce HE rotations. Compared to BOLT~\cite{pang2023bolt} and PowerFormer~\cite{park2024powerformer}, our protocol achieves a $29\times$ and $8\times$ reduction in HE rotations, respectively.

\subsection{Contributions}
Our contributions can be summarized as follows:
% \ml{Make each item a sentence.}
\begin{enumerate}[9]
    \item[\ding{182}] We observe that truncations and HE/MPC conversions account for major communication overhead and propose a novel framework, dubbed~\method, that breaks the layer barrier to enable fine-grained linear operator fusion for communication-efficient private Transformer inference.
    % that features fine-grained fusion of linear operators for communication-efficient private Transformer inference.
    % A new private inference framework~\method~using operator-wise evaluation with fusion offers new insights into the design of the private inference protocol.
    \item[\ding{183}] \method~proposes the first secure CKKS and MPC conversion protocol to enable CKKS-based computation of fused linear operators and reduce the ciphertext bit width increase.
    % We propose the first secure hybrid CKKS and MPC framework with secure and efficient conversion protocols to reduce the ciphertext bit widths to improve communication and computation efficiency for fused linear operators.
    % The first secure hybrid CKKS and MPC framework with secure and efficient conversion protocols.
    \item[\ding{184}] We propose a rotation-efficient MatMul protocol for fused computations, reducing $8\sim 29\times$ HE rotations compared with prior-arts~\cite{pang2023bolt,park2024powerformer}.
    % Compact packing algorithms for ciphertext-ciphertext matrix multiplication, reducing $50\sim 90$\% HE rotations compared with BOLT~\cite{pang2023bolt}.
    \item[\ding{185}] We evaluate BLB on various Transformer models, including BERT-base, BERT-large, and GPT2-base. With extensive experiments, we demonstrate $21\times$ and $2\times$ communication reduction compared to BOLT~\cite{pang2023bolt} and Bumblebee~\cite{lu2023bumblebee}, respectively. With GPU-based HE computation, BLB achieves \delete{$14\times$ and $2\times$}\revise{$13\times$ and $1.8\times$} latency reduction compared to BOLT~\cite{pang2023bolt} and Bumblebee~\cite{lu2023bumblebee}, respectively.
    % Significant improvements in both communication and latency. We evaluate BLB on various Transformer models, including BERT-base, BERT-large, and GPT2-base, showing a $21\times$ reduction (or $-95\%$) in communication compared to BOLT~\cite{pang2023bolt} and a $2\times$ reduction compared to Bumblebee~\cite{lu2023bumblebee}. With GPU acceleration for HE computation, BLB achieves $17\times$ and $2\times$ latency reduction compared to BOLT~\cite{pang2023bolt} and Bumblebee~\cite{lu2023bumblebee}, respectively.
\end{enumerate}

\subsection{Comparison with FHE Frameworks}

% \ml{Add more references to FHE frameworks.}\xts{done}
Another important variant of the private inference framework is based on Fully Homomorphic Encryption (FHE)~\cite{lee2022low,kim2023optimized_fhe,ju2023neujeans,cryptoeprint:2024/136NEXUS,park2024powerformer,cryptoeprint:2024/1881THOR,chen2024secure}. As FHE-based frameworks evaluate the entire neural network (NN) on encrypted data in a non-interactive manner, their communication costs are usually lower compared to the hybrid HE/MPC frameworks. However, FHE-based frameworks still struggle with accurately evaluating complex nonlinear layers, e.g., Softmax, GeLU, etc, in Transformer models \cite{cryptoeprint:2024/136NEXUS,chen2024secure}. For example, using the open-source FHE framework NEXUS~\cite{cryptoeprint:2024/136NEXUS}, we find the nonlinear layer approximation exhibits a maximum relative error of 297\%, which significantly impacts the overall inference accuracy and requires extensive fine-tuning. In contrast, hybrid HE/MPC frameworks \cite{pang2023bolt,lu2023bumblebee} often exhibit much better inference accuracy at the cost of higher communication costs.  Therefore, in this paper, we focus on the hybrid HE/MPC protocol and propose \method~to improve its communication efficiency.

\section{Preliminaries}\label{sec:pre}

% \ml{I do not see a related work section.}
% \xts{do we need an intro to what HE is packing?}
\subsection{Notations}
\begin{table}[!tb]
    \centering
    \caption{Notations used in the paper.}
    \label{tab:notation}
    \resizebox{0.89\linewidth}{!}{
    \begin{tabular}{c|c}
    \toprule
    Notations & Meanings \\
    \midrule
    $\mathbb{C,R,Z}$ & the sets of complex, real and integral numbers \\
    \midrule
    $\lceil \cdot \rceil$,$\lfloor\cdot \rfloor$,$\lfloor\cdot \rceil$ & ceiling, flooring, and rounding operations \\
    % \midrule
    % $\lceil x \rceil_q$,$\lfloor x \rfloor_q$,$\lfloor x \rceil_q$ & $\lceil x \rceil$ mod $q$, $\lfloor x \rfloor$ mod $q$, $\lfloor x \rceil$ mod $q$\\
    \midrule
    $\mathbb{Z}_q$ & $\mathbb{Z}_q=\mathbb{Z}\cap [-\lceil q/2 \rceil ,\lceil q/2 \rceil]$\\
    \midrule
    $[n]$ & $\{0,..,n-1\}$ for a non-negative integer $n$\\
    \midrule
    $N$ & polynomials degree \\
    \midrule
    $\mathbb{A}_{N,q}$ & \makecell{the set of integer polynomials \\ $\mathbb{A}_{N}=\mathbb{Z}_q[x]/(x^N+1)$}\\
    \midrule
    $\odot$ & element-wise multiplication \\
    \midrule
    $q$  & ciphertext modulus, $q$ is a prime\\
    % \hline
    % $\boxplus,\boxminus,\boxtimes $  & \makecell{Homomorphic addition, subtraction and multiplication} \\
    \bottomrule
    \end{tabular}
    \vspace{-7pt}
    }
\end{table}
Table~\ref{tab:notation} summarizes the notations used in this paper. We represent vectors with lower-case letters (e.g., $x$) and matrices with upper-case letters (e.g., $X$). We use lower-case letters with a ``hat'' symbol (e.g., $\hat{x}$) to represent a polynomial. We use $\mbm{1}\{\mathcal{P}\}$ to denote the indicator function, which is 1 when $\mathcal{P}$ is true and 0 otherwise.
% \xts{need double check here when packing is ok.}

\subsection{Transformer Model}\label{subsec:pre_transformer}
\begin{figure}[!tb]
    \centering
    \includegraphics[width=0.73\linewidth]{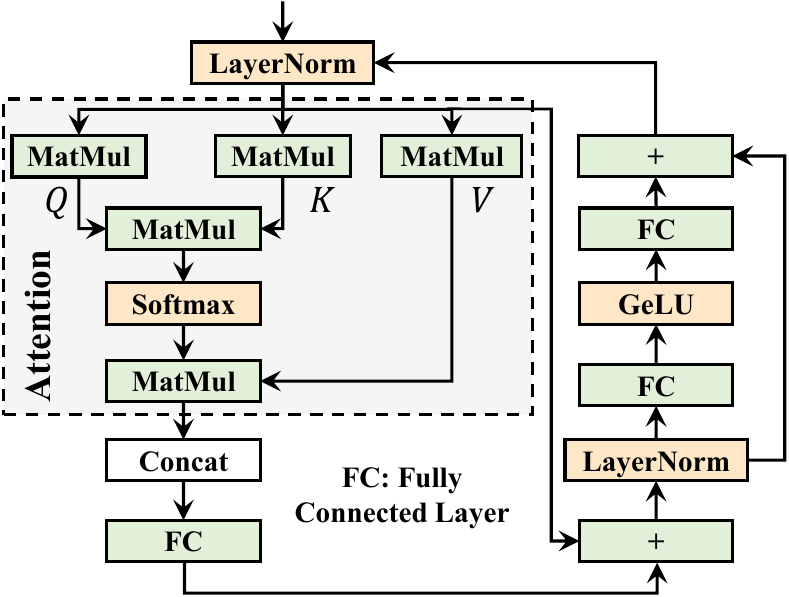}
    \caption{A Transformer block in BERT~\cite{kenton2019bert}.}
    \label{fig:transformer_block}
    \vspace{-7pt}
\end{figure}
Figure~\ref{fig:transformer_block} shows the structure of a classical Transformer block, which is used in BERT~\cite{kenton2019bert} models. The multi-head attention is the key component in the Transformer block. Concretely, an input $X\in \mathbb{R}^{m\times d}$ is multiplied with three weight matrices to produce a query matrix $Q=XW_Q$, a key matrix $K=XW_K$, and a value matrix $V=XW_V$, where $W_Q,W_K,W_V\in \mathbb{R}^{d\times d}$, and $m,d$ are the sequence length and the hidden dimension, respectively. $Q,K,V$ are then divided into H heads: $Q_h,K_h,V_h\in \mathbb{R}^{m\times \frac{d}{H}}$, where $h\in [H]$, and $H$ is the number of heads. Then, the multi-head attention is computed by:
% \xts{what needs to be introduced depends on the packing section}\ml{Need to emphasize multi-head attention here}\xts{improved}:
\vspace{-0.5em}
\begin{equation}\label{eq:attention}
    \operatorname{Att}_h=\operatorname{Softmax}(\frac{Q_hK_h^T}{\sqrt{d/H}})V_h
\end{equation}
% \vspace{-0.5em}
The $H$ resulting matrices $\operatorname{Att}_h$ are concatenated and projected by $W_O$, i.e., $\operatorname{Concat}(\operatorname{Att}_h)W_O$.
\subsection{Threat Model}\label{subsec:threat_model}
\method~works in a general private inference scenario that involves two parties, i.e., server and client. The server holds the proprietary NN model, and the client owns private input~\cite{Juvekar_Vaikuntanathan_gazelle_2018,Mishra_Delphi_2020,huang2022cheetah,hao2022iron,pang2023bolt,lu2023bumblebee}.~\method~enables the client to obtain the inference results while keeping the server's model weights and the client's input private. Consistent with previous works~\cite{Juvekar_Vaikuntanathan_gazelle_2018,rathee2020cryptflow2,rathee2021sirnn, huang2022cheetah, pang2023bolt,lu2023bumblebee},~\method~assumes the NN architecture is known to both sides and adopts an \textit{honest-but-curious} security model in which both parties follow the specification of the protocol but also try to learn more than allowed. More security discussion is provided in Appendix~\ref{app:security}.
% \xts{todo: add malicious adversary}

% \subsection{Fixed-point representation}

% \ml{Why is this section necessary?}

% CKKS HE and MPC typically operate on fixed-point numbers on which an absolute value $x$ is converted to $\lfloor x\cdot 2^s \rfloor \mod 2^l$, where $l,s$ are the bit width and scale. 

% \textbf{Fixed-point addition.} Two fixed point numbers with the same scale can be added by simply adding the two numbers.

% \textbf{Fixed-point multiplication.} Multiplication will increase the scale of the result by the sum of the scales of the two operands. To avoid the scale growth, there exists a truncation after the multiplication, i.e., $\lfloor x\cdot 2^s \rfloor \cdot \lfloor y\cdot 2^s \rfloor / 2^s \mod 2^l$.

\subsection{Cryptographic Primitives}\label{subsec:pre_crpto}
% \ml{The introduction here is too complex and redundant.}
\textbf{CKKS HE Scheme.}
We now present the key implementation characteristics of CKKS. Full details can be found in~\cite{cheon2017homomorphic,cheon2019full}.
\begin{enumerate}[9]
    \item[$\bullet$] \textbf{CKKS Plaintext and Ciphertext.} In CKKS, a plaintext $m$ is represented as a vector of $N/2$ fixed-point numbers. The scale parameter $s$ determines the width of the fractional part of each element. 
    After encryption, each ciphertext $\operatorname{ct}\in \mathbb{A}^2_{N, q}$ is a pair of polynomials. Each polynomial has $N$ integer coefficients modulo a large prime $q$.
    % \item[$\bullet$] \textbf{HE Packing Algorithm.} The HE packing algorithm encodes two-dimensional matrices in Transformer models into one-dimensional plaintexts. This is necessary because the CKKS scheme~\cite{} only supports Single Instruction Multiple Data (SIMD) operations on one-dimensional plaintexts. HE Packing directly decides the efficiency of the HE inference.
    \item[$\bullet$] \textbf{Rescaling.} A rescale operation reduces the ciphertext modulus $q$ \newtext{by a factor $d$} and the scale by an additive factor factor $\log_2 d$. Multiplying two ciphertexts with scale $s$ produces a result with scale \delete{$s^2$}\newtext{$2s$}. By rescaling with $\log_2 d \approx s$ after multiplication, the scale is approximately reset to $s$.
    \item[$\bullet$] \textbf{Homomorphic Addition ($\boxplus$) and Multiplication ($\boxtimes$).} Given two ciphertexts or a ciphertext and a plaintext, the operations $\boxplus$ and $\boxtimes$ produce encryptions of their element-wise sum and product, respectively.
    \item[$\bullet$] \textbf{Rotation.} The left-rotation \delete{$\operatorname{Rot}_l^s(m)$}\newtext{$\operatorname{Rot}_l^t(m)$} left-rotates the vector $m$ by \delete{$s$}\newtext{$t$} slots. The right-rotation \delete{$\operatorname{Rot}_r^s(m)$}\newtext{$\operatorname{Rot}_r^t(m)$} right-rotates the vector $m$ by \delete{$s$}\newtext{$t$} slots. By default, we use left rotation.
\end{enumerate}

\noindent\textbf{Additive Secret Share.}
We use a 2-out-of-2 additive secret share to keep the input data private throughout inference. We denote two parties by $P_0$ and $P_1$, where $P_0$ is the client and $P_1$ is the server. We use $[\![ x ]\!]^{M}$ to denote an additive share of $x$ over $\mathbb{Z}_M$. We write $[\![ x ]\!]^{M}=([\![ x ]\!]^{M}_0,[\![ x ]\!]^{M}_1)$ where $P_0$ holds $[\![ x ]\!]^{M}_0$ and $P_1$ holds $[\![ x ]\!]^{M}_1$, such that $[\![ x ]\!]^{M}_0+[\![ x ]\!]^{M}_1=x\bmod M$. In particular, we use $[\![ x ]\!]^B$ to represent bool sharing where $x$ is a 1-bit value. 
% We also use $[\![ \cdot;2^s ]\!]$ to explicitly denote the the shared value with scale $s$.
% These sharings are chosen randomly, for instance, by first choosing $[\![ x ]\!]^{q}_0$ uniformly at random from $\mathbb{Z}_q$, and then assigning $[\![ x ]\!]^{q}_1=x-[\![ x ]\!]^{q}_0 \mod q$. 
Moreover, we designate the shares over $M=2^l$ as ring shares and those as prime shares when $M=q$ is a prime. \newtext{Besides, the MPC protocols used in this paper are based on Oblivious Transfer~\cite{rathee2021sirnn,pang2023bolt,lu2023bumblebee}.}

% \textbf{Basic MPC Protocol.} This paper involves an extension protocol $\Pi_{\mathrm{Ext}}^{M,2^l}$~\cite{rathee2021sirnn} that extends $[\![ x ]\!]^M$ to $[\![ x ]\!]^{2^l}$, where $2^l>M$.
% \ml{Use a table to summarize the basic MPC protocols below.}
% \begin{enumerate}[9]
%     \item[$\bullet$] .
%     \item[$\bullet$] $\Pi_{\mathrm{Reduce}}^{{l_1},{l_2}}$. Reduce $[\![ x ]\!]^{2^{l_1}}$ to $[\![ x ]\!]^{2^{l_2}}$, where $l_1>l_2$.
% \end{enumerate} 

% \subsection{Hybrid HE/MPC Inference}
% \subsection{Prior-art nonlinear layer protocols for Transformers}
% \ml{For the method section, do we need an overview?}

% \ml{Need to use Theorem to formalize our theories and use Remarks to  highlight our findings and conclusions.}

\section{BLB Framework Overview}
\label{sec:overview}
% \xts{need to improve at last!}

\begin{figure}[!tb]
    \centering
    \includegraphics[width=0.95\linewidth]{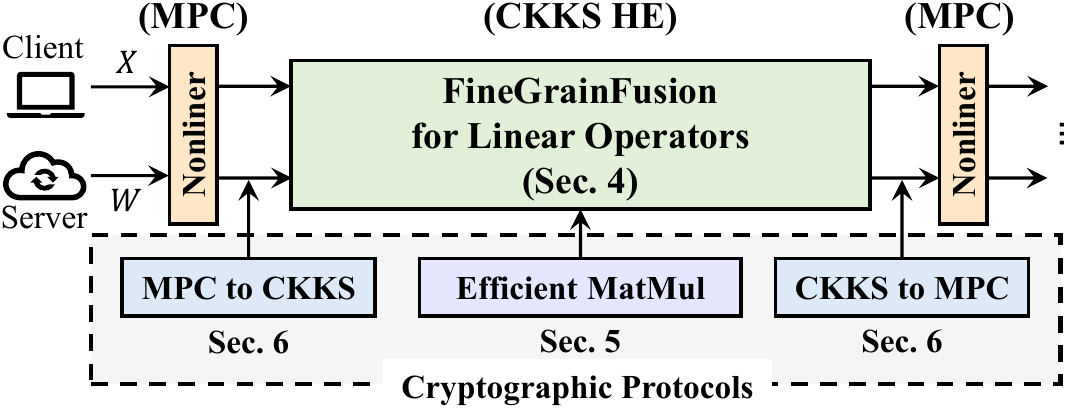}
    \caption{Overview of~\method. 
    % \ml{Overview figure can be improved.}
    }
    \label{fig:overview}
\end{figure}

Figure~\ref{fig:overview} provides an overview of the~\method~framework, which integrates MPC and CKKS HE schemes to evaluate linear and nonlinear operators, respectively. \method~features three key components: firstly, \method~proposes FineGrainFusion paradigm to systematically explore the fusion patterns of diverse linear operators in the Transformer models, as explained in \textsection~\ref{sec:FGF}; secondly, to accommodate the packing requirements of fused MatMuls, efficient MatMul protocols are proposed in \textsection~\ref{sec:encode} to reduce the required HE rotations drastically; thirdly, to overcome the significant increase of ciphertext bit widths after fusion, \textsection~\ref{sec:conversion} proposes the first secure conversion protocol between CKKS and MPC, enabling secure yet efficient transitions between linear and nonlinear operators.

% We introduce the FineGrainFusion paradigm in \textsection~\ref{sec:FGF}, which systematically explores fusion patterns of linear operators in Transformer models. To accommodate the packing requirements of FineGrainFusion, a series of efficient MatMul protocols are proposed in \textsection~\ref{sec:encode}. Finally, \textsection~\ref{sec:conversion} presents the first secure conversion protocol between CKKS and MPC, enabling transitions between linear and nonlinear operators.

\section{\fusion~Paradigm}
\label{sec:FGF}

% \ml{When do we use ``MatMul'' instead of matrix multiplication?}\xts{I change all matrix multiplication to MatMul in Sec.5/6}

As diverse linear operators exist in Transformer models, we propose the \fusion~paradigm in this section to analyze the fusion patterns systematically. We begin by classifying operators into different categories and then propose a formal analysis of fusion patterns for linear operators of different categories. We also optimize the HE packing algorithms based on the categorization.

\subsection{Operator Categorization}
\label{subsec:operators}

\begin{table}[!tb]
    \centering
    \huge
    \caption{Definition of operators. $\operatorname{ct-ct}$ means the two inputs are both ciphertexts, and $\operatorname{ct-pt}$ means one input is ciphertext and the other is plaintext. 
    % \ml{Should we add a figure of the Transformer block in the preliminary and point out different types of operators in that figure?}\xts{that figure will be huge, fig.9, in pre, we haven't defined these operators yet.}
    }
    \label{tab:opertors}
    \resizebox{1.0\linewidth}{!}{
    \begin{tabular}{c|c|c}
    \toprule
    \textbf{Type} & \textbf{Name} & \textbf{Description} \\
    \midrule
    \rowcolor{gray!20}
    \multicolumn{3}{c}{\textit{Linear operators}} \\
    \midrule
    \multirow{2}{*}{Identity}&$\operatorname{ewadd_{cc}, ewadd_{cp}}$&Element-wise addition of $\operatorname{ct-ct}, \operatorname{ct-pt}$ \\
    \cmidrule{2-3}
    &$\operatorname{ewmul_{cc}, ewmul_{cp}}$&Element-wise multiplication of $\operatorname{ct-ct}, \operatorname{ct-pt}$ \\
    \midrule
    \multirow{2}{*}{Expansion}&$\operatorname{sadd_{cc}, sadd_{cp}}$&Scalar addition of $\operatorname{ct-ct}, \operatorname{ct-pt}$ \\
    \cmidrule{2-3}
    &$\operatorname{smul_{cc}, smul_{cp}}$&Scalar multiplication of $\operatorname{ct-ct}, \operatorname{ct-pt}$ \\
    \midrule
    Reduction&$\operatorname{sum}$&Sum in a specified dimension of the input\\
    \midrule
    Transformation&$\operatorname{matmul_{cc}, matmul_{cp}}$&MatMul of $\operatorname{ct-ct}, \operatorname{ct-pt}$ \\
    \midrule
    \rowcolor{gray!20}
    \multicolumn{3}{c}{\textit{Nonlinear operators}} \\
    \midrule
    \multicolumn{2}{c|}{$\cmp$}& $[\![ \mbm{1}\{x<y \} ]\!]^B \leftarrow \operatorname{cmp}([\![ x ]\!]^M, [\![ y ]\!]^M)$\\
    \midrule
    \multicolumn{2}{c|}{$\operatorname{mux}$}& $[\![ b\cdot x ]\!]^M \leftarrow \operatorname{mux}([\![ b ]\!]^B, [\![ x ]\!]^M)$\\
    \midrule
    \multicolumn{2}{c|}{$\operatorname{rec, rsqrt}$}&Reciprocal, Reciprocal Sqrt~\cite{rathee2021sirnn}\\
    % \midrule
    % \multicolumn{2}{c|}{$\operatorname{softmax}$}&Softmax function\\
    \bottomrule
    \end{tabular}
    }
\end{table}

% \ml{``operations'' or ``operators''?}\xts{must be operators}

As shown in Table~\ref{tab:opertors}, we classify the operators in a Transformer model into linear and nonlinear. Nonlinear operators mainly include comparison, multiplexer, reciprocal, and reciprocal square root. All nonlinear operators are evaluated with MPC protocols in the hybrid HE/MPC framework following~\cite{rathee2020cryptflow2,rathee2021sirnn}.

Linear operators in a Transformer model include element-wise addition and multiplication, scalar addition and multiplication, summation, and MatMuls between ciphertexts (denoted as ct-ct) or between ciphertexts and plaintexts (denoted as ct-pt). Note that it is often expensive to change the packing of ciphertexts as it involves costly homomorphic operations, e.g., multiplications and rotations~\cite{pang2023bolt}. In contrast, changing the packing of plaintexts incurs negligible overhead. 
% it is usually free to change the packing of plaintexts. 
% \ml{Why do we need the following sentence?}\xts{since scalar operation is not defined yet?}
Scalar addition and multiplication are implemented via broadcasting, where the scalar is expanded to match the dimensions of the target matrix, followed by element-wise operations.

To facilitate fusion analysis, we further classify the linear operators into four high-level abstract categories, including Identity, Expansion, Reduction, and Transformation, based on the dimension transformation between the operator's input and output. Table~\ref{tab:opertors} shows the details of the operator classification. Consider the inputs and intermediate outputs in a Transformer model as 2D tensors of dimensions $(L, D)$, where $L$ denotes the \textit{spatial} dimension along which computations are conducted in parallel and $D$ denotes the \textit{reduce} dimension along which aggregation is conducted. Then, the four categories can be defined as follows.
% \ml{Tensor or ciphertexts? very confusing here.}
\begin{enumerate}[9]
    \item[$\bullet$] Identity: The input and output dimensions are identical, with element-wise operators that do not alter the packing.
    \item[$\bullet$] Expansion: The dimensions change from $(L,1)$ to $(L,D)$, where $D$ is expanded, or broadcasted from $1$ to $D$. Scalar addition and multiplication are examples of this type.
    \item[$\bullet$] Reduction: The dimensions change from $(L,D)$ to $(L,1)$, where $D$ is aggregated to $1$, as in summation.
    \item[$\bullet$] Transformation: The dimensions change from $(L,D)$ to $(L,D')$, with the reduce dimension $D$ transformed to $D'$, as in MatMul.
\end{enumerate}
Operators in the Identity, Expansion, and Reduction categories can share a common packing method. However, for the Transformation type, $\operatorname{matmul_{cp}}$ and $\operatorname{matmul_{cc}}$ require separate packing algorithms and should be analyzed individually.

\subsection{FineGrainFusion Patterns}\label{subsec:fusion_patterns}

% \ml{In the previous section, how fusion may impact the packing algorithm is not yet explained.}\xts{improved above}
With operator categorization, we develop fusion patterns for various type combinations.
% Linear operator fusion challenges the HE packing algorithm, where we encode two-dimensional matrices into one-dimensional vectors.
To ensure both correctness and efficiency for each fusion pattern, we propose two additional requirements:  
\ding{182} \textbf{Consistent packing for each operator.} To simplify the packing algorithms after fusion, the input and output of all linear operators should have the same packing structure.  
\ding{183} \textbf{Tailored packing algorithms for each fusion pattern.} For adjacent operators eligible for fusion, the packing algorithm must be tailored based on the specific combination of operator types.
\begin{figure}[!tb]
    \centering
    \includegraphics[width=1.0\linewidth]{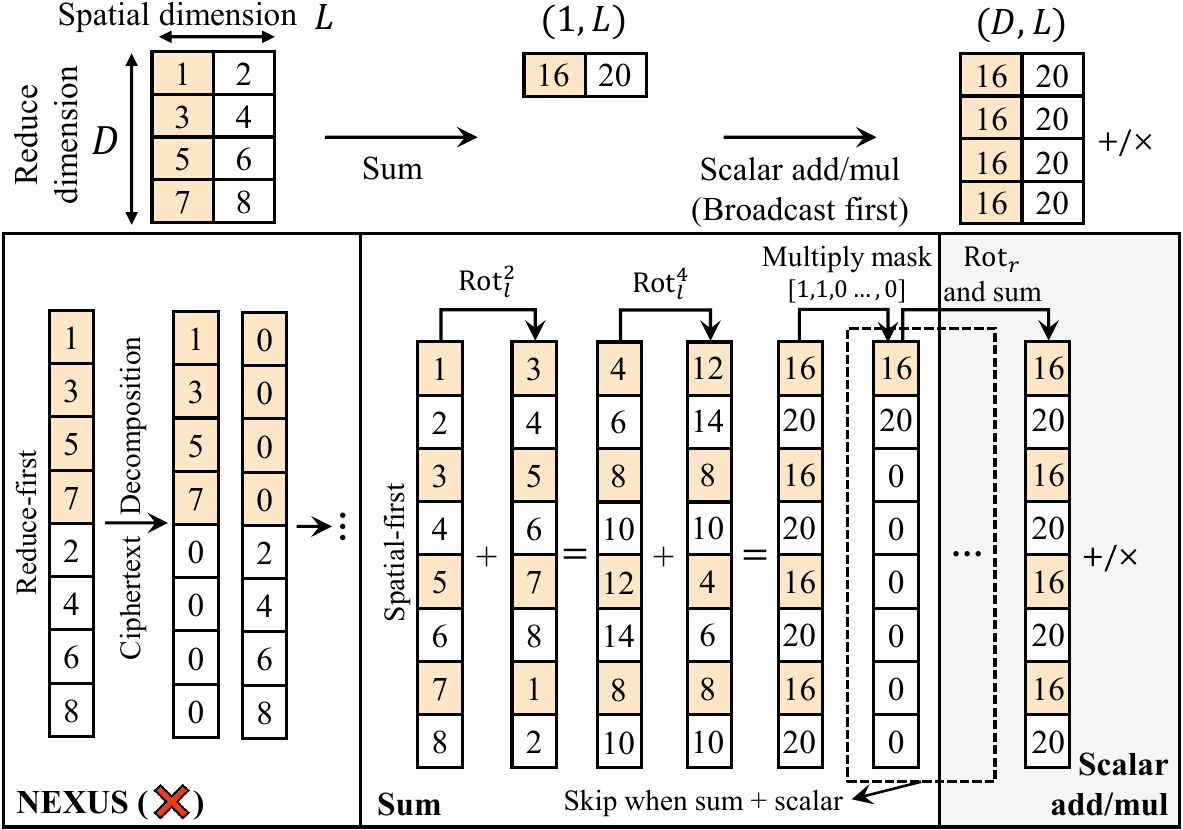}
    \caption{A toy example comparing NEXUS's packing algorithm with ours for summation and scalar operators.}
    \label{fig:encode1}
\end{figure}  
\begin{table}[!tb]
    \centering
    % \huge
    \caption{Complexity comparison of packing algorithms for the summation operator followed by the scalar operator.}
    \label{tab:packing_scalar}
    \resizebox{0.8\linewidth}{!}{
    \begin{tabular}{c|c|c|c}
    \toprule
    &\#Rot&\#Mul&Mul. Depth\\
    \midrule
    NEXUS~\cite{cryptoeprint:2024/136NEXUS} & $3L+L\log_2D$ & $2L+1$ & 3  \\
    \midrule
    \method~(w/o fusion) & $2\log_2D$ & 1 & 1 \\
    \midrule
    \method~(w/ fusion) & $\log_2D$ & 0 & 0 \\
    \bottomrule
    \end{tabular}
    }
\end{table}
\subsubsection{Consistent Packing for Single Operator}
Each operator must adhere to a consistent \textit{spatial-first} packing rule for input and output. In this context, the \textit{spatial-first} packing means the elements along the spatial dimension are placed consecutively before moving to the next row in the reduce dimension.
Identity operators naturally satisfy this rule. For Transformation operators, both $\operatorname{matmul_{cp}}$ and $\operatorname{matmul_{cc}}$ meet the rule following the protocols proposed in BOLT~\cite{pang2023bolt} and Jiang et al.~\cite{jiang2018secure}.
% Notably, $\operatorname{matmul_{cc}}$ demands additional HE rotations.

% if the operand to be broadcast or reduced is plaintext, they can be converted into element-wise operations. Otherwise, packing algorithms must be designed.
For Expansion and Reduction operators, 
existing packing algorithms in NEXUS~\cite{cryptoeprint:2024/136NEXUS} modify the packing and increase the number of ciphertexts, leading to considerably more complex packing algorithms. The primary issue lies in NEXUS's adoption of \textit{reduce-first} packing, which proves unsuitable for these operators.
We propose a simple yet efficient packing algorithm for summation and scalar addition/multiplication, as illustrated in Figure~\ref{fig:encode1}. These operators are implemented through a rotate-and-sum process defined as follows:
% \begin{equation*}
\begin{gather*}
    \color{black}
    \operatorname{sum}(m)=[\underbrace{1,\ldots,1}_L,0,\ldots,0] \odot m_l^{\log_2 D} \\
    \color{black}
    \operatorname{broadcast}(m)=m_r^{\log_2 D} \\
    \color{black}
    m_l^{0}=m, m_l^{i} = m_l^{i-1}+\operatorname{Rot}^{2^{i-1} L}_l(m_l^{i-1})\ \text{for } i = 1, \ldots, \log_2 D \\
    \color{black}
    m_r^{0}=m, m_r^{i} = m_r^{i-1}+\operatorname{Rot}^{2^{i-1} L}_r(m_r^{i-1})\ \text{for } i = 1, \ldots, \log_2 D
\end{gather*}
We perform a broadcast for scalar addition and multiplication and then conduct element-wise operations. When a summation operator precedes a scalar operator, the masking and the broadcast steps can be omitted, as shown in Figure~\ref{fig:encode1}. Table~\ref{tab:packing_scalar} compares the theoretical complexity with NEXUS~\cite{cryptoeprint:2024/136NEXUS}. Our packing algorithms need only $\log_2D$ rotations while preserving a consistent packing for input and output.

\subsubsection{Tailored Packing Algorithms for Fusion Patterns}
% \begin{table}[!tb]
%     \centering
%     \caption{xxx.}
%     \huge
%     \label{tab:patterns}
%     \resizebox{1.0\linewidth}{!}{
%     \begin{tabular}{c|c|c|c|c}
%     \toprule
%     \diagbox{\textbf{First op}}{\textbf{Second op}} & \textbf{Identity} & \textbf{Expansion} & \textbf{Reduction} & \textbf{Transformation} \\
%     \midrule
%     \textbf{Identity} & \cellcolor{mygreen}Identity& \cellcolor{mygreen}Expansion&\cellcolor{mygreen} Reduction&\cellcolor{mygreen}Transformation  \\
%     \midrule
%     \textbf{Expansion} & \cellcolor{mygreen}Expansion &\cellcolor{myred}$\mbm{\times}$ & \cellcolor{myorange}Identity &\cellcolor{myorange}Expansion \\
%     \midrule
%     \textbf{Reduction} & \cellcolor{mygreen}Reduction &\cellcolor{myorange}Transformation & \cellcolor{myred}$\mbm{\times}$ &\cellcolor{myred}$\mbm{\times}$ \\
%     \midrule
%     \textbf{Transformation} & \cellcolor{mygreen}Transformation &\cellcolor{myred}$\mbm{\times}$ & \cellcolor{mygreen}Reduction &\cellcolor{myorange}Transformation \\
%     \bottomrule
%     \end{tabular}
%     }
% \end{table}

\begin{table}[!tb]
    \centering
    \caption{Fusion pattern analysis. The first column and the first row show the mapping types of the first and second operators. The colored cells show the mapping type of the operator after fusion. Green cells indicate legal fusions that do not require new packing algorithms. Orange cells represent legal fusions that necessitate new packing algorithms. Red cells denote fusion patterns that do not occur in Transformer models.}
    \huge
    \label{tab:patterns}
    \resizebox{1.0\linewidth}{!}{
    \begin{tabular}{c|c|c|c|c}
    \toprule
    \diagbox{\textbf{First op}}{\textbf{Second op}} & \textbf{Identity} & \textbf{Expansion} & \textbf{Reduction} & \textbf{Transformation} \\
    \midrule
    \textbf{Identity} & \cellcolor{mygreen}Identity& \cellcolor{mygreen}Expansion&\cellcolor{mygreen} Reduction&\cellcolor{mygreen}Transformation  \\
    \midrule
    \textbf{Expansion} & \cellcolor{mygreen}Expansion &\cellcolor{myred}$\mbm{\times}$ & \cellcolor{myorange}Identity &\cellcolor{myorange}Expansion \\
    \midrule
    \textbf{Reduction} & \cellcolor{mygreen}Reduction &\cellcolor{myorange}Identity & \cellcolor{myred}$\mbm{\times}$ &\cellcolor{myred}$\mbm{\times}$ \\
    \midrule
    \textbf{Transformation} & \cellcolor{mygreen}Transformation &\cellcolor{myred}$\mbm{\times}$ & \cellcolor{myorange}Reduction &\cellcolor{myorange}Transformation \\
    \bottomrule
    \end{tabular}
    }
    \vspace{-3pt}
\end{table}
After establishing packing algorithms for individual operators, we analyze the fusion patterns of adjacent operator pairs. For two candidate operators, we can: \ding{182} infer the type of the resulting fused operator and \ding{183} determine whether the fusion imposes new constraints on packing, ensuring compatibility and efficiency of this fusion. 

Table~\ref{tab:patterns} provides the detailed fusion patterns. The first column and the first row represent the types of the first and the second operators to be fused, and the colored cells indicate the resulting operator type. The fusion patterns are categorized into green, orange, and red.
% \ml{Move to the Caption of the Table.}\xts{done}
% As we can see, 
% The fusion flexibility varies among the four operator types. 
% Identity operators are the most flexible and can be fused with any other type without altering the packing rules. Expansion and Reduction operators exhibit moderate flexibility, as we do not need to change their packing algorithm. The Transformation operators are the least flexible, particularly $\operatorname{matmul_{cc}}$, where both input matrices are ciphertexts, making it difficult to change their packing rule.
Below, we elaborate on the green and orange fusion patterns in Table~\ref{tab:patterns}.
% \xts{this part is useless}
\begin{enumerate}[9]
    \item[$\bullet$] The identity operator can be freely fused with any other operator type. Expansion and Reduction operators can be fused with each other using the packing algorithms in Figure~\ref{fig:encode1}. Additionally, the Expansion operator can be fused with the Transformation operator by applying the respective packing algorithms for each.
    % \item[$\bullet$] Expansion with others. The expansion operator can be fused with the subsequent Reduction operator using the packing algorithm in Figure~\ref{fig:encode1}. When fused with Transformation operators, fusion can also be directly applied using the packing algorithms for each operator.
    % \item[$\bullet$] Reduction with others. The Reduction operator can be fused with the Expansion operator as shown in Figure~\ref{fig:encode1}. 
    \item[$\bullet$] Transformation with others. The Transformation operator can be fused with the subsequent Reduction operator, representing the combination of MatMul and summation. When fused with Transformation operators, it represents the fusion of two consecutive MatMuls. For $\operatorname{matmul_{cp}}+\operatorname{matmul_{cp}}$, fusion is straightforward. As for $\operatorname{matmul_{cp}}+\operatorname{matmul_{cc}}$ and $\operatorname{matmul_{cc}}+\operatorname{matmul_{cp}}$, although fusion is feasible following the packing algorithms proposed in~\cite{pang2023bolt,park2024powerformer}, the associated computational overhead is substantial.
    % a typical scenario is $Q_hK_h^T$ in attention layer where $Q_h,K_h$ are obtained from a $\operatorname{matmul_{cp}}$.
    % Existing packing algorithms~\cite{pang2023bolt,park2024powerformer} introduce substantial computational costs due to the inherent challenges in modifying the ciphertext packing. Another challenging case is fusing $\operatorname{matmul_{cc}}+\operatorname{matmul_{cp}}$, which is used in $\operatorname{Softmax}(\cdot)V_h$ followed by an output projection $\operatorname{matmul_{cp}}$. Existing methods~\cite{pang2023bolt,park2024powerformer} also requires much more rotations compared to $\operatorname{matmul_{cp}}$.
    To alleviate the computational cost, we developed an efficient MatMul protocol for the fused computations, which will be detailed in \textsection~\ref{sec:encode}.
    % \xts{need final revision when section~\ref{sec:encode} is OK}
\end{enumerate}
With the FineGrainFusion paradigm, NN layers are broken into basic operators, creating an operator-level computation graph. Fusion patterns are matched to merge compatible linear operators, reducing communication overhead while ensuring correctness.
% \subsubsection{Private Inference Flow and HE Parameter Selection}
% The FGF paradigm decomposes network layers into basic operators, forming an operator-level computation graph. The graph is analyzed to identify and match fusion patterns. Operators satisfying the fusion patterns are merged to reduce communication overhead while maintaining correctness.

% Fused linear operators have varying multiplication depths, which affect the HE parameters. Manually designing these parameters is complex, so we use the CKKS compiler EVA~\cite{dathathri2020eva} to generate the required HE parameters for fused operators. Detailed HE parameters will be provided in \textsection~\ref{sec:exp}.  

% \ml{MatMul or matrix multiplication?}\xts{I change all matrix multiplication to MatMul in Sec.5/6. While Sec.7 is still matrix multiplication for clarity}

\section{Efficient Protocol for Fused MatMuls}
\label{sec:encode}
% \xts{all symbols should be aligned at last, including: $m,n,d_1,d_2,L,D$, etc.}

\begin{figure*}[t]
    \centering
    \includegraphics[width=1\linewidth]{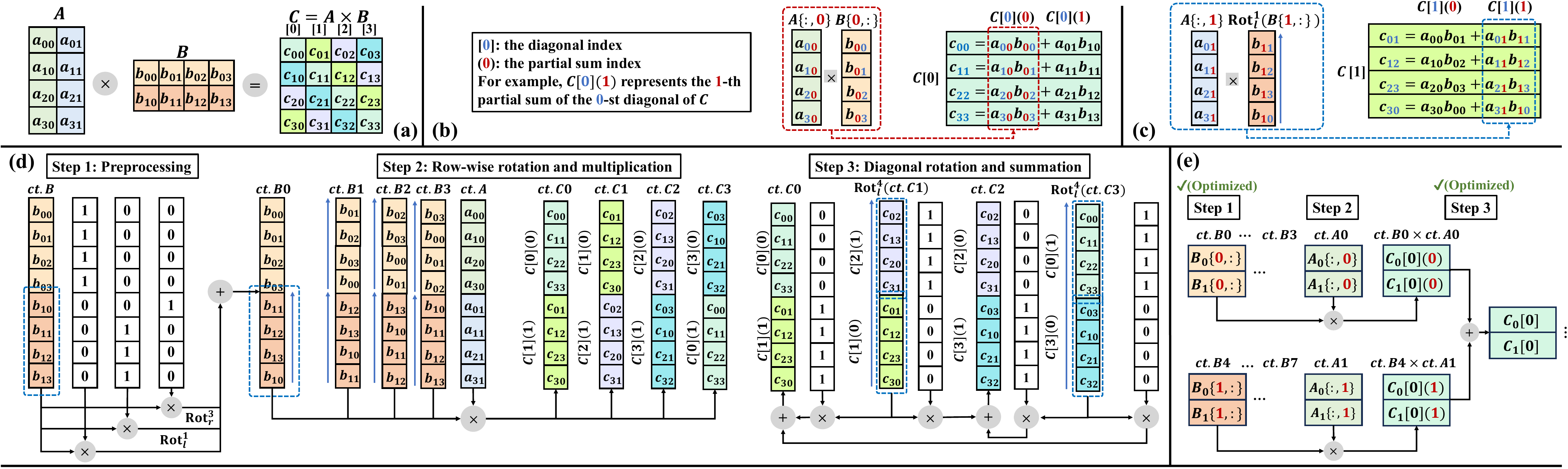}
    \caption{ A toy example for $\operatorname{matmul_{cc}}$ : (a) Multiplying matrices $A$ and $B$ in plaintext; (b) Observation 1: Element-wise multiplication of $A$'s columns and $B$'s rows generates the partial sums of $C$'s diagonals. We use [i] to denote the index of the diagonal and (j) to denote the index of the partial sum; (c) Observation 2: Rotating $B\{k,:\}$ left by $t$ steps, the multiplication produces the $k$-th partial sum of the $t$-th diagonal of $C$;
    % The left diagrams show the multiplication of $A$'s columns with the rotation of $B$'s rows to produce diagonal sums of $C$. The right diagram highlights two key properties of column-row multiplications. 
    (d) Illustration of \method's protocol, where $\operatorname{Rot}_r$ and $\operatorname{Rot}_l$ denote right-rotate and left-rotate, respectively; (e) Multi-head optimization.
    % \xts{can be improved} \ml{Rotations in (d) are not explained yet.}\xts{delete bolt}
    }
    \label{fig:matmul_cc}
\end{figure*}

In this section, we introduce \method's protocol for fused MatMuls. The multi-head attention computation in a Transformer model involves consecutive MatMuls with fusion opportunities, i.e., $Q_hK_h^T$ (where $Q_h, K_h$ are produced by $Q = XW_Q$ and $K = XW_K$) and $\operatorname{Concat}(\operatorname{Att}_h)W_O$ (where $\operatorname{Att}_h = \operatorname{Softmax}(\cdot) \times V_h$), as shown in Equation~\ref{eq:attention}. As $Q_h$ and $K_h$ are ciphertexts generated through the $\mmcp$ protocol, $Q_hK_h^T$ requires the $\mmcc$ protocol. Similarly, $\operatorname{Concat}(\operatorname{Att}_h)W_O$ and $\operatorname{Att}_h$ are computed through the $\mmcp$ and $\mmcc$, respectively.

Fusing these operators is not straightforward, because upon fusion, the packing of intermediate ciphertexts, e.g., $Q_h$, $K_h$, $\operatorname{Att}_h$, etc, is limited by the previous MatMul protocol while modifying the packing requires costly homomorphic operations. As both inputs of $\mmcc$ are ciphertexts, the $\mmcc$ protocol often becomes the bottleneck due to the packing constraints. For example, BOLT~\cite{pang2023bolt} only optimizes $\mmcp$ and overlooks $\mmcc$, leading to 20$\times$ more rotations. Powerformer~\cite{park2024powerformer} optimizes $\mmcc$ but still suffers from high cost for non-square matrices. We propose a rotation-efficient protocol for $\mmcc$ with further multi-head and BSGS optimizations, which reduces rotations by $8 \sim 29 \times$ compared to BOLT and Powerformer.

\subsection{Building Blocks}\label{subsec: ccmatmul}
\textbf{Rotation-Efficient $\operatorname{ct-ct}$ MatMul.}
We begin with a $\operatorname{ct-ct}$ MatMul, \( C = A \times B \) where \( A \in \mathbb{R}^{L \times D} \) and \( B \in \mathbb{R}^{D \times L} \). Matrix $A$ is in a spatial-first packing, and $B$ is in a reduce-first packing. Let \( A\{i,j\} \) denote the element in the \( i \)-th row and \( j \)-th column of \( A \). \( A\{i,:\} \) denotes the \( i \)-th row, and \( A\{:,j\} \) denotes the \( j \)-th column of \( A \).
Figure~\ref{fig:matmul_cc} shows an example where $L=4,D=2$. 
% that multiply a column-wise packed ciphertext matrix with a row-wise packed ciphertext matrix. We will use the example \( C = A \times B \), where \( A \in \mathbb{R}^{n \times m} \) and \( B \in \mathbb{R}^{m \times n} \) throughout the rest of this subsection for illustration.
% Figure~\ref{fig: matmul_basic} illustrates our rotation-efficient protocol, which is based on the following two observations:
Our rotation-efficient protocol is based on the following two observations:

% \noindent
\ding{182} \textbf{Element-wise multiplication of $A\{:,k\}$ and $B\{k,:\}$ gives the $k$-th partial sum of the main diagonal of $C$}. For example in Figure~\ref{fig:matmul_cc} (b), multiplying $A\{:,0\}$ with $B\{0,:\}$ produces the 0-th partial sum of $C$'s main diagonal, denoted as $C[0](0)$.

% \noindent
\ding{183} \textbf{Rotating $B\{k,:\}$ left by $t$ steps, the multiplication produces the $k$-th partial sum of the $t$-th diagonal of $C$, denoted as $C[t](k)$}. After rotating $B\{k,:\}$ by $t$ steps, the $i$-th element of $B\{k,:\}$, denoted $B\{k,i\}$, becomes $B\{k,i+t\}$ \newtext{in the original matrix}. Multiplying $A\{i,k\}$ by $B\{k,i+t\}$ produces the $k$-th partial sum of $C\{i,i+t\}$, corresponding to the $i$-th element in the $t$-th diagonal of $C$. As illustrated in Figure~\ref{fig:matmul_cc}(c), rotating $B\{1,:\}$ by 1 step and multiply $A\{:,1\}$ with it, we will obtain $C[1](1)$.

With these two observations, we can perform MatMuls by multiplying $A\{:,k\}$ with rotated $B\{k,:\}$ to obtain partial sums of all diagonals and sum all partial sums to get the final result. 
% However, a direct use of this property, like BOLT, will result in a large number of rotations. Figure \ref{fig:matmul_cc} (d) presents an example of BOLT's protocol. Given $A$ and $B$, in step 1, BOLT directly multiplies these two ciphertexts and get the different partial result of the \textbf{same} diagonal. Then in step 2, BOLT uses a large number of rotations to sum together all partial sums, which is rotation-heavy.
% Figure~\ref{fig:matmul_cc}(e) provides an example of our protocol. To reduce numerous rotations in step 2, partial sums of different diagonals should be generated. The key insight of our method is based on the second observation, which is to first rotate within each row of $B$ and then multiply it with $A$. Since multiplication between each rotated row of $B$ and the column of $A$ results in different diagonals and thus reduces the rotations needed for summation.
Figure~\ref{fig:matmul_cc}(d) illustrates our protocol which involves three steps: \ding{182} \textbf{Preprocessing.} This step rotates $B\{t,:\}, \forall t \in [D]$ by $t$ steps. After preprocessing, $B$ remains row-packed, but the elements within each row are rotated. This enables the multiplication with $A$ to produce different diagonals in a single ciphertext, eliminating the need for accumulating elements within the same ciphertext. In contrast, BOLT~\cite{pang2023bolt} does not incorporate this preprocessing, causing partial sums from the same diagonal to be packed into a single ciphertext. This requires $\log L$ rotations per ciphertext to accumulate the partial sums, significantly increasing the number of rotations.

Additionally, direct row rotation is not natively supported but can be achieved through masking followed by rotation, named \textit{inner rotation}. As shown in Figure~\ref{fig:matmul_cc} (d), to rotate $B\{1,:\}$ one step to the left, we apply three masks to extract $[b_{00},b_{01},b_{02},b_{03},0,\ldots]$, $[\ldots,0,b_{11},b_{12},b_{13}]$, and $[\ldots,b_{10},\dots]$. We then rotate $[\ldots,0,b_{11},b_{12},b_{13}]$ left by one step and $[\ldots,b_{10},\dots]$ right by three steps. Adding the three results gives the final rotated output. Step 1 requires $D$ inner rotations, each involving two rotations and $\operatorname{ct-pt}$ multiplications.
\ding{183} \textbf{Row-wise rotation and multiplication.} We apply inner rotations within each row of $B$ by $t$ steps ($\forall t \in [L]$), generating $L$ ciphertexts. These ciphertexts are multiplied by $A$ to obtain the partial sums for all diagonals. This step requires $L-1$ inner rotations and $\operatorname{ct-ct}$ multiplications.
\ding{184} \textbf{Diagonal rotation and summation.}
After step 2, we obtain $L$ ciphertexts, each containing the partial sums of different diagonals. We then rotate across diagonals to align their positions, extract the partial sums from each diagonal, and sum them to obtain the final result. This step requires $L$ rotations and $2L$ $\operatorname{ct-pt}$ multiplications.
% \xts{all done}
% we perform rotations across the partial sums to align the diagonals, and then add them together to get the diagonals. 

% We directly multiply the preprocessed $B$ with $A$ to obtain the first set of partial sums of the diagonals, which is the $(0)$-th partial sum of diagonal $[0]$ and the $(1)$-st partial sum of diagonal $[1]$ in Figure \ref{fig:matmul_cc} (e). We then keep rotating $B$ within rows and perform multiplications to obtain the the rest of partial sums. After rotating $L-1$ times, we have all the partial sums for diagonals, and rotations across diagonals are required to align these partial sums before the final summation. 

\textbf{Multi-Head Optimization.}
In multi-head attention, there are $H$ heads that can be computed in parallel as $H$ separate MatMuls.  Formally, we need to compute \( H \) parallel products \( C_h = A_h \times B_h \), where \( A_h \in \mathbb{R}^{L \times D} \) and \( B_h \in \mathbb{R}^{D \times L} \) for \( h \in [H] \). We observe that \textbf{different heads are independent}, allowing us to pack multiple heads into a single ciphertext. Specifically, instead of packing the entire matrix \( A_h \) into a ciphertext, we pack \( D/H \) columns of all \( A_h \) matrices into one ciphertext. This results in a ciphertext containing a tensor of dimensions \( (L, H, D/H) \), as shown in Figure~\ref{fig:matmul_cc} (e) for $H=2$.
This multi-head packing (MHP) reduces the rotations in steps 1 and 3 of our MatMul protocol. In step 1, each row of $B$ needs an inner rotation. With MHP, we can rotate $H$ heads in parallel, reducing the row count from $D$ to $D/H$. As shown in Figure~\ref{fig:matmul_cc} (e), step 1 can be skipped since $D/H$ is exactly 1. Moreover, after step 2, the partial sums of the same diagonal are aligned at the same position. This enables us to sum these ciphertext results first in Step 3, then apply the rotation. In contrast, without MHP, each ciphertext would need to be rotated individually.

To achieve MHP in $\operatorname{matmul_{cc}}$, the packing of neighboring $\operatorname{matmul_{cp}}$ needs to change. Specifically, for $\operatorname{matmul_{cp}} + \operatorname{matmul_{cc}}$, we need to reorder the columns of the weight matrix \newtext{in plaintext} in $\operatorname{matmul_{cp}}$ to match the MHP format of $\operatorname{matmul_{cc}}$'s inputs. For $\operatorname{matmul_{cc}} + \operatorname{matmul_{cp}}$, we need to reorder the rows of the weight matrix \newtext{in plaintext} to align with the MHP format of $\operatorname{matmul_{cc}}$'s output. Note that such reordering does not affect the computational complexity of $\mmcp$.

\textbf{BSGS Optimization.} \delete{The number of rotations in step 1 and step 2 can be further reduced using BSGS~\cite{ju2023neujeans} with details provided in Appendix~\ref{app:bsgs}.}\newtext{
The BSGS optimization splits rotations into baby-step and giant-step phases and is widely used to reduce the number of rotations in matrix multiplications~\cite{pang2023bolt,xu2024privcirnet}. In our Fused MatMuls protocol, the rotations in steps 1 and 2 can be further reduced using BSGS with a detailed explanation provided in Appendix~\ref{app:bsgs}.} The multiplication depth of our protocol is 4, the same as Powerformer~\cite{park2024powerformer}. Although BOLT has a depth of 2, its overall latency is much higher due to excessive rotations, as shown in \textsection~\ref{sec:exp}. 

\begin{table}[!tb]
    \centering
    \huge
    \caption{Theoretical complexity comparison of $\mmcc$ protocol across prior works. ``CMult.'' refers to the $\operatorname{ct-ct}$ multiplications. $m,d,H$ are defined in \textsection~\ref{subsec:pre_transformer}. The data is based on the dimensions in BERT-large, with $s=16384$ indicating the length of a plaintext vector.
    % \ml{What about the number of HE multiplications? Need to mention. Also, use a specific layer in BERT or GPT to show the actual quantitative comparison.}\xts{done}
    }
    \label{tab:complexity_compare}
    \resizebox{1.0\linewidth}{!}{
    \begin{tabular}{c|c|c|c|c}
    \toprule
    \multicolumn{2}{c|}{$\operatorname{matmul_{cc}}$}&BOLT~\cite{pang2023bolt}&Powerformer~\cite{park2024powerformer}& BLB\\
    \midrule
    \multirow{4}{*}{\makecell{$Q_hK_h^T$\\$h\in [H]$}}&\multirow{2}{*}{\# Rot} & $O(\frac{md}{s}(m+\log_2m))$ & $O(\frac{md}{s}(\frac{5}{2}m+4\sqrt{m}))$  & $O(\frac{md}{s}(\frac{\sqrt{d}}{H}+\sqrt{m})+m)$ \\
    & & 18432 & 5856 & 640 \\
    \cmidrule{2-5}
    & \multirow{2}{*}{\# CMult.} & $O(\frac{m^2d}{s})$ &$O(\frac{m^2d}{s})$ &$O(\frac{m^2d}{s})$  \\
    & & 2048 & 1024 & 1024 \\
    \midrule
    \multirow{4}{*}{\makecell{$\operatorname{Softmax}\times V_h$\\$h\in [H]$}}&\multirow{2}{*}{\# Rot} & $O(\frac{m^2H}{s}(m+\log_2m))$ & $O(\frac{m^2H}{s}(\frac{5}{2}m+4\sqrt{m}))$  & $O(\frac{m^2H}{s}(\sqrt{\frac{m}{H}}+\sqrt{m})+m)$ \\
    & & 28560 & 6508 & 1056 \\
    \cmidrule{2-5}
    & \multirow{2}{*}{\# CMult.} & $O(\frac{m^2d}{s})$ &$O(\frac{m^3H}{s})$ &$O(\frac{m^3H}{s})$  \\
    & & 1024 & 2048 & 2048 \\
    \bottomrule
    \end{tabular}
    }
\end{table}
\subsection{Efficient Fused Computations in Multi-Head Attention} 
We explain how our protocol is applied to fused computations in multi-head attention. For \( Q_hK_h^T,\forall h \in [H] \), we first reorder the columns of \( W_Q \) and \( W_K \), then apply BOLT's protocol to compute \( Q_h \) and \( K_h^T \). \( Q_h \) and $K_h$ are both in spatial-first packing, thus \( K_h^T \) is in reduce-first packing. These are then multiplied using our \( \operatorname{ct-ct} \) MatMul protocol, resulting in \( Q_hK_h^T \) in diagonal packing. Since the next operator is nonlinear, we convert \( Q_hK_h^T \) to secret-share form and repack it.

For $\operatorname{Concat}(\operatorname{\operatorname{Softmax}(\cdot)\times V}_h)W_O$, $\operatorname{Softmax}(\cdot)$ is already in secret-share form, so we can directly repack it to the multi-head spatial-first packing. We then convert $V_h$ to secret-share form and repack it to the multi-head reduce-first packing. Our $\operatorname{matmul_{cc}}$ protocol requires the input dimension and the output dimension to be the same, so we pad with zeros if $\operatorname{Softmax}(\cdot)\times V_h$ does not meet this condition.
After $\operatorname{matmul_{cc}}$, we get $\operatorname{Att}_h$ in multi-head diagonal packing. To fuse with the next $\operatorname{matmul_{cp}}$, we reorder the rows of $W_O$ and apply a modified $\mmcp$ protocol to get the result in a spatial-first packing.
The details are in Appendix~\ref{app: diagcp}, and the complexity of $\operatorname{matmul_{cp}}$ remains the same as in BOLT.

\textbf{Theoretical complexity.} Table~\ref{tab:complexity_compare} shows the complexity comparison of~\method~with prior-art methods in the number of rotations and $\operatorname{ct-ct}$ multiplications. Our technique saves about $29\times $ and $8\times $ rotations compared to BOLT and Powerformer.

\section{Secure Conversion between CKKS and MPC}\label{sec:conversion}
To mitigate the growth in ciphertext bit width after fusion,~\method~utilizes the CKKS HE for evaluating linear operators. 
However, we identify a critical security flaw in the previously proposed CKKS-to-MPC conversion protocol~\cite{boemer2020mp2ml}.
This section analyzes the security issues in the existing protocol~\cite{boemer2020mp2ml} and presents our secure conversion protocol.

\subsection{Problem Statement}
To understand the security issue of~\cite{boemer2020mp2ml}, we first illustrate the detailed mechanisms of the CKKS scheme.

\subsubsection{CKKS Approximate HE and Message Encoding} 
The message space of the CKKS scheme is $\pring$, a polynomial ring of coefficient modulus $q$.
Let $\poly{x}, \poly{y} \in \pring$ being two messages.
The CKKS scheme supports 
approximate homomorphic addition
$\Dec(\Enc(\poly{x}) \HomAdd \Enc(\poly{y})) \approx \poly{x} + \poly{y} \in \pring$ 
and
approximate homomorphic multiplication
$\Dec(\Enc(\poly{x}) \HomMul \Enc(\poly{y})) \approx \poly{x} \cdot \poly{y} \in \pring$.

% \subsubsection{The CKKS Message Encoding}
One of the most important features of the CKKS scheme is supporting computation on ciphertexts of real values.
Particularly, the CKKS scheme utilizes a function
$\pi: \pring \mapsto \mathbb{R}^{N/2}$  to achieve this feature.
The CKKS scheme selects a positive real value $\Delta$ called scaling factor and defines the encoding and decoding functions as follows\footnote{Indeed, the CKKS encoding supports a complex space $\mathbb{C}^{N/2}$. We omit the imaginary part and consider $\mathbb{R}^{N/2}$ to simplify the presentation.}.
\begin{equation}
\label{eq:ckks_encode}
\begin{aligned}
    &\Encode(m \in \mathbb{R}^{N/2}) = \lfloor \Delta \cdot \pi^{-1}(m) \rceil \in \pring , \\
    &\Decode(\hat{m} \in \pring) = \pi(\lfloor \Delta^{-1} \cdot \hat{m}\rceil) \in \mathbb{R}^{N/2}.
\end{aligned}
\end{equation}
The rounding $\lfloor \cdot \rceil$ is applied to each polynomial coefficient.
Indeed, the mapping $\pi^{-1}$ is realized via a fast Fourier transform (FFT). 
% By selecting a proper size of $\Delta$, we can limit the magnitude $\lVert \Encode(m) \lVert_\infty < q/2$, and view the encoding result as an element of $\pring$.
% {\color{red} Lu: Should we simplify the presentation and consider $\mathbb{R}^{N/2}$ only?}

\subsubsection{From CKKS to MPC}
The conversion from CKKS to MPC can be described as an interactive two-party protocol that takes 
$\Enc(\Encode(m))$ a CKKS ciphertext of $m\in \mathbb{R}^{N/2}$ as the input, and generates
two integer vectors $u_0, u_1 \in \mathbb{Z}_q^{N/2}$ such that
$\lfloor\Delta \cdot m \rceil \approx  u_0 + u_1 \bmod q$.
% \footnote{ For a complex vector $m \in \mathbb{C}^{N/2}$, we write $\lceil m \rfloor_c \in \mathbb{Z}^N$ to denote $\lceil m.{\rm real} \rfloor \lVert \lceil m.{\rm img} \rfloor$, i.e., concatenating the real part and the imaginary part.}. 
The approximation here is due to the approximate property of the CKKS scheme. 
In other words, the integer vectors $u_0, u_1$ can be viewed as the additive secret shares of the real vector $m$ over $\mathbb{Z}_q$ with the fixed-point scale $s\approx \log_2\Delta$. 
Both $q$ and $\Delta$ are parameters of the CKKS scheme.

\begin{figure}[t]
    \centering
    \includegraphics[width=1.0\linewidth]{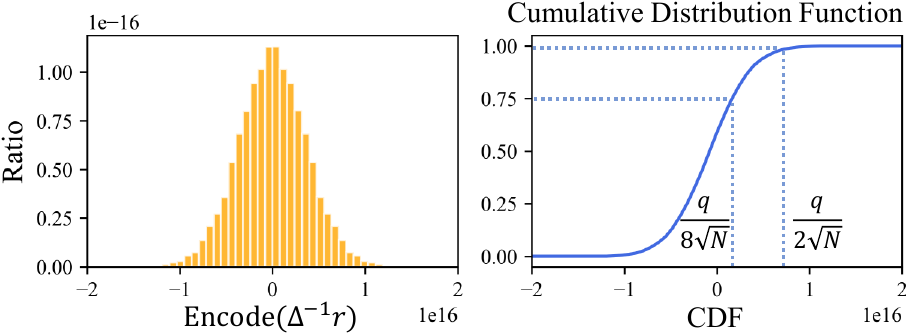}
    \caption{This example demonstrates the distribution of the coefficients of $\Encode(\Delta^{-1}r)\in \pring$ for uniform random vector $r \in \mathbb{Z}_q^{N/2}$ with $N = 8192$ and $q\approx 2^{60}$.
    From this example, we can see that the output distribution is highly concentrated within the range $[-q/(8\sqrt{N}), q/(8\sqrt{N})]$.
    }
    \label{fig:distribution}
\end{figure}
\subsection{The Pitfalls of Previous Conversion}

MP2ML~\cite{boemer2020mp2ml} proposed a CKKS-to-MPC conversion protocol, which has been utilized by many other works such as~\cite{balla2023heliks,singh2024hyena}. 
Their protocol involves the following steps:
% However, we will demonstrate that the conversion in~\cite{boemer2020mp2ml} is \textbf{insecure}. 
% Figure~\ref{fig:secure}(a) illustrates the MP2ML conversion protocol. 
\begin{enumerate}[9]
    \item[$\bullet$] Server samples uniformly random vectors $r \in \mathbb{Z}^{N/2}_q$ and encodes it as $\poly{r} = \operatorname{Encode}(\Delta^{-1} \cdot r) \in \pring$.
    \item[$\bullet$] Server then masks the ciphertext through homomorphic addition: $C = \Enc(\Encode(m)) \HomAdd \poly{r}$.
    \item[$\bullet$] The masked ciphertext $C$ is then sent to the client, who obtains his share
    $u_0 = \lfloor \Delta\cdot \Decode(\Dec(C))\rceil \bmod q$ after decryption.
    The server's share is defined as $u_1 = -r\bmod q$.
    % $\Encode(x\odot w)-\hat{r}+e\cdot \hat{w}$, as derived in Remark~\ref{remark:2}.
\end{enumerate}
 % To satisfy semantic security~\cite{shannon1949communication}, the client's share must not reveal any information about $x$ or $w$. 
However, although the vector $r$ is sampled uniformly at random, the distribution of the encoding $\hat{r}$ is narrowly concentrated around zero due to the properties of FFT, as demonstrated in Figure~\ref{fig:distribution}. 
Particularly, the values of $\poly{r}$ majorly range over $[-q/(8\sqrt{N}), q/(8\sqrt{N})]$. 
Since a relatively large dimension $N$ is commonly used for the CKKS scheme, such as $N \ge 2^{13}$, 
this bell-shaped distribution fails to mask the high-end bits of the message $m$.

% Moreover, the message $m$ is multiplied by a scaling factor $\Delta > 0$ within the encoding.
In private transformer inference, the server starts with a CKKS ciphertext of $\Encode(x \odot w)$ where $x$ is the client's input and $w$ is the server's model weight.
The server aims to convert the encryption of $\Encode(x\odot w)$ to the additive secret share form by adding the mask $\Encode(x\odot w) + \poly{r}$.
However, the masking polynomial $\poly{r}$ distributes narrowly around zero while the distribution expectation of $\Encode(x\odot w)$  depends on the message $x\odot w$.
Thus, the zero-centered and bell-shaped masking $\poly{r}$ may fail to mask the messages when the expectation of $x\odot w$ is far from zero.
More importantly, when the product magnitude $\lVert \Delta\cdot x\odot w\lVert_\infty$ is relatively large, such as $> q/8$, some high-end bits of $w$ will be leaked to the client.
% when using the conversion method~\cite{boemer2020mp2ml}. 
All these above lead to the conclusion that the narrowly distributed random masking used by MP2ML~\cite{boemer2020mp2ml} is insecure.

Beyond security concerns, the previous conversion suffers from limited efficiency since it only supports MPC protocols that operate on $\mathbb{Z}_q$. In contrast, MPC protocols over $\mathbb{Z}_{2^l}$ are significantly more communication-efficient~\cite{rathee2020cryptflow2,rathee2021sirnn}. Thus, a modulus conversion protocol between $\mathbb{Z}_q$ and $\mathbb{Z}_{2^l}$ is essential to enable lightweight MPC protocols.

% In summary, the flaws of the previous conversion come from: \ding{182} Insecurity due to improper random masking, \ding{183} Limited efficiency due to the lack of modulus conversions, as we marked in Figure~\ref{fig:secure}.
% The conversion protocol starts with the server holding the ciphertext of $x\odot w$. 

% Next, we show how this vulnerability allows the client to infer information about $w$. Figure~\ref{fig:distribution} (b) and (c) show the distribution of the client's share when $w$ follows a normal distribution with variances of $10^6$ and $10^7$, respectively. As the variance increases, the distribution approaches a random distribution due to the modulus operation, and the masking polynomial $\hat{r}$ fails to hide the weight information.

\begin{figure}[t]
\revise{
\fbox{
\centering
\begin{minipage}{0.46\textwidth}
  \centering
  \underline{\centering Functionality $\mathcal{F}_{C2M}$ }\\[0pt]
  Receive a CKKS encryption  $\Enc(\hat{a}) \in \mathbb{A}^2_{N, q}$ from $P_1$ and receive the decryption key from $P_0$,
  this functionality
  \begin{enumerate}
  \setlength{\itemsep}{2pt}
  \setlength{\topsep}{2pt}
  \setlength{\parsep}{0pt}
  \setlength{\parskip}{0pt}
  \item Samples $\poly{r} \in \mathbb{A}_{N, q}$ uniformly at random.
  \item Decrypt $\Enc(\hat{a})$ with decryption key and gets $\hat{a}$.
  \item Gives $\lceil \Decode(\hat{a} + \hat{r} \bmod 2^l) \rfloor \bmod 2^l \in \mathbb{Z}_{2^l}^{N/2}$ to $P_0$.
  \item Gives $\lceil \Decode(-\hat{r}\bmod 2^l) \rfloor \bmod 2^l \in \mathbb{Z}_{2^l}^{N/2}$ to $P_1$.
  \end{enumerate}
\end{minipage}
}
\caption{CKKS-to-MPC Conversion Functionality} \label{func:ckks2mpc}
}
\end{figure}

\begin{algorithm}[t]
    \DontPrintSemicolon
        \SetAlgoLined
        \KwIn{$P_1$ holds $\Enc(\Encode({x}))\in \mathbb{A}^2_{N,q}$.}
        \KwOut{$\forall b \in \{0, 1\}$, $P_b$ gets $[\![ {x}]\!]_b^{2^l}$}

        $P_1$ samples ${\hat{r}} \in \mathbb{A}_{N,q}$ uniformly at random and computes $\Enc(\Encode({x}))\boxplus \hat{r}$. Then $P_1$ sends the masked ciphertexts to $P_0$ and sets $[\![ \newtext{{\tt tmp}} ]\!]^q_1={-\hat{r}}$.

        $P_0$ decrypts to get $[\![ {\tt tmp} ]\!]^{q}_0=\Encode({x})+{\hat{r}}$.

        $P_0$ and $P_1$ invoke $[\![ {\tt tmp} ]\!]^{2^l} = \operatorname{Field-to-Ring}([\![ \newtext{{\tt tmp}} ]\!]^{q})$.

        $\forall b \in \{0, 1\}$, $P_b$ locally evaluates the CKKS decoding and learn $[\![ {x}]\!]_b^{2^l}= \lceil \Decode([\![ {\tt tmp} ]\!]_b^{2^l})\rfloor $.

        \caption{Secure CKKS to MPC conversion}
        \label{alg:ckks2mpc}
\end{algorithm}

\begin{algorithm}[t]
    \DontPrintSemicolon
        \SetAlgoLined
        \KwIn{$\forall b \in \{0, 1\}$, $P_b$ holds $[\![ {x}]\!]_b^{2^l}$.}
        \KwOut{$P_1$ gets $\Enc(\Encode({x}))\in \mathbb{A}^2_{N,q}$.}

        $\forall b \in \{0, 1\}$, $P_b$ locally evaluates the CKKS encoding and learn $[\![ {\tt tmp}]\!]_b^{2^l}=\Encode([\!
        [ {x} ]\!]_b^{2^l})$.

        $P_0$ and $P_1$ invoke $[\![ {\tt tmp} ]\!]^{q}=\operatorname{ring-to-field}([\![ {\tt tmp} ]\!]^{2^l})$.

        $P_0$ encrypts $[\![ {\tt tmp} ]\!]^{q}_0$ as $\Enc([\![ {\tt tmp} ]\!]^{q}_0)$ and sends it to $P_1$.

        $P_1$ conduct homomorphic addition to get $\Enc(\Encode({x}))=\Enc([\![ {\tt tmp} ]\!]^{q}_0)\boxplus [\![ {\tt tmp} ]\!]^{q}_1$.
        \caption{Secure MPC to CKKS conversion}
        \label{alg:mpc2ckks}
\end{algorithm}

% Beyond security concerns, the previous conversion protocol also suffers from limited efficiency since it only supports MPC protocols operating over $\mathbb{Z}_q$, as shown in Figure~\ref{fig:secure} (a). In contrast, MPC protocols over $\mathbb{Z}_{2^l}$ are significantly more communication-efficient~\cite{rathee2020cryptflow2,rathee2021sirnn}. Thus, a modulus conversion protocol between $\mathbb{Z}_p$ and $\mathbb{Z}_{2^l}$ is essential to enable lightweight MPC protocols.

% In summary, the failure of the previous conversion protocol comes from: \ding{182} Insecurity due to improper random masking, \ding{183} Limited efficiency due to the lack of modulus conversions, as we marked in Figure~\ref{fig:secure}.

\subsection{Secure and Efficient Conversion Protocols}
\textbf{Secure Masking.}  
% To solve the security issue, 
We suggest sampling the masking polynomial $\poly{r}$ directly from $\pring$ uniformly at random rather than computing it via the CKKS encoding function, as described in Theorem~\ref{rmk:mask}. \revise{The proof is available in Appendix~\ref{app:security}.}

% \begin{remark}%\label{rmk:mask}
\delete{
Given an CKKS encryption $C = \Enc(\Encode(x))$, the decryption $\poly{d} = \Dec(C \HomAdd \poly{r})$ is uniform random over $\pring$ when the polynomial $\poly{r}$ is sampled uniformly from $\pring$ at random.
The masking polynomial $\poly{r}$ and the decryption $\poly{d}$ can be viewed as an additive share of $\Encode(x)$. 
That is $(-\poly{r}) + \poly{d} \bmod q \approx \Encode(x) \in \pring$.
}
% \end{remark}

\revise{
\begin{theorem}\label{rmk:mask}\label{thm:mask}
The CKKS-to-MPC conversion in Algorithm~\ref{alg:ckks2mpc} securely realizes the $\mathcal{F}_{C2M}$ functionality in Figure~\ref{func:ckks2mpc} against honest-but-curious adversary. 
\end{theorem}
}

In the context of private transformer inference, the client first obtains a masked polynomial $\poly{d} = \Encode(x\odot w) + \poly{e} + \poly{r} \bmod q$ 
for a noisy term $\poly{e}$ after the decryption. This masked polynomial $\poly{d}$ distributes uniformly over $\pring$ since $\poly{r} \in \pring$ is a uniform random polynomial. 
No information about $w$ (i.e., server's input) and $\poly{e}$ is revealed to the client.

\textbf{Efficient and Accurate Conversion Protocol.}
After secure masking, we obtain the shares of $\Encode(x\odot w)$ over $\mathbb{Z}_q$. To enable lightweight MPC protocols over $\mathbb{Z}_{2^l}$, we utilize the $\operatorname{field-to-ring}$ and $\operatorname{ring-to-field}$ protocols from~\cite{pang2023bolt,rathee2021sirnn} for switching between $\mathbb{Z}_q$ and $\mathbb{Z}_{2^l}$, as detailed in Appendix~\ref{app:ring-field}.

Subsequently, both parties perform local decoding in the fixed-point representation over $\mathbb{Z}_{2^l}$ to get the share of $x\odot w$. The decoding involves multiplying by a public FFT matrix. 
We use the $O(N\log_2N)$ FFT algorithm to compute the matrix multiplication, which demands $O(\log_2N)$ multiplicative depths. To avoid extra communications due to fixed-point truncations, we extend the shares to a larger ring, e.g., $\mathbb{Z}^{l+40}$, and conduct local truncations with an error rate below $2^{-40}$. Further details are provided in Appendix~\ref{app:accurate_encoding}.

\textbf{Putting It Together.} With these improvements, we present the full CKKS-to-MPC protocol in Algorithm~\ref{alg:ckks2mpc}. The reverse conversion, MPC-to-CKKS, follows a similar process outlined in Algorithm~\ref{alg:mpc2ckks}. In the MPC-to-CKKS protocol, both parties first perform local encoding for a shared vector $\ashare{{ x}}^{2^\ell}$, then switch the secret share modulus from $2^\ell$ to $q$ using $\operatorname{ring-to-field}$. Finally, the client encrypts its share and sends the ciphertext to the server, which can reconstruct the CKKS encryption of $\Encode({ x})$ via one homomorphic addition.

\section{Evaluation}\label{sec:exp}
\subsection{Experimental Setup}
\begin{figure}[!t]
    \centering
    \includegraphics[width=1.0\linewidth]{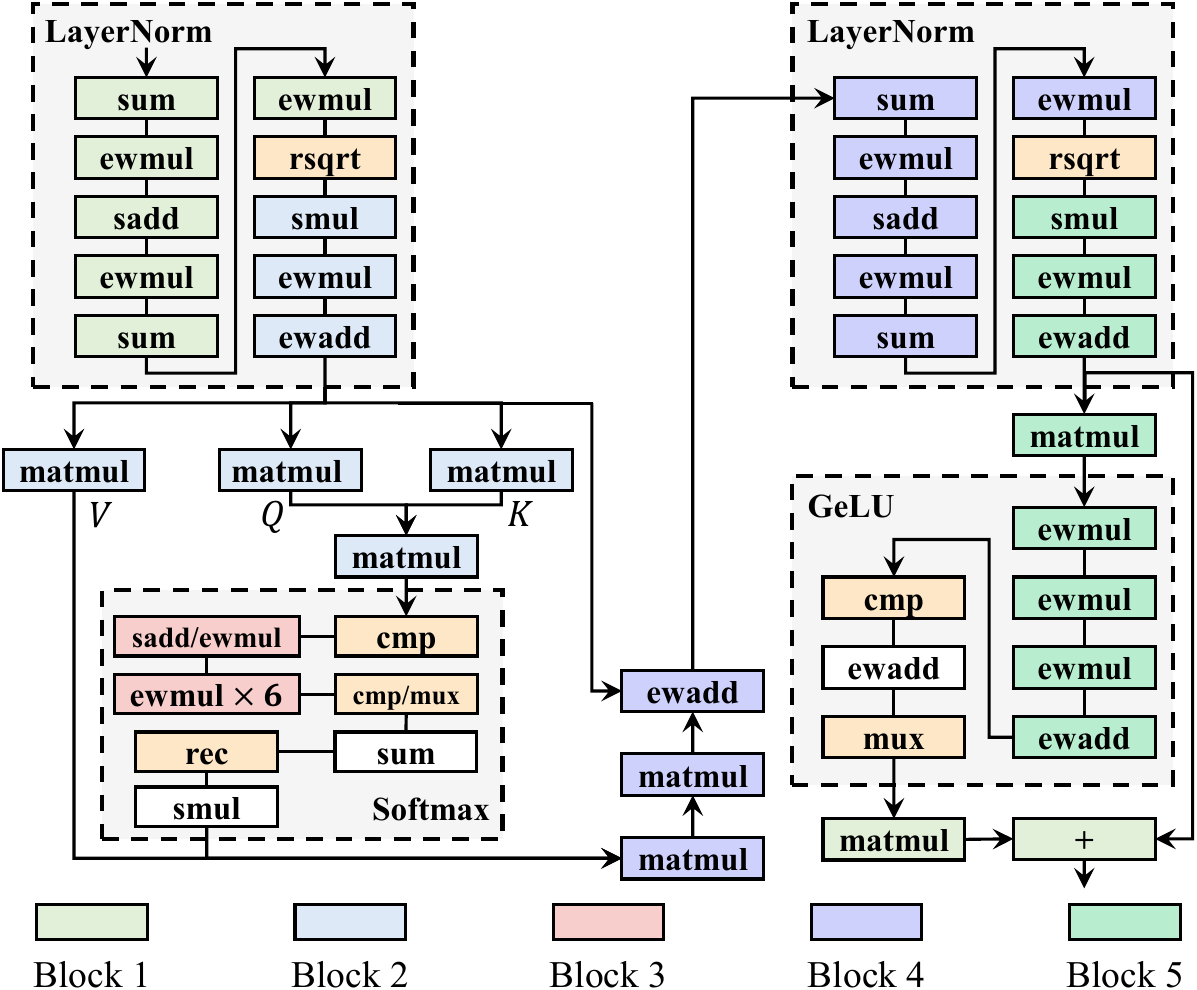}
    \caption{Five blocks of fused linear operators in a Transformer layer using~\method. All the operators shown in the figure are defined in Table~\ref{tab:opertors}.}
    \label{fig:fusion_pattern}
\end{figure}

\textbf{Implementation.} We implement \method~based on the SEAL~\cite{sealcrypto} library and the Ezpc library~\cite{chandran2017ezpc,kumar2020cryptflow} in C++ \delete{on CPU}. The GPU acceleration \delete{for~\method~}was developed based on PhantomFHE\cite{PhantomFHE_BFV}.
We set the security parameter $\lambda=128$. We use a bit width of $l=43$ and scale $s=13$ for nonlinear operators. We apply \method~to a Transformer block, and after the fine-grained fusion, there are in total 5 blocks of fused linear operators, marked with different colors in Figure~\ref{fig:fusion_pattern}. To determine the HE parameters (i.e., ciphertext bit width and scale) for each block, we apply the EVA CKKS compiler~\cite{dathathri2020eva} and require the Mean Square Error (MSE) of CKKS-based computation to be $<10^{-11}$ compared to the plaintext counterparts. The detailed parameters are shown in Table~\ref{tab:eva_params}.
% The computation graphs of these blocks are input to the EVA CKKS compiler~\cite{dathathri2020eva} to generate their HE parameters where we require that the Mean Square Error (MSE) between ciphertext and plaintext computation results is below $10^{-11}$. 
% We apply~\method~to a Transformer block and identify five linear fusion blocks, marked with different colors in Figure~\ref{fig:fusion_pattern}.
% \ml{Add seal and PhantomFHE?}\xts{done}
\begin{table}[!t]
    \centering
    % \huge
    \caption{CKKS HE parameters for different blocks.}
    \label{tab:eva_params}
    \resizebox{1.0\linewidth}{!}{
    \begin{threeparttable}
    \begin{tabular}{c|c|c|c|c|c}
    \toprule
    \textbf{Block} & \textbf{Mul. Depth} & \textbf{Degree $N$} & \textbf{Ciphertext bit width}\tnote{*} & \textbf{Scale} & \textbf{MSE} \\
    \midrule
    1 & 4 & 16384 & $\{60, 60, 60, 60, 60\}$ & 38 & $7.2\times 10^{-13}$ \\
    2 & 7 & 32768 & $\{60, 35,60, 60, 60, 60, 60, 60\}$ & 44 & $3.4\times 10^{-12}$ \\
    3 & 7 & 32768 & $\{60, 40, 40, 40, 40, 40, 40, 40, 60\}$ & 40 & $5.2\times 10^{-13}$ \\
    4 & 7 & 32768 & $\{60, 60, 60, 60, 60, 60, 60\}$ & 43 & $2.2\times 10^{-13}$ \\
    5 & 6 & 32768 & $\{60, 32, 60, 60, 60, 60, 60, 60\}$ & 46 & $6.9\times 10^{-14}$ \\
    \bottomrule
    \end{tabular}
    \begin{tablenotes}
        % \small
        \item[*] The ciphertext bit width is in the Residue Number System (RNS) representation.
    \end{tablenotes}
    \end{threeparttable}
    }
\end{table}

\noindent \textbf{Oblivious Transfer (OT) Primitives.} Following~\cite{huang2022cheetah,pang2023bolt,lu2023bumblebee}, our nonlinear protocols are implemented with OT. There are two main types of OT primitives, i.e., the IKNP-style OT~\cite{kolesnikov2013improvedIKNP} used in BOLT~\cite{pang2023bolt} and the VOLE-style OT~\cite{boyle2019efficientVOLE} used in Bumblebee~\cite{lu2023bumblebee}. For a fair comparison, we evaluate~\method~with both OT primitives. Note that OT primitives do not affect the protocol construction of linear and nonlinear operators.

%, and there are two main types of OT primitives: the IKNP-style OT~\cite{kolesnikov2013improvedIKNP} and the VOLE-style OT~\cite{boyle2019efficientVOLE}. IKNP is more computationally efficient but has a higher communication cost, while VOLE has a lower communication cost but is less efficient in computation. Previous work BOLT~\cite{pang2023bolt} uses the IKNP-style OT, while Bumblebee~\cite{lu2023bumblebee} uses the VOLE-style OT. For a fair comparison, we evaluate~\method~under both OT primitives, named~\method~(IKNP) and~\method~(VOLE). Note that OT primitives do not affect the upper layer protocol construction.

\noindent \textbf{Protocols for Nonlinear Layers.} We use Bumblebee's protocol for $\operatorname{LayerNorm}$ and $\operatorname{Softmax}$, and BOLT's protocol for $\operatorname{GeLU}$. These nonlinear layers are further decomposed into operators and then processed using~\method. Appendix~\ref{app:nonlinear_protocols} provides a detailed review of these protocols.

\noindent \textbf{Test Environment.} 
\revise{All} experiments are conducted on machines with a 2.7 GHz Intel Xeon(R) Platinum 8558P CPU and 64 threads. We run the GPU experiments on an NVIDIA A100 GPU. Wireless network conditions are simulated using Linux's traffic control command: for the Local Area Network (LAN), we use 1 Gbps bandwidth and a 0.3ms roundtrip time. For the Wide Area Network (WAN), we simulate three settings: WAN$_1$ (400Mbps, 4ms), WAN$_2$ (100Mbps, 4ms)~\revise{and WAN$_3$ (100Mbps, 80ms)}. 

\noindent \textbf{Models and Datasets.} Following~\cite{pang2023bolt,lu2023bumblebee}, we evaluate~\method~on BERT-base~\cite{kenton2019bert}, BERT-large, and GPT2-base~\cite{radford2019languagegpt2} across four datasets from the GLUE benchmarks~\cite{wang2019gluemultitaskbenchmarkanalysis}: MRPC, RTE, SST2, and QNLI. We use the pre-trained models from huggingface~\cite{huggingface} \textbf{without fine-tuning}.
% and \textbf{do not fine-tune} them.
% \ml{Why are these datasets selected? Following BOLT?}

\noindent \textbf{Baselines.} We compare~\method~with prior-art private Transformer inference frameworks, including hybrid HE/MPC frameworks Iron~\cite{hao2022iron}, BOLT~\cite{pang2023bolt}, and Bumblebee~\cite{lu2023bumblebee}; 3PC frameworks SIGMA~\cite{gupta2023sigma} and MPCFormer~\cite{li2022mpcformer}; and the FHE-based framework NEXUS~\cite{cryptoeprint:2024/136NEXUS}.

\subsection{Microbenchmarks}
\begin{table}[t]
    \centering
    \huge
    \caption{The number of HE rotations and latency comparison for $Q_hK_h^T$ and $\operatorname{Softmax}\times V_h $ with previous methods.}
    \label{tab:exp_micro}
    \resizebox{0.9\linewidth}{!}{
    \begin{threeparttable}
    \begin{tabular}{c|c|c|c|c}
    \toprule
    \textbf{Model} & \textbf{Operator} & \textbf{Method} & \textbf{\# Rot} &\textbf{Latency (s)}\\
    \midrule
    \multirow{6}{*}{\makecell{BERT-large \\ 128 input tokens}} &\multirow{3}{*}{\makecell{$Q_hK_h^T $ \\ $h\in [H]$}} & BOLT~\cite{pang2023bolt}&14302& 24.0\\
    & & Powerformer~\cite{park2024powerformer}&5856& 8.0\\
    & & \textbf{BLB}&\textbf{640}& \textbf{1.7}\\
    \cmidrule{2-5}
    & \multirow{3}{*}{\makecell{$\operatorname{Softmax}\times V_h $ \\ $h\in [H]$}} & BOLT~\cite{pang2023bolt}&14332& 25.0\\
    & & Powerformer~\cite{park2024powerformer}&6508& 9.4\\
    & & \textbf{BLB}&\textbf{1056}& \textbf{2.9}\\
    \midrule 
    \multirow{6}{*}{\makecell{GPT2-base \\ 64 input tokens}} & \multirow{3}{*}{\makecell{$Q_hK_h^T $ \\ $h\in [H]$}} & BOLT~\cite{pang2023bolt} & 5356 & 5.0\\
    & & Powerformer~\cite{park2024powerformer} & 2304 & 2.5\\
    & & \textbf{BLB}&\textbf{488}& \textbf{1.4}\\
    \cmidrule{2-5}
    & \multirow{3}{*}{\makecell{$\operatorname{Softmax}\times V_h $ \\ $h\in [H]$}} & BOLT~\cite{pang2023bolt}&5374& 5.1\\
    & & Powerformer~\cite{park2024powerformer}&2496& 2.6\\
    & & \textbf{BLB}&\textbf{488}& \textbf{1.4}\\
    \bottomrule
    \end{tabular}
    % \begin{tablenotes}
    %     \footnotesize
    %     \item[*] BLB requires fewer rotations with a batch size of 16 than with 12, as the latter needs splitting into groups of 8 and 4, resulting in more rotations.
    % \end{tablenotes}
    \end{threeparttable}
    }
\end{table}

\noindent \textbf{MatMul Protocol Comparison.}
% \ml{Maybe move the micro benchmarks to the ablation study?}
% \ml{Also, why do we compare this batched MatMul? Should not we compare the efficiency for $X W_Q W_K^T X$ directly?}\xts{because we only improve $QK^T$ and $S\cdot V$, we do not improve $W_QX$.}
In Table~\ref{tab:exp_micro}, we benchmark~\method~on the $Q_hK_h^T$ and $\operatorname{Softmax}\times V_h $ from BERT-large and GPT2-base on CPU. Notably,~\method~reduces the number of rotations substantially, resulting in $6.9\times$ and $3.1\times$ lower latency than BOLT and Powerformer, respectively.
% \subsubsection{Profile of a Transformer block\xts{need to mention which protocol we use for nonlinear layers}}
% To demonstrate the advantages of the~\method~framework, we profile the communication and latency of a Transformer block in detail and compared~\method~with previous ``LW'' frameworks BOLT~\cite{} and bumblebee~\cite{lu2023bumblebee}. The results are shown in Figure~\ref{fig:micro_profile}. We can see that: \textbf{\underline{1)}} 

\subsection{End-to-End Evaluation}
% \subsubsection{Accuracy}
\begin{table}[!tb]
    \centering
    % \huge
    \caption{Accuracy comparison between plaintext and~\method~inference.}
    \label{tab:exp_acc}
    \resizebox{1.0\linewidth}{!}{
    \begin{tabular}{c|c|c|c|c|c|c}
    \toprule
    \multirow{2}{*}{}&  \multicolumn{2}{c|}{\textbf{Bert-base}} &  \multicolumn{2}{c|}{\textbf{Bert-large}} & \multicolumn{2}{c}{\textbf{GPT2-base}} \\
    \cmidrule{2-7}
    & \textbf{Plaintext} & \textbf{\method} & \textbf{Plaintext} & \textbf{\method} & \textbf{Plaintext} & \textbf{\method} \\
    \midrule
    MRPC & 88.8 & 88.5 & 89.2  & 89.1 & 82.7 & 82.5\\
    RTE & 69.3 & 69.7 & 70.4 & 70.0 & 66.1 & 66.4\\
    SST2 & 93.2 & 93.0 & 92.5 & 92.4 & 89.7 & 89.9\\
    QNLI & 91.3 & 91.2 & 91.9 & 91.7 & 88.9 & 88.9\\
    \bottomrule
    \end{tabular}
    }
\end{table}

\noindent \textbf{Accuracy Comparison.}
Table~\ref{tab:exp_acc} shows the private inference accuracy of~\method. We can see that~\method~achieves comparable levels of accuracy with plaintext inference. It should be emphasized that the main errors come from the piecewise linear approximations for nonlinear layers~\cite{lu2023bumblebee}, and we \textbf{do not perform any fine-tuning} on the models.~\revise{On certain datasets, approximation slightly improves accuracy. We attribute this to its implicit regularization effect, which helps alleviate overfitting by introducing controlled noise. Similar observations have been reported in Figure 1 in PUMA~\cite{dong2023puma} and Table 2 in BOLT~\cite{pang2023bolt}.}

\noindent \textbf{Communication Breakdown Analysis.}
\begin{figure*}[h]
    \centering
    \includegraphics[width=0.95\linewidth]{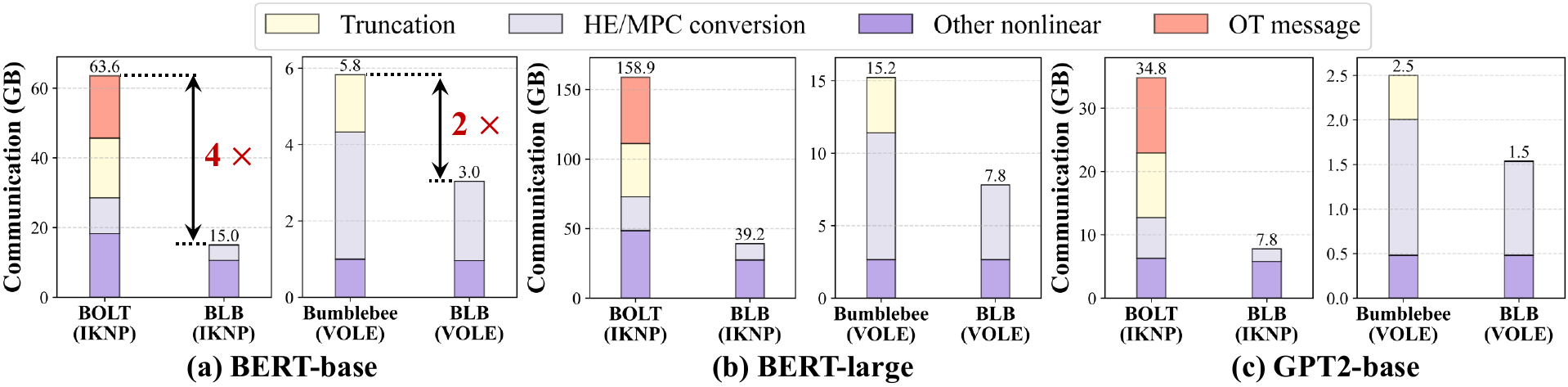}
    \caption{Communication comparison between~\method, BOLT and Bumblebee. The ones using the same OT primitive are grouped.}
    \label{fig:exp_comm}
\end{figure*}
In Figure~\ref{fig:exp_comm}, we show the communication breakdown of~\method~and compare it with BOLT and Bumblebee. 
% For a fair comparison,~\method~uses the same OT primitive as BOLT and Bumblebee.
As can be observed, \textbf{\underline{1)}} \method~reduces communication by approximately $4\times$ and $2\times$ across three models compared to BOLT and Bumblebee, respectively. \textbf{\underline{2)}} The improvement stems from the elimination of truncations and a reduction in HE/MPC conversions. \textbf{\underline{3)}} BOLT relies on OT for MatMul operators in nonlinear layers, which introduces extra ``OT message'' communication. In contrast, Bumblebee and~\method~use HE for all linear operators, removing the need for ``OT message'' communication. Besides, the ``other nonlinear'' parts including  $\operatorname{cmp}$, $\operatorname{rec}$ and $\operatorname{rsqrt}$ operators.

\noindent \textbf{Latency Breakdown Analysis.}
\begin{figure*}[h]
    \centering
    \includegraphics[width=0.95\linewidth]{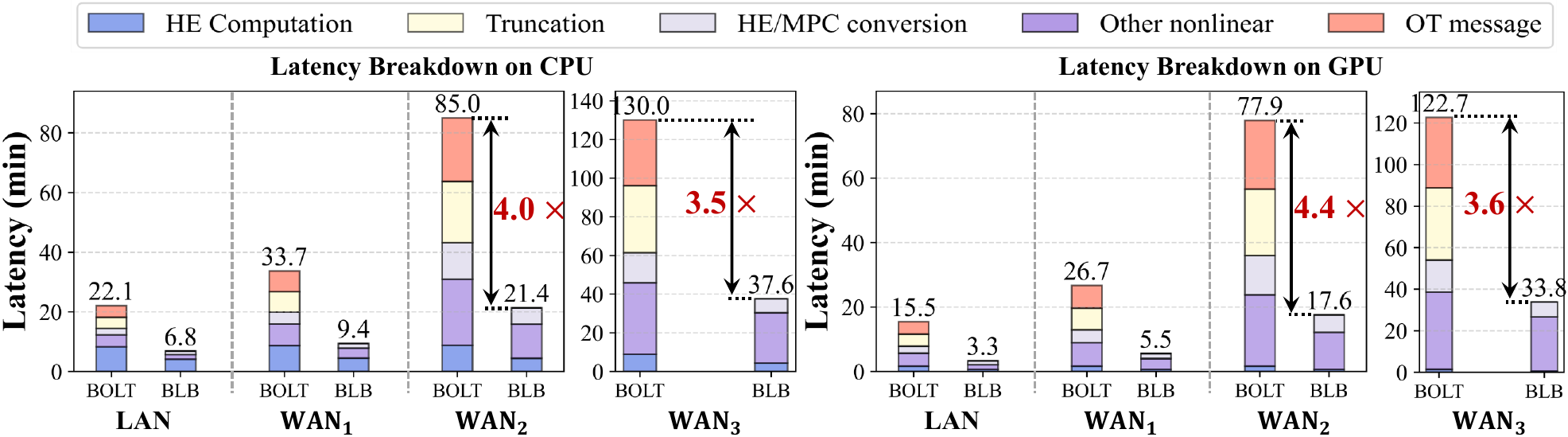}
    \caption{\revise{Latency comparison between~\method~and BOLT on BERT-base model on CPU and GPU. The bandwidths and round-trip times of four network conditions are: LAN:\{ 1Gbps, 0.3ms\}, WAN$_1$:\{400Mbps, 4ms\}, WAN$_2$:\{100Mbps, 4ms\}, WAN$_3$:\{100Mbps, 80ms\}.}}
    \label{fig:exp_time_bolt}
    \vspace{-5pt}
\end{figure*}
\begin{figure*}[h]
    \centering
    \includegraphics[width=0.94\linewidth]{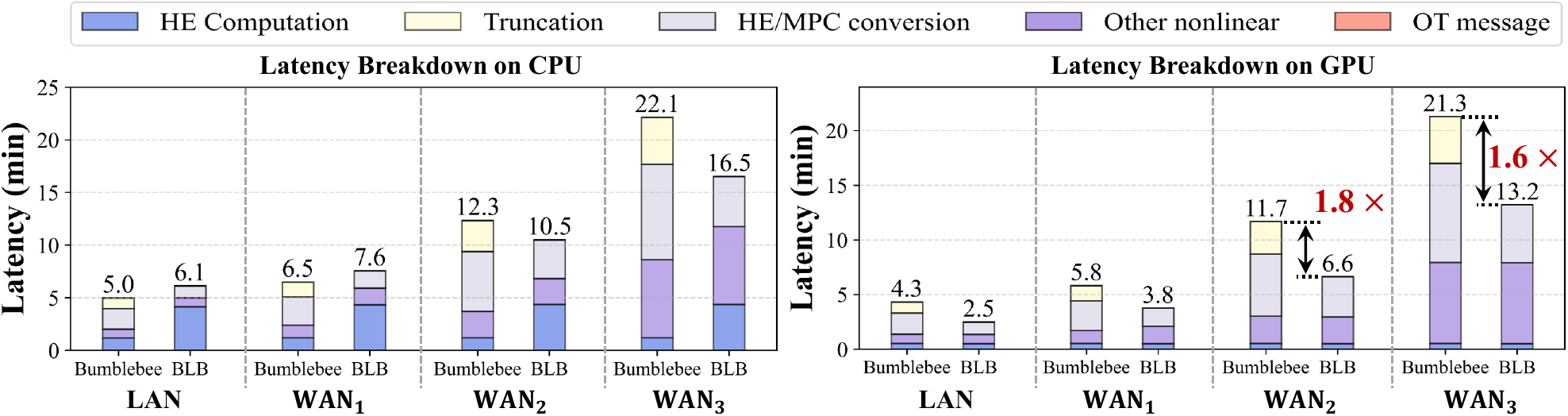}
    \caption{\revise{Latency comparison between~\method~and Bumblebee on BERT-base model on CPU and GPU. The bandwidths and round-trip times of four network conditions are:  LAN:\{ 1Gbps, 0.3ms\}, WAN$_1$:\{400Mbps, 4ms\}, WAN$_2$:\{100Mbps, 4ms\}, WAN$_3$:\{100Mbps, 80ms\}.}}
    \label{fig:exp_time_bumble}
\end{figure*}
\delete{
Figure~\ref{fig:exp_time} demonstrates the latency breakdown of~\method, BOLT, and Bumblebee across three network settings using the BERT-base model. Similarly,~\method~uses the same OT primitive as BOLT and Bumblebee. Key observations include: \textbf{\underline{1)}} As the network conditions change from LAN to WAN$_2$,~\method~consistently outperforms BOLT and Bumblebee. \textbf{\underline{2)}} Compared to BOLT,~\method~(IKNP) achieves $3.3\sim 4.0\times$ latency reduction. With GPU acceleration, the improvements are further amplified to $4.8\sim 6.7\times$. \textbf{\underline{3)}} Compared to Bumblebee, there exists \textbf{a communication and computation trade-off}. After fusion, the HE \textit{level} increases, resulting in more computation in~\method~compared to Bumblebee. However, we adhere to an optimization principle of \textbf{``accuracy > communication > computation''}. Under the premise of ensuring accuracy, reducing communication is prioritized, as advances in systems and hardware allow for computation to be accelerated more easily and parallelized~\cite{samardzic2021f1,samardzic2022craterlake,PhantomFHE_BFV}. As shown in Figure~\ref{fig:exp_time}, with GPU acceleration,~\method~achieves approximately $2\times$ latency reduction compared to Bumblebee.
}
\revise{Figure~\ref{fig:exp_time_bolt} and Figure~\ref{fig:exp_time_bumble} present the latency breakdown of~\method, BOLT, and Bumblebee across four network settings using the BERT-base model on both CPU and GPU. Notably,~\method~uses the same OT primitive as both BOLT and Bumblebee. The key observations are as follows:
\textbf{\underline{1)}} Compared to BOLT,~\method~(IKNP) achieves a $3.3\sim 4.0\times$ latency reduction on CPU and a $3.6\sim 4.9\times$ reduction on GPU.
\textbf{\underline{2)}} Compared to Bumblebee, there exists a clear trade-off between communication and computation. After fusion, the HE \textit{level} increases, which leads to higher computational cost and consequently increased latency under LAN compared to Bumblebee. Nevertheless, our design adheres to the optimization principle of \textbf{``accuracy > communication > computation''}. That is, while ensuring accuracy, communication overhead is minimized even at the expense of increased computation. This prioritization is justified by the fact that computation can be more readily accelerated and parallelized through hardware and system-level improvements~\cite{samardzic2021f1,samardzic2022craterlake,PhantomFHE_BFV}. As shown in Figure~\ref{fig:exp_time_bumble}, on GPU,~\method~achieves a $1.5\sim 1.8\times$ latency reduction compared to Bumblebee.}
% \subsubsection{Comparison with Existing Frameworks}

\noindent \textbf{Comparison with Existing Frameworks.}
\begin{table}[!tb]
    \centering
    % \huge
    \begingroup
    \color{black}
    \caption{\revise{End-to-end comparison with existing private Transformer inference frameworks. Iron and NEXUS results are taken from their papers due to a lack of end-to-end implementations. For other baselines, including SIGMA~\cite{gupta2023sigma}, MPCFormer~\cite{li2022mpcformer}, BOLT~\cite{pang2023bolt} and Bumblebee~\cite{lu2023bumblebee}, we re-run their codes on the same machine as~\method. For SIGMA, we report the latency of both the offline key transmission and online inference, whereas the original SIGMA paper only reports the latency of the online phase. We use VOLE-OT for~\method.}}
    \label{tab:exp_end2end_new}
    \resizebox{1.0\linewidth}{!}{
    % \begin{threeparttable}
    \begin{tabular}{ccSSSS}
    \toprule
    \multicolumn{1}{c}{\multirow{2}{*}{{Model}}} & \multicolumn{1}{c}{\multirow{2}{*}{Framework}} & \multicolumn{3}{c}{{Latency (min)}} & \multicolumn{1}{c}{\multirow{2}{*}{\makecell{Comm. \\ (GB)}}} \\
    \cmidrule{3-5}
    & & {LAN} & {WAN$_2$} & {WAN$_3$} & \\
    \midrule
    \multirow{8}{*}{\makecell{GPT2-base \\ 64 input tokens}} 
    & SIGMA~\cite{gupta2023sigma}${}^*$ & 7.7  & 25.1 & 38.5  & 28.7  \\
    & MPCFormer~\cite{li2022mpcformer}${}^*$ & 2.1  & 7.2 & 10.5  & 7.3 \\
    & BOLT~\cite{pang2023bolt} & 9.7  & 34.0 & 53.6  & 34.8 \\
    & Bumblebee~\cite{lu2023bumblebee} & \textbf{2.9}  & \textbf{5.2} & 12.3  & 2.5 \\
    & \method & {4.0}  & {6.8} &\textbf{10.2\ \ } & \textbf{1.5} \\
    \cmidrule{2-6}
    & BOLT (GPU)~\cite{pang2023bolt} & 7.6  & 31.8 & 51.6  & 34.8 \\
    & Bumblebee (GPU)~\cite{lu2023bumblebee} & 2.6  & 4.9 & 12.0  & 2.5 \\
    & \method~(GPU) & \textbf{2.0}  & \textbf{3.9} & \textbf{8.1}  & \textbf{1.5} \\
    \midrule
    \multirow{11}{*}{\makecell{BERT-base \\ 128 input tokens}} 
    & NEXUS~\cite{cryptoeprint:2024/136NEXUS} & \multicolumn{1}{c}{457}  & \multicolumn{1}{c}{458} & \multicolumn{1}{c}{470}  & 0.2 \\
    & Iron~\cite{hao2022iron} & 66.7  & \multicolumn{1}{c}{/} & / & 76.5 \\
    & SIGMA~\cite{gupta2023sigma}${}^*$ & 7.8  & 30.4 & 48.6 & 34.4 \\
    & MPCFormer~\cite{li2022mpcformer}${}^*$ & 5.5  & 18.0 & 27.5 & 12.1 \\
    & BOLT~\cite{pang2023bolt} & 22.1  & 85.0& 130.0 & 63.6 \\
    & Bumblebee~\cite{lu2023bumblebee} & \textbf{5.0}  & 12.3 & 22.1 & 5.8 \\
    & \method & 6.1  & \textbf{10.5\ \ } & \textbf{17.2\ \ } & \textbf{3.0} \\
    \cmidrule{2-6}
    & NEXUS (GPU)~\cite{cryptoeprint:2024/136NEXUS} & 19.9  & 20.8 & 32.5  & 0.2 \\
    & BOLT (GPU)~\cite{pang2023bolt} & 15.5  & 77.9& 122.7 & 63.6 \\
    & Bumblebee (GPU)~\cite{lu2023bumblebee} & 4.3  & 11.6 & 21.3 & 5.8 \\
    & \method~(GPU) & \textbf{2.5}  & \textbf{6.6} & \textbf{13.2}\ \    & \textbf{3.0} \\
    \midrule
    \multirow{9}{*}{\makecell{BERT-large \\ 128 input tokens}} 
    & Iron~\cite{hao2022iron} & 92.0  & \multicolumn{1}{c}{/}& / & 220.0 \\
    & SIGMA~\cite{gupta2023sigma}${}^*$ & 20.5  & 102.5&155.5 & 92.8 \\
    & MPCFormer~\cite{li2022mpcformer}${}^*$ & 7.7  & 34.1 & 52.0  & 32.6 \\
    & BOLT~\cite{pang2023bolt} & 57.6  & 222.0&274.4 & 158.9 \\
    & Bumblebee~\cite{lu2023bumblebee} & \textbf{10.6\ \ }  & 32.4 &51.3 & 15.2 \\
    & \method & 15.1  & \textbf{29.0}\ \  & \textbf{37.7}\ \   & \textbf{7.8} \\
    \cmidrule{2-6}
    & BOLT (GPU)~\cite{pang2023bolt} & 43.8  & 208.2&260.6 & 158.9 \\
    & Bumblebee (GPU)~\cite{lu2023bumblebee} & 9.7  & 30.8 &49.2 & 15.2 \\
    & \method~(GPU) & \textbf{6.6}  & \textbf{16.2 \ } & \textbf{24.9 \ }  & \textbf{7.8} \\
    \bottomrule
    \multicolumn{5}{l}{${}^*$ Frameworks that require three parties.} \\
    \end{tabular}
    % \begin{tablenotes}
    %     \footnotesize
    %     \item[*] Frameworks that require three parties.
    %     \item[$\top$] FHE-based framework with huge accuracy loss.
    % \end{tablenotes}
    % \end{threeparttable}
    }
    \endgroup
    \vspace{-5pt}
\end{table}
We present the end-to-end comparisons of~\method~with prior-art private inference frameworks across three Transformer models in Table~\ref{tab:exp_end2end_new}.
\delete{In summary,~\method~achieves SOTA performance in both latency and communication. Specifically,~\method~demonstrates a $1.7 \sim 29 \times$ communication reduction and a $1.5 \sim 75\times$ latency reduction. While NEXUS~\cite{cryptoeprint:2024/136NEXUS} exhibits lower communication, it suffers from significant accuracy degradation and is still less efficient than~\method~in terms of latency.}
\revise{In summary,~\method~achieves state-of-the-art communication and latency performance, with superior GPU efficiency. Specifically,~\method~reduces communication by $1.7 \sim 29\times$ and latency by $1.4 \sim 13\times$ on GPU. 
% On CPU,~\method~outperforms all but Bumblebee: Bumblebee is better under good network conditions, while BLB achieves lower latency in poor networks like WAN$_3$ due to reduced communication overhead.
Although NEXUS~\cite{cryptoeprint:2024/136NEXUS} achieves lower communication, it remains less efficient than~\method~in terms of latency due to its reliance on heavy bootstrapping.}
\subsection{Ablation Study}
\begin{table}[!tb]
    \centering
    \caption{Comparison of ciphertext bit width, degree, communication, and latency between BFV and CKKS.}
    \label{tab:exp_ablation_bfv}
    \resizebox{1.0\linewidth}{!}{
    \begin{threeparttable}
    \begin{tabular}{c|cc|cc|cc|cc}
    \toprule
    \multirow{2}{*}{\textbf{Block idx}} & \multicolumn{2}{c|}{\makecell{\textbf{Ciphertext} \\ \textbf{Bit Width (bits)}}} & \multicolumn{2}{c|}{\textbf{Degree}} & \multicolumn{2}{c|}{\makecell{\textbf{Comm.}\\ \textbf{(MB)}}} & \multicolumn{2}{c}{\makecell{\textbf{Latency} \\ \textbf{(s)}}} \\
    \cmidrule(lr){2-9}
    & \textbf{BFV} & \textbf{CKKS} & \textbf{BFV} & \textbf{CKKS} & \textbf{BFV} & \textbf{CKKS} & \textbf{BFV} & \textbf{CKKS} \\
    \midrule
    1 & 414 & 300 & 16384 & 16384 & 55 & 31 & 10 & 6 \\
    2 & 675 & 455 & 32768 & 32768 & 62 & 23 & 12 & 7 \\
    3 & 1400 & 400 & 65536\tnote{*} & 32768 & / & 46 & / & 2 \\
    4 & 636 & 420 & 32768 & 32768 & 81 & 40 & 14 & 8 \\
    5 & 902 & 452 & 65536\tnote{*} & 32768 & / & 26 & / & 5 \\
    \bottomrule
    \end{tabular}
    \begin{tablenotes}
        \footnotesize
        \item[*] Degrees larger than 32768 are not supported in the SEAL library.
    \end{tablenotes}
    \end{threeparttable}
    }
\end{table}
\textbf{Comparison with BFV.} In this study, we assess the performance of the CKKS scheme relative to the BFV scheme. The results are summarized in Table~\ref{tab:exp_ablation_bfv}~\revise{where block idx refers to the five blocks of fused linear operators in a Transformer layer, as shown in Figure~\ref{fig:fusion_pattern}.} The key observations are as follows: \textbf{\underline{1)}} The BFV scheme requires a larger ciphertext bit width due to its greater scale. This necessitates a higher polynomial degree to achieve the desired security level $\lambda$, a requirement that is not supported by certain homomorphic encryption libraries, such as SEAL~\cite{sealcrypto}. \textbf{\underline{2)}} The communication overhead of BFV is approximately twice that of CKKS, primarily because of its larger plaintext bit width, which increases the ciphertext size. In contrast, CKKS mitigates this issue by controlling the scale expansion, thereby reducing the communication cost. \textbf{\underline{3)}} Given its larger ciphertext bit width, BFV incurs significantly higher computational costs compared to CKKS, resulting in substantially increased latency. These observations highlight the advantages of the CKKS scheme, particularly in the efficient evaluation of consecutive linear operators.

\begin{table}[!tb]
    \centering
    \huge
    \begingroup
    \color{black}
    \caption{\revise{Ablation study on different components.}}
    \label{tab:exp_components}
    \resizebox{1.0\linewidth}{!}{
    % \begin{threeparttable}
    \begin{tabular}{cccSSS}
    \toprule
    \multicolumn{1}{c}{\multirow{2}{*}{{Model}}} & \multicolumn{1}{c}{\multirow{2}{*}{Framework}} & \multicolumn{1}{c}{\multirow{2}{*}{\makecell{Comm. \\ (GB)}}} & \multicolumn{3}{c}{{Latency (min)}}  \\
    \cmidrule{4-6}
    & & & LAN & {WAN$_2$} & {WAN$_3$} \\
    \midrule
    % \cmidrule{3-4}
    \multirow{4}{*}{\makecell{GPT2-base \\ 64 input tokens}}& baseline (BFV+MPC) & 2.5 & 2.6 & 4.9 & 12.0\\
    \cmidrule{2-6}
    & \makecell{+FineGrainFusion/ \\ CKKS-MPC Conversion} & 1.5 & 2.8 & 4.7 & 9.0 \\
    \cmidrule{2-6}
    & +Fused MatMul Protocol & 1.5 & 2.0 & 3.9 & 8.1 \\
    \midrule
    \multirow{4}{*}{\makecell{BERT-base \\ 128 input tokens}}& baseline (BFV+MPC) & 5.8 & 4.3 & 11.6&21.3\\
    \cmidrule{2-6}
    & \makecell{+FineGrainFusion/ \\ CKKS-MPC Conversion} & 3.0& 4.1 & 8.2 & 14.8 \\
    \cmidrule{2-6}
    & +Fused MatMul Protocol & 3.0 & 2.5 & 6.6 & 13.2 \\
    \midrule
    \multirow{4}{*}{\makecell{BERT-large \\ 128 input tokens}}& baseline (BFV+MPC) & 15.2 & 9.7 & 30.8 & 49.2\\
    \cmidrule{2-6}
    & \makecell{+FineGrainFusion/ \\ CKKS-MPC Conversion} & 7.8 & 10.7 & 20.4 &29.1 \\
    \cmidrule{2-6}
    & +Fused MatMul Protocol & 7.8 & 6.6 & 16.2 &24.9 \\
    \bottomrule
    \end{tabular}
    }
    \endgroup
    \vspace{-5pt}
\end{table}
\revise{
\noindent\textbf{Effectiveness of Different Components.} We present ablation studies across three models in Table~\ref{tab:exp_components}, using Bumblebee's protocol with BFV and MPC as a strong baseline without fusion. Applying FineGrainFusion and our hybrid CKKS-MPC protocol achieves around a $2\times$ reduction in communication, with some increase in HE computation. This trade-off becomes more favorable under poor network conditions. Under WAN$_3$, FineGrainFusion reduces latency by $1.3\sim 1.8\times$. Incorporating the fused MatMul protocol further lowers HE computation, leading to a $1.4\sim 1.6\times$ reduction in end-to-end latency under LAN.
}

\noindent \revise{\textbf{Comparison with NEXUS under Batched Inputs.} We provide this experiment in Appendix~\ref{app:exp}. It should be emphasized that batching is challenging in private inference because data from different users \textbf{cannot} be packed into the same ciphertext, as each user's data is encrypted under a different secret key. As a result, batching is only applicable to multiple queries from the same user, which significantly limits its use.}
\section{Related works}\label{sec:related}
\textbf{Private Transformer Inference. }
With the proliferation of ChatGPT, significant efforts have been made to enable private Transformer inference \cite{hao2022iron,pang2023bolt,lu2023bumblebee,huang2024secbert,luo2024secformer,zeng2024securegpt,xu2024privcirnet,cryptoeprint:2024/136NEXUS,park2024powerformer,cryptoeprint:2024/1881THOR,li2022mpcformer,akimoto2023privformer,gupta2023sigma,dong2023puma,chen2024accelerating,li2024nimbus,he2024rhombus}.
Iron~\cite{hao2022iron} is the first 2PC-based framework that optimizes the packing for $\operatorname{matmul_{cp}}$ and nonlinear layer protocols. Building on this, BOLT~\cite{pang2023bolt} and Bumblebee~\cite{lu2023bumblebee} further optimize the packing strategies and propose accurate piecewise linear approximations for nonlinear layers without fine-tuning. FHE-based frameworks have also been explored for private Transformer inference. NEXUS~\cite{cryptoeprint:2024/136NEXUS} is the first FHE-based framework that features novel HE-based computation of linear and nonlinear Transformer operators. However, NEXUS requires high-order polynomial approximation for nonlinear layers and thus suffers from high computational costs. Powerformer~\cite{park2024powerformer} improves upon NEXUS by incorporating the $\mmcc$ protocol from~\cite{jiang2018secure} but still suffers from high cost for non-square MatMuls in Transformers. THOR~\cite{cryptoeprint:2024/1881THOR} proposes a new diagonal packing algorithm to further reduce the HE rotations but at the cost of higher multiplicative depth of both $\mmcp$ and $\mmcc$. Additionally, protocols such as PUMA~\cite{dong2023puma}, MPCFormer~\cite{li2022mpcformer}, PrivFormer~\cite{akimoto2023privformer}, and SIGMA~\cite{gupta2023sigma} have also studied three-party computation schemes for private Transformer inference.

% that features various novel packing methods for FHE but incurs high computational costs. PowerFormer~\cite{park2024powerformer} reduced HE rotation costs by employing ciphertext-ciphertext matrix multiplication algorithms from Jiang et al~\cite{jiang2018secure}. 
% The concurrent work THOR~\cite{cryptoeprint:2024/1881THOR} proposes using a diagonal representation for encoding, which reduces the number of rotations but increases the multiplicative depth of both $\operatorname{ct-pt}$ and $\operatorname{ct-ct}$ MatMuls.
% THOR~\cite{cryptoeprint:2024/1881THOR} introduces a series of new packing methods, trading higher multiplication depth for fewer HE rotations. THOR effectively reduces latency because, in the FHE framework, the additional multiplication depth in linear layers has minimal impact, as the majority of the multiplication depth is concentrated in nonlinear functions.
% However, THOR's method is unsuitable for hybrid HE/MPC frameworks, as increasing the multiplication depth significantly inflates the ciphertext bit width, leading to higher communication and computation costs\xts{need double check about THOR}.
\textbf{Layer Fusion Optimization. }
Layer fusion has been proposed in several works to reduce the cost of private inference. PrivCirNet~\cite{xu2024privcirnet} fuses consecutive convolution layers in MobileNetV2~\cite{sandler2018mobilenetv2}. Zhang et al.~\cite{zhang2024individual} jointly optimize the evaluation of the convolution layer and ReLU layer in CNNs to reduce HE rotations and multiplications. For Transformer, BOLT~\cite{pang2023bolt} fuses two consecutive MatMuls in the attention layers. All these approaches are based on layer-wise evaluation, which limits their performance improvements.
% A contemporaneous work, FASTLMPI~\cite{chen2024accelerating}, explored collaborative computation using HE and SS, replacing the OT protocol for linear operators in BOLT's nonlinear layers with an HE-based implementation. However, Bumblebee~\cite{lu2023bumblebee} had already adopted this approach, and FASTLMPI does not address layer fusion optimizations.
\section{Conclusion}\label{sec:conclusion}
% We propose~\method~, the first secure hybrid CKKS and MPC framework for private Transformer inference.~\method~decomposes the layers into individual operators and fuses adjacent linear operators. Through a systematic analysis of linear operator fusion and an optimized MatMul protocol for fused computations,~\method~achieves significant improvements in both communication efficiency and latency compared to prior-art approaches, marking a significant step towards practical secure inference for Transformers.

We propose a hybrid HE/MPC framework, dubbed \method, to remodel private Transformer inference. \method~breaks the layer barrier in Transformers to allow fine-grained fusion of linear operators, drastically reducing the communication induced by HE/MPC conversion. \method~employs CKKS-based computation of fused linear operators to overcome the ciphertext bit width increase, enabled by our first secure CKKS/MPC conversion protocol. \method~further designs efficient packing algorithms for fused matrix multiplications with multi-head and BSGS optimizations. Our experiments demonstrate up to 21$\times$ communication reduction compared to existing methods, leading to up to \delete{14$\times$}\revise{13$\times$} latency reduction.

% To overcome the bit width increase, we propose the first secure yet efficient CKKS/MPC conversion protocol to enable CKKS-based computation of fused operators. 
% \method~incorporates efficient packing algorithms for fused matrix multiplications in Transformers leveraging the batch processing nature of the multi-head attention.
% To reduce the bit width increase induced by operator fusion, we propose the first secure yet efficient CKKS/MPC conversion protocol to enable CKKS-based computation of fused operators. 

% \clearpage
\section{Ethical Considerations}
This work focuses on improving the efficiency of privacy-preserving inference frameworks. All datasets and models used are publicly available; no proprietary, sensitive, or personal data is involved. We followed strict ethical standards to ensure no privacy violations occurred during experimentation or could result from our proposed methods.

Our contributions aim to strengthen secure computation without introducing new risks to data security or user privacy. By enhancing the practicality and scalability of privacy-preserving machine learning, this work encourages responsible adoption of privacy-centric technologies.

The ethical impact has been carefully evaluated. Since the methods do not involve sensitive data and are designed to bolster privacy and security in machine learning applications. We believe this work aligns with principles of privacy, security, and responsible innovation.

\section{Open Science}
To promote availability, all datasets and pre-trained models used in this work are publicly accessible through their sources, as referenced in the paper. Upon acceptance, we will release the experimental logs and testing code in a public repository to enable immediate availability of our results.
Further organization and release of the full source code, including implementation details, will follow in subsequent updates. The repository will be accompanied by documentation describing the experimental setup, including hardware configurations, software dependencies, and parameter settings. By adhering to open science principles, we aim to support validation efforts and foster further research in privacy-preserving inference.
% \clearpage
%-------------------------------------------------------------------------------

\bibliographystyle{plain}
\bibliography{usenix}

\begin{thebibliography}{10}

\bibitem{sealcrypto}
{M}icrosoft {SEAL} (release 3.6).
\newblock \url{https://github.com/Microsoft/SEAL}, November 2020.

\bibitem{akimoto2023privformer}
Yoshimasa Akimoto, Kazuto Fukuchi, Youhei Akimoto, and Jun Sakuma.
\newblock Privformer: Privacy-preserving transformer with mpc.
\newblock In {\em 2023 IEEE 8th EuroS\&P}, pages 392--410. IEEE, 2023.

\bibitem{balla2023heliks}
Shashank Balla and Farinaz Koushanfar.
\newblock Heliks: He linear algebra kernels for secure inference.
\newblock In {\em Proceedings of the 2023 ACM SIGSAC Conference on Computer and Communications Security}, pages 2306--2320, 2023.

\bibitem{chandran2017ezpc}
Nishanth Chandran and Divya~Gupta et~al.
\newblock Ezpc: Programmable, efficient, and scalable secure two-party computation for machine learning.
\newblock {\em Cryptology ePrint Archive}, 2017.

\bibitem{chen2024accelerating}
Yuntian Chen and Zhanyong~Tang et~al.
\newblock Accelerating private large transformers inference through fine-grained collaborative computation.
\newblock {\em arXiv preprint arXiv:2412.16537}, 2024.

\bibitem{cheng2022personal}
Peng Cheng and Utz Roedig.
\newblock Personal voice assistant security and privacy—a survey.
\newblock {\em Proceedings of the IEEE}, 110(4):476--507, 2022.

\bibitem{cheon2019full}
Jung~Hee Cheon, Kyoohyung Han, Andrey Kim, Miran Kim, and Yongsoo Song.
\newblock A full rns variant of approximate homomorphic encryption.
\newblock In {\em Selected Areas in Cryptography--SAC 2018: 25th International Conference}, pages 347--368. Springer, 2019.

\bibitem{cheon2017homomorphic}
Jung~Hee Cheon, Andrey Kim, Miran Kim, and Yongsoo Song.
\newblock Homomorphic encryption for arithmetic of approximate numbers.
\newblock In {\em Advances in Cryptology--ASIACRYPT 2017: 23rd International Conference on the Theory and Applications of Cryptology and Information Security}, pages 409--437. Springer, 2017.

\bibitem{dathathri2020eva}
Roshan Dathathri and Blagovesta~Kostova et~al.
\newblock Eva: An encrypted vector arithmetic language and compiler for efficient homomorphic computation.
\newblock In {\em Proceedings of the 41st ACM SIGPLAN Conference on Programming Language Design and Implementation}, 2020.

\bibitem{dong2023puma}
Ye~Dong and Wen jie Lu~et al.
\newblock Puma: Secure inference of llama-7b in five minutes.
\newblock {\em arXiv preprint arXiv:2307.12533}, 2023.

\bibitem{radford2019languagegpt2}
Alec~Radford et~al.
\newblock Language models are unsupervised multitask learners.
\newblock {\em OpenAI blog}, 1(8):9, 2019.

\bibitem{zeng2024securegpt}
Chenkai~Zeng et~al.
\newblock Securegpt: A framework for multi-party privacy-preserving transformer inference in gpt.
\newblock {\em IEEE Transactions on Information Forensics and Security}, 2024.

\bibitem{rathee2020cryptflow2}
Deevashwer~Rathee et~al.
\newblock Cryptflow2: Practical 2-party secure inference.
\newblock In {\em Proceedings of the 2020 ACM SIGSAC Conference on Computer and Communications Security}, pages 325--342, 2020.

\bibitem{boyle2019efficientVOLE}
Elette~Boyle et~al.
\newblock Efficient two-round ot extension and silent non-interactive secure computation.
\newblock In {\em Proceedings of the 2019 ACM SIGSAC Conference on Computer and Communications Security}.

\bibitem{boemer2020mp2ml}
Fabian~Boemer et~al.
\newblock Mp2ml: A mixed-protocol machine learning framework for private inference.
\newblock In {\em Proceedings of the 15th International Conference on Availability, Reliability and Security}, pages 1--10, 2020.

\bibitem{touvron2023llama}
Hugo~Touvron et~al.
\newblock Llama: Open and efficient foundation language models.
\newblock {\em arXiv preprint arXiv:2302.13971}, 2023.

\bibitem{he2024rhombus}
Jiaxing~He et~al.
\newblock Rhombus: Fast homomorphic matrix-vector multiplication for secure two-party inference.
\newblock In {\em CCS}, 2024.

\bibitem{chen2024secure}
Jingwei~Chen et~al.
\newblock Secure transformer-based neural network inference for protein sequence classification.
\newblock {\em Cryptology ePrint Archive}, 2024.

\bibitem{santos2024curl}
Manuel B~Santos et~al.
\newblock Curl: Private llms through wavelet-encoded look-up tables.
\newblock {\em Conference on Applied Machine Learning for Information Security}, 2024.

\bibitem{samardzic2022craterlake}
Nikola~Samardzic et~al.
\newblock Craterlake: a hardware accelerator for efficient unbounded computation on encrypted data.
\newblock In {\em Proceedings of the 49th Annual International Symposium on Computer Architecture}, 2022.

\bibitem{Mishra_Delphi_2020}
Pratyush~Mishra et~al.
\newblock Delphi: A cryptographic inference service for neural networks, Jan 2020.

\bibitem{singh2024hyena}
Sarabjeet~Singh et~al.
\newblock Hyena: Balancing packing, reuse, and rotations for encrypted inference.
\newblock In {\em 2024 IEEE Symposium on Security and Privacy (SP)}, pages 107--107. IEEE Computer Society, 2024.

\bibitem{hussain2021coinn}
Siam Umar~Hussain et~al.
\newblock Coinn: Crypto/ml codesign for oblivious inference via neural networks.
\newblock In {\em Proceedings of the 2021 ACM SIGSAC Conference on Computer and Communications Security}, pages 3266--3281, 2021.

\bibitem{xu2024privcirnet}
Tianshi~Xu et~al.
\newblock Privcirnet: Efficient private inference via block circulant transformation.
\newblock {\em Neural Information Processing Systems (NeurIPS)}, 2024.

\bibitem{wang2022characterization}
Yongqin~Wang et~al.
\newblock Characterization of mpc-based private inference for transformer-based models.
\newblock In {\em 2022 IEEE ISPASS}, pages 187--197.

\bibitem{li2024nimbus}
Zhengyi~Li et~al.
\newblock Nimbus: Secure and efficient two-party inference for transformers.
\newblock {\em NeurIPS}, 2024.

\bibitem{samardzic2021f1}
Samardzic~Nikola et~el.
\newblock F1: A fast and programmable accelerator for fully homomorphic encryption.
\newblock In {\em MICRO-54: 54th Annual IEEE/ACM International Symposium on Microarchitecture}, pages 238--252, 2021.

\bibitem{goldreich1998secure}
Oded Goldreich.
\newblock Secure multi-party computation.
\newblock {\em Manuscript. Preliminary version}, 78(110):1--108, 1998.

\bibitem{DBLP:books/cu/Goldreich2004}
Oded Goldreich.
\newblock {\em The Foundations of Cryptography - Volume 2: Basic Applications}.
\newblock Cambridge University Press, 2004.

\bibitem{gupta2023sigma}
Kanav Gupta, Neha Jawalkar, Ananta Mukherjee, Nishanth Chandran, Divya Gupta, Ashish Panwar, and Rahul Sharma.
\newblock Sigma: Secure gpt inference with function secret sharing.
\newblock {\em Cryptology ePrint Archive}, 2023.

\bibitem{hao2022iron}
Meng Hao, Hongwei Li, Hanxiao Chen, Pengzhi Xing, Guowen Xu, and Tianwei Zhang.
\newblock Iron: Private inference on transformers.
\newblock In {\em Advances in Neural Information Processing Systems}, 2022.

\bibitem{huang2024secbert}
Hai Huang and Yongjian Wang.
\newblock Secbert: Privacy-preserving pre-training based neural network inference system.
\newblock {\em Neural Networks}, 172:106135, 2024.

\bibitem{huang2022cheetah}
Zhicong Huang, Wen-jie Lu, Cheng Hong, and Jiansheng Ding.
\newblock Cheetah: Lean and fast secure $\{$Two-Party$\}$ deep neural network inference.
\newblock In {\em USENIX Security Symposium 2022}, pages 809--826, 2022.

\bibitem{huggingface}
Huggingface.
\newblock \url{https://huggingface.co/}.

\bibitem{jiang2018secure}
Xiaoqian Jiang, Miran Kim, Kristin Lauter, and Yongsoo Song.
\newblock Secure outsourced matrix computation and application to neural networks.
\newblock In {\em Proceedings of the 2018 ACM SIGSAC conference on computer and communications security}, pages 1209--1222, 2018.

\bibitem{ju2023neujeans}
Jae~Hyung Ju, Jaiyoung Park, Jongmin Kim, Donghwan Kim, and Jung~Ho Ahn.
\newblock Neujeans: Private neural network inference with joint optimization of convolution and bootstrapping.
\newblock {\em The ACM Conference on Computer and Communications Security (CCS)}, 2024.

\bibitem{Juvekar_Vaikuntanathan_gazelle_2018}
Chiraag Juvekar, Vinod Vaikuntanathan, and AnanthaP. Chandrakasan.
\newblock {GAZELLE}: A low latency framework for secure neural network inference, Jan 2018.

\bibitem{kenton2019bert}
Jacob Devlin Ming-Wei~Chang Kenton and Lee~Kristina Toutanova.
\newblock Bert: Pre-training of deep bidirectional transformers for language understanding.
\newblock In {\em Proceedings of naacL-HLT}, volume~1, page~2, 2019.

\bibitem{kim2023optimized_fhe}
Dongwoo Kim and Cyril Guyot.
\newblock Optimized privacy-preserving cnn inference with fully homomorphic encryption.
\newblock {\em IEEE Transactions on Information Forensics and Security}, 18:2175--2187, 2023.

\bibitem{kolesnikov2013improvedIKNP}
Vladimir Kolesnikov and Ranjit Kumaresan.
\newblock Improved ot extension for transferring short secrets.
\newblock In {\em Advances in Cryptology--CRYPTO 2013: 33rd Annual Cryptology Conference, 2013.}, pages 54--70, 2013.

\bibitem{kumar2020cryptflow}
Nishant Kumar, Mayank Rathee, Nishanth Chandran, Divya Gupta, Aseem Rastogi, and Rahul Sharma.
\newblock Cryptflow: Secure tensorflow inference.
\newblock In {\em 2020 IEEE Symposium on Security and Privacy (SP)}, 2020.

\bibitem{lee2022low}
Eunsang Lee, Joon-Woo Lee, Junghyun Lee, Young-Sik Kim, Yongjune Kim, Jong-Seon No, and Woosuk Choi.
\newblock Low-complexity deep convolutional neural networks on fully homomorphic encryption using multiplexed parallel convolutions.
\newblock In {\em International Conference on Machine Learning}, pages 12403--12422. PMLR, 2022.

\bibitem{li2022mpcformer}
Dacheng Li, Rulin Shao, Hongyi Wang, Han Guo, Eric~P Xing, and Hao Zhang.
\newblock Mpcformer: fast, performant and private transformer inference with mpc.
\newblock {\em arXiv preprint arXiv:2211.01452}, 2022.

\bibitem{li2023efficient}
Yun Li, Yufei Duan, Zhicong Huang, Cheng Hong, Chao Zhang, and Yifan Song.
\newblock Efficient 3pc for binary circuits with application to maliciously-secure dnn inference.
\newblock In {\em 32nd USENIX Security Symposium (USENIX Security 23)}, pages 5377--5394, 2023.

\bibitem{lu2023bumblebee}
Wen-jie Lu, Zhicong Huang, Zhen Gu, Jingyu Li, Jian Liu, Kui Ren, Cheng Hong, Tao Wei, and WenGuang Chen.
\newblock Bumblebee: Secure two-party inference framework for large transformers.
\newblock {\em Network and Distributed System Security (NDSS)}, 2025.

\bibitem{luo2024secformer}
Jinglong Luo and Yehong~Zhang et~al.
\newblock Secformer: Fast and accurate privacy-preserving inference for transformer models via smpc.
\newblock In {\em Findings of the Association for Computational Linguistics ACL}, 2024.

\bibitem{mohassel2017secureml}
Payman Mohassel and Yupeng Zhang.
\newblock Secureml: A system for scalable privacy-preserving machine learning.
\newblock In {\em 2017 IEEE symposium on security and privacy (SP)}, pages 19--38. IEEE, 2017.

\bibitem{cryptoeprint:2024/1881THOR}
Jungho Moon, Dongwoo Yoo, Xiaoqian Jiang, and Miran Kim.
\newblock {THOR}: Secure transformer inference with homomorphic encryption.
\newblock Cryptology {ePrint} Archive, Paper 2024/1881, 2024.

\bibitem{pang2023bolt}
Q.~Pang, J.~Zhu, H.~Möllering, W.~Zheng, and T.~Schneider.
\newblock Bolt: Privacy-preserving, accurate and efficient inference for transformers.
\newblock In {\em 2024 IEEE Symposium on Security and Privacy (SP)}, pages 133--133, Los Alamitos, CA, USA, may 2024. IEEE Computer Society.

\bibitem{park2024powerformer}
Dongjin Park, Eunsang Lee, and Joon-Woo Lee.
\newblock Powerformer: Efficient privacy-preserving transformer with batch rectifier-power max function and optimized homomorphic attention.
\newblock {\em Cryptology ePrint Archive}, 2024.

\bibitem{rathee2021sirnn}
Deevashwer Rathee, Mayank Rathee, Rahul Kranti~Kiran Goli, Divya Gupta, Rahul Sharma, Nishanth Chandran, and Aseem Rastogi.
\newblock Sirnn: A math library for secure rnn inference.
\newblock In {\em 2021 IEEE Symposium on Security and Privacy (SP)}, pages 1003--1020. IEEE, 2021.

\bibitem{sandler2018mobilenetv2}
Mark Sandler and Andrew~Howard et~al.
\newblock Mobilenetv2: Inverted residuals and linear bottlenecks.
\newblock In {\em Proceedings of the IEEE conference on computer vision and pattern recognition}, pages 4510--4520, 2018.

\bibitem{sarker2024transformer}
Prodip~Kumar Sarker, Qingjie Zhao, and Md~Kamal Uddin.
\newblock Transformer-based person re-identification: a comprehensive review.
\newblock {\em IEEE Transactions on Intelligent Vehicles}, 2024.

\bibitem{shamshad2023Transformers}
Fahad Shamshad and Salman~Khan et~al.
\newblock Transformers in medical imaging: A survey.
\newblock {\em Medical Image Analysis}, 88:102802, 2023.

\bibitem{PhantomFHE_BFV}
Shiyu Shen and Hao~Yang et~al.
\newblock Leveraging gpu in homomorphic encryption: Framework design and analysis of bfv variants.
\newblock {\em IEEE Transactions on Computers}, 73(12):2817--2829, 2024.

\bibitem{wang2019gluemultitaskbenchmarkanalysis}
Alex Wang, Amanpreet Singh, Julian Michael, Felix Hill, Omer Levy, and Samuel~R. Bowman.
\newblock Glue: A multi-task benchmark and analysis platform for natural language understanding, 2019.

\bibitem{zeng2022mpcvit}
Wenxuan Zeng, Meng Li, Wenjie Xiong, Wenjie Lu, Jin Tan, Runsheng Wang, and Ru~Huang.
\newblock Mpcvit: Searching for mpc-friendly vision transformer with heterogeneous attention.
\newblock {\em arXiv preprint arXiv:2211.13955}, 2022.

\bibitem{cryptoeprint:2024/136NEXUS}
Jiawen Zhang and Xinpeng~Yang et~al.
\newblock Secure transformer inference made non-interactive.
\newblock Network and Distributed System Security (NDSS), 2025.

\bibitem{zhang2024individual}
Qiao Zhang, Tao Xiang, Chunsheng Xin, and Hongyi Wu.
\newblock From individual computation to allied optimization: Remodeling privacy-preserving neural inference with function input tuning.
\newblock In {\em 2024 IEEE Symposium on Security and Privacy (SP)}, pages 101--101.

\end{thebibliography}
% \usepackage[backend=biber,style=numeric,maxnames=1,minnames=1]{biblatex}
% \addbibresource{usenix.bib}
% \printbibliography

% \clearpage
\appendix
\section{Security Discussion}\label{app:security}
\paragraph{Security of \method} In our threat model, we assume that both the client and the server behave \textit{honestly-but-curiously}. The methods introduced in \textsection~\ref{sec:FGF} and \textsection~\ref{sec:encode} are built upon well-established cryptographic foundations, including CKKS and MPC protocols, which have been thoroughly validated as secure~\cite{cheon2017homomorphic,goldreich1998secure}. Additionally, the security of the proposed CKKS and MPC conversion protocol is verified below.
\vspace{-5pt}

\paragraph{\revise{Proof of Theorem~\ref{thm:mask}}}
\revise{
\begin{proof}
Our proof follows the simulation-based proof~\cite{DBLP:books/cu/Goldreich2004}.
Denote $\mathcal{O}_0, \mathcal{O}_1 \leftarrow \mathcal{F}_{C2M}(\Enc(\Encode(x))$ as the outputs of the functionality.
Let ${\tt view}_b, {\tt out}_b$ be the view and output of the party $P_b$ in the execution of Algorithm~\ref{alg:ckks2mpc}.
We construct simulators $\mathcal{S}_0$ and $\mathcal{S}_1$ such that:
\vspace{-4pt}
\begin{align*}
&\{\mathcal{S}_0( \text{Key},\mathcal{O}_{0}), \mathcal{O}_{0}, \mathcal{O}_{1}\} \equiv^c \{{\tt view}_0, {\tt out}_{0}, {\tt out}_{1}\}\\
&\{\mathcal{S}_1( \Enc(\Encode(x),\mathcal{O}_{1}), \mathcal{O}_{0}, \mathcal{O}_{1}\} \equiv^c \{{\tt view}_1, {\tt out}_{0}, {\tt out}_{1}\}
\end{align*}
\vspace{-0pt}
where $\equiv^c$ denotes computationally indistinguishable.

The view of $P_0$ in the execution of the real protocol includes: 1) the secret key, 2) the masked CKKS ciphertext, and 3) its view in the $\operatorname{Field-to-Ring}$ sub-protocol~\cite{rathee2021sirnn}.
$\mathcal{S}_0$ simulates these messages by:

\ding{182} $S_0$ inputs with random generated $\text{key}$ and $O_0$ where $O_0$ is the decoded result of a uniform plaintext polynomial. Then, $S_0$ encodes $O_0$ to the uniform plaintext polynomial and then encrypts it with $\text{Key}$ to get a CKKS ciphertext which is indistinguishable from the 2nd message in ${\tt view}_0$.

\ding{183} Calling the underlying simulator from \cite{rathee2021sirnn}
% \footnote{We adapt the signed ring-to-ring conversion protocol in~\cite{rathee2021sirnn} to field-to-ring conversion and will be detailed in \textsection \ref{app:ring-field}.}
.

For the output part, recall that the functionality $\mathcal{F}_{C2M}$ outputs $O_0, O_1$, an additive share of a uniformly masked CKKS decoded plaintext. Thus, $O_0, O_1$ are both decoded results of the uniformly distributed polynomials over $\mathbb{A}_{N, 2^l}$. These are exactly how Algorithm~\ref{alg:ckks2mpc} computes
${\tt output}_0 = \Decode(\ashr{{\tt tmp}}^{2^l}_0),{\tt output}_1 = \Decode(\ashr{{\tt tmp}}^{2^l}_1)$. 
Therefore, by the semantic security of CKKS and the uniformity of the additive share construction, the joint distribution of $\{\mathcal{S}_0( \text{Key},\mathcal{O}_{0}), \mathcal{O}_{0}, \mathcal{O}_{1}\}$ is indistinguishable from $ \{{\tt view}_0, {\tt output}_{0}, {\tt output}_{1}\}$.

The view of $P_1$ in the execution of the real protocol includes 1) Its input $\Enc(\Encode(x))$, 2) $\poly{r}$, a randomly sampled polynomial,
3) its view in the $\operatorname{Field-to-Ring}$ sub-protocol.
$\mathcal{S}_1$ simulates these messages by:

\ding{182} $S_1$ inputs with a ciphertext $\Enc(\Encode(x))$ and $O_1$. The ciphertext $\Enc(\Encode(x))$ is computationally indistinguishable from the 1st message in ${\tt view}_1$.

\ding{183} Sampling polynomial from $\mathbb{A}_{N, q}$ uniformly at random. 
The distribution of this message is identical to the 2nd message in ${\tt view}_1$.

\ding{184} Calling the underlying simulator from \cite{rathee2021sirnn}. 

For the output part, $O_0,O_1$ are both decoding results of uniform random polynomials, which is exactly how Algorithm~\ref{alg:ckks2mpc} computes
${\tt output}_0 = \Decode(\ashr{{\tt tmp}}^{2^l}_0),{\tt output}_1 = \Decode(\ashr{{\tt tmp}}^{2^l}_1)$. 
Thus, the joint distribution of $\{\mathcal{S}_1( \Enc(\Encode(x),\mathcal{O}_{1}), \mathcal{O}_{0}, \mathcal{O}_{1}\}$ is indistinguishable from $\{{\tt view}_1, {\tt output}_{0}, {\tt output}_{1}\}$.
\end{proof}
}
\vspace{-1em}
\section{Protocols for Nonlinear Layers}\label{app:nonlinear_protocols}

The nonlinear layers in Transformer models mainly include $\operatorname{LayerNorm}$, $\operatorname{Softmax}$, and $\operatorname{GeLU}$~\cite{kenton2019bert,radford2019languagegpt2}. For $\operatorname{LayerNorm}$ and $\operatorname{Softmax}$, we leverage Bumblebee's protocols~\cite{lu2023bumblebee}, while for $\operatorname{GeLU}$, we employ BOLT's protocol~\cite{pang2023bolt}.
\begin{algorithm}[!tb]
    % \DontPrintSemicolon
    % \SetAlgoLined
    \KwIn{$P_0\ \&\ P_1$ hold $[\![ {X} ]\!]$ where $X\in \mathbb{Z}^{L\times D}$. $P_1$ holds weights $\gamma,\beta \in \mathbb{Z}^{D}$.}
    \KwOut{$P_0\ \&\ P_1$ get $[\![ \operatorname{LayerNorm}({X})]\!]$}
    \caption{Secure LayerNorm, $\Pi_{\text{LayerNorm}}$}
    \label{alg:layernorm}
    \SetNlSty{}{\color{green}}{}
    $P_0$ and $P_1$ invoke $[\![ {X}_{sum}]\!]=\operatorname{sum}_j([\![ {X}]\!]_{i,j})$.
    
    \SetNlSty{}{\color{green}}{}
    $P_0$ and $P_1$ invoke $[\![ {\mu} ]\!]=
    \operatorname{ewmul_{cp}}([\![ {X}_{sum}]\!],\frac{1}{D})$.

    \SetNlSty{}{\color{green}}{}
    $P_0$ and $P_1$ invoke $[\![ {X}_{\mu}]\!]=\operatorname{sadd_{cc}}([\![ {X}]\!],-[\![ {\mu}]\!])$.

    \SetNlSty{}{\color{green}}{}
    $P_0$ and $P_1$ invoke $[\![ {X}_{\mu}^2 ]\!]=\operatorname{ewmul_{cc}}([\![ {X}_{\mu} ]\!],[\![ {X}_{\mu} ]\!])$.

    \SetNlSty{}{\color{green}}{}
    $P_0$ and $P_1$ invoke $[\![ {X}_{tmp} ]\!]=\operatorname{sum}_j([\![ {X}_{\mu}^2 ]\!]_{i,j})$.

    \SetNlSty{}{\color{green}}{}
    $P_0$ and $P_1$ invoke $[\![ {\sigma}^2]\!]=\operatorname{ewmul_{cp}}([\![ {X}_{tmp} ]\!], \frac{1}{D})$.

    \SetNlSty{}{\color{red}}{}
    $P_0$ and $P_1$ invoke $[\![ \frac{1}{\sigma}]\!]=\operatorname{rsqrt}([\![ {\sigma}^2]\!])$.

    \SetNlSty{}{\color{green}}{}
    $P_0$ and $P_1$ invoke $[\![ \frac{{X}_{\mu}}{\sigma}]\!]=\operatorname{smul_{cc}}([\![ {X}_{\mu} ]\!],[\![ \frac{1}{\sigma} ]\!])$.

    \SetNlSty{}{\color{green}}{}
    $P_0$ and $P_1$ invoke $[\![ \frac{{X}_{\mu}}{\sigma}\cdot \gamma ]\!]=\operatorname{ewmul_{cp}}([\![ \frac{{X}_{\mu}}{\sigma} ]\!],\gamma)$.

    \SetNlSty{}{\color{green}}{}
    $P_0$ and $P_1$ invoke $[\![ \operatorname{LayerNorm}({X})]\!]=\operatorname{ewadd_{cp}}([\![ \frac{{X}_{\mu}}{\sigma}\cdot \gamma ]\!],\beta)$.
\end{algorithm}
\begin{algorithm}[!tb]
    % \DontPrintSemicolon
    \small
    % \SetAlgoLined
    \KwIn{$P_0\ \&\ P_1$ hold $[\![ x]\!]$.}
    \KwOut{$P_0\ \&\ P_1$ get $[\![ \operatorname{GeLU}(x)]\!]$}
    \caption{Secure GeLU, $\Pi_{\operatorname{GeLU}}$}
    \label{alg:gelu}
    \SetNlSty{}{\color{green}}{}
    $P_0$ and $P_1$ invoke $[\![ x^2 ]\!]=\operatorname{ewmul_{cc}}(\ashare{x},\ashare{x})$.

    $P_0$ and $P_1$ invoke $[\![ x^3 ]\!]=\operatorname{ewmul_{cc}}([\![ x^2 ]\!],\ashare{x})$ and $[\![ x^4 ]\!]=\operatorname{ewmul_{cc}}([\![ x^2 ]\!],[\![ x^2 ]\!])$.

    $P_0$ and $P_1$ evaluate $[\![ ax^4]\!], [\![ bx^3]\!], [\![ cx^2]\!], [\![ (0.5+d)x]\!]$ and $[\![ (0.5-d)x]\!]$ through $\operatorname{ewmul_{cp}}$.

    $P_0$ and $P_1$ evaluate two polynomials $[\![ F_0(x)]\!]=[\![ax^4-bx^3+cx^2+(0.5-d)x+e]\!]$ and $[\![ F_1(x)]\!]=[\![ax^4+bx^3+cx^2+(0.5+d)x+e]\!]$ through $\operatorname{ewadd_{cc}}$ and $\operatorname{ewadd_{cp}}$.

    \SetNlSty{}{\color{red}}{}
    $P_0$ and $P_1$ invoke $[\![ {b_0} ]\!]^B=\operatorname{cmp}(\ashare{x},-2.7), [\![ {b_1} ]\!]^B=\operatorname{cmp}(\ashare{x}, 0)$ and $[\![ {b_2} ]\!]^B=\operatorname{cmp}(2.7, \ashare{x})$.

    \SetNlSty{}{\color{green}}{}
    $\forall b\in \{0,1\}, P_b$ locally compute $[\![ {z_0} ]\!]^B=[\![ {b_0} ]\!]^B \oplus [\![ {b_1} ]\!]^B, [\![ {z_1} ]\!]^B=[\![ {b_1} ]\!]^B \oplus [\![ {b_2} ]\!]^B \oplus b$ and $[\![ {z_2} ]\!]^B=[\![ {b_2} ]\!]^B$. Note $z_0={1}\{-2.7 < x \le 0\}$, $z_1={1}\{0<x\le 2.7\}$ and $z_2 = {1}\{2.7<x\}$

    \SetNlSty{}{\color{red}}{}
    $P_0$ and $P_1$ invoke $[\![ \operatorname{GeLU}(x) ]\!]=\operatorname{mux}([\![ F_0(x) ]\!], [\![ {z_0} ]\!]^B) + \operatorname{mux}([\![ F_1(x) ]\!], [\![ {z_1} ]\!]^B) + \operatorname{mux}(x, [\![ {z_2} ]\!]^B)$.
\end{algorithm}
\begin{algorithm}[!tb]
    % \DontPrintSemicolon
    \small
    % \SetAlgoLined
    \KwIn{$P_0\ \&\ P_1$ hold $[\![ X]\!]$ where $X\in \mathbb{Z}^{L\times L}$.}
    \KwOut{$P_0\ \&\ P_1$ get $[\![ \operatorname{Softmax}(X)]\!]$}
    \caption{Secure Softmax, $\Pi_{\operatorname{Softmax}}$}
    \label{alg:softmax}
    \SetNlSty{}{\color{red}}{}
    $P_0$ and $P_1$ evaluate $[\![ \bar{X} ]\!]$ where $\bar{X}_i=\operatorname{max}_j(\ashare{x}_{i,j})$ through $L$ times $\operatorname{cmp}$.

    \SetNlSty{}{\color{green}}{}
    $P_0$ and $P_1$ invoke $[\![ X_{tmp1} ]\!]=\operatorname{sadd_{cc}}(\ashare{x},-[\![ \bar{X} ]\!])$.

    $P_0$ and $P_1$ invoke $[\![ \frac{X_{tmp1}}{2^6} ]\!] = \operatorname{ewmul_{cp}}([\![X_{tmp1} ]\!], \frac{1}{2^6})$.

    $P_0$ and $P_1$ invoke $[\![ \exp(X_{tmp1}) ]\!]=[\![ (1+\frac{X_{tmp1}}{2^6})^{2^6} ]\!]$ through one $\operatorname{ewadd_{cp}}$ and six times $\operatorname{ewmul_{cc}}$.

    \SetNlSty{}{\color{red}}{}
    $P_0$ and $P_1$ invoke $[\![ b ]\!]^B=\operatorname{cmp}(T_{\exp}, \ashare{x})$.

    $P_0$ and $P_1$ invoke $[\![ X_{\exp} ]\!]=\operatorname{mux}([\![ \exp(X_{tmp1}) ]\!], [\![ b ]\!]^B)$.

    \SetNlSty{}{\color{green}}{}
    $P_0$ and $P_1$ invoke $[\![X_{tmp2} ]\!]=\operatorname{sum}_j([\![ X_{\exp} ]\!]_{i,j})$.

    \SetNlSty{}{\color{red}}{}
    $P_0$ and $P_1$ invoke $[\![\frac{1}{X_{tmp2}} ]\!]=\operatorname{rec}([\![ \frac{1}{X_{tmp2}} ]\!])$.

    \SetNlSty{}{\color{green}}{}
    $P_0$ and $P_1$ invoke $[\![ \operatorname{Softmax}(X) ]\!]=\operatorname{smul_{cc}}([\![ X_{\exp} ]\!],[\![ \frac{1}{X_{tmp2}} ]\!])$.
\end{algorithm}

\textbf{LayerNorm.} Given an input matrix ${X} \in \mathbb{R}^{L\times D}$, where $L$ is the sequence length and $D$ is the hidden dimension, the LayerNorm operation is defined as:
$\operatorname{LayerNorm}({X})_{i,j}=\gamma_j\cdot \frac{{X}_{i,j}-\mu_i}{\sigma_i}+\beta_j$,
where $\mu_i=\frac{1}{D}\sum_{j=1}^{D}{X}_{i,j}$ is the mean and $\sigma_i=\sqrt{\frac{1}{D}\sum_{j=1}^{D}({X}_{i,j}-\mu_i)^2}$ is the variance, $\gamma_j$ and $\beta_j$ are learnable parameters.

\textbf{GeLU.} Given an input element $x \in \mathbb{R}$, prior-art works approximate GeLU function using piecewise functions. We adopt the approximation proposed in BOLT~\cite{pang2023bolt}:

\begin{equation*}
\setlength\abovedisplayskip{-15pt}
    \operatorname{ApproxGELU}(x) = 
    \begin{cases}
        x & \text{if } x > 2.7, \\
        \begin{aligned}
            & a|x|^4 + b|x|^3 + c|x|^2 \\
            & \quad + d|x| + e + 0.5x
        \end{aligned} & \text{if } |x| \leq 2.7, \\
        0 & \text{if } x < -2.7.
    \end{cases}
\end{equation*}
The concrete parameters can be found in BOLT~\cite{pang2023bolt}.
% \begin{equation*}
%     \begin{aligned}
%         a &= 0.020848611754127593,
%         b = -0.18352506127082727, \\
%         c &= 0.5410550166368381, 
%         d = -0.03798164612714154, \\
%         e &= 0.001620808531841547.
%     \end{aligned}
% \end{equation*}
  
\textbf{Softmax.} Given an input matrix ${X} \in \mathbb{R}^{L\times L}$, the Softmax function is defined as:
$\operatorname{Softmax}({X})_{i, j}=\frac{e^{{X}_{i, j}-\bar{X}_i}}{\sum_{j \in[L]} e^{{X}_{i, j}-\bar{X}_i}}, i \in[L], j \in[L]$,
where $\bar{X}_i=\max_{j\in[L]}{X}_{i,j}$ is the maximum value in the $i$-th row. Since the input of the exponentiation is negative, prior-art protocol computes the exponentiation using the Taylor series with a simple clipping~\cite{lu2023bumblebee}:
\begin{equation}
    \operatorname{neg} \operatorname{Exp}(x)= \begin{cases}0, & x<T_{\exp } \\ \left(1+\frac{x}{2^t}\right)^{2^t}, & x \in\left[T_{\exp }, 0\right]\end{cases}
\end{equation}
We set $l=6$ and $T_{\exp }=-13$ to ensure accuracy.

Next, we present the operator-wise decomposition of these nonlinear layers. The $\operatorname{LayerNorm}$ decomposition is shown in Algorithm~\ref{alg:layernorm}, $\operatorname{GeLU}$ in Algorithm~\ref{alg:gelu}, and $\operatorname{Softmax}$ in Algorithm~\ref{alg:softmax}.
All operators used in these algorithms are defined in Table~\ref{tab:opertors}. Additionally, we transform $\operatorname{sadd_{cp}}$ and $\operatorname{smul_{cp}}$ to $\operatorname{ewadd_{cp}}$ and $\operatorname{ewmul_{cp}}$ respectively, as plaintext can be repacked freely without incurring additional costs. Each operator in these algorithms requires a private protocol for evaluation, and all intermediate results are in secret-shared form.
% And we omit the detailed description of the protocols for brevity.
\vspace{-1em}
\section{Detailed Protocols}\label{app:detail_protocol}
\begin{figure*}[t]
    \centering
    \includegraphics[width=1\linewidth]{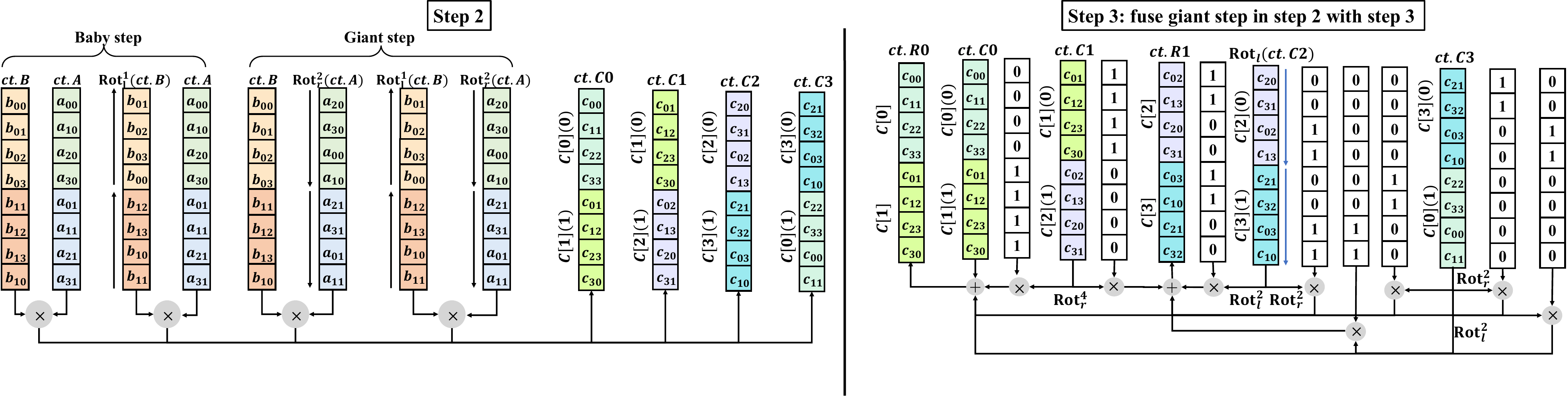}
    \caption{Applying BSGS optimization to the step 2 of our $\operatorname{matmul_{cc}}$ protocol.}\label{fig:bsgs}  
\end{figure*}
\begin{figure*}[t]
    \centering
    \includegraphics[width=1\linewidth]{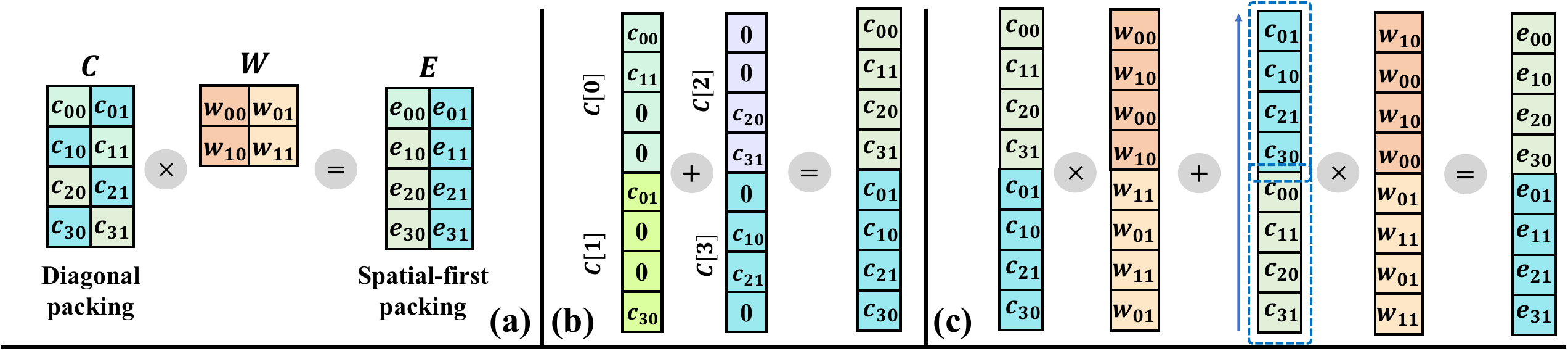}
    \caption{An example for \method's modified \( \operatorname{ct-pt} \) MatMul protocol for $\operatorname{Att}\times W_O$: (a) Multiplying matrices $C$ and $W$ in plaintext. (b) If the output dimension of $V_h$ in the previous step $\operatorname{Softmax}(\cdot)\times V_h$ is half the input dimension, we pad $V_h$ with zeros and perform the $\mmcc$. Afterward, we can add adjacent ciphertext results to obtain a dense diagonal packing of the output. (c) With $C$ in diagonal packing, we adjust the weight matrix $W$ to a duplicated spatial-first packing. We then rotate $C$ and multiply it with $W$ to produce the output $E$ in spatial-first packing.}
    \label{fig:diag}
    \vspace{-5pt}
\end{figure*}
\subsection{BSGS Optimization}\label{app:bsgs}
BSGS algorithm reduces the number of rotations during the multiply-accumulate process and has been widely adopted~\cite{pang2023bolt,ju2023neujeans,xu2024privcirnet,park2024powerformer}. For the computation $\sum_{i=0}^R A\times\operatorname{Rot}^i(B)$, instead of rotating each input individually, the BSGS algorithm divides the rotations into two steps: baby-step and giant-step, which can be formulated as:
$\sum_{j=0}^{G-1}\operatorname{Rot}^{j\cdot B}\left(\sum_{i=0}^{B-1} {\operatorname{Rot}^{-j\cdot B}(A)} \times {\operatorname{Rot}^i(B)}\right)$, 
where $G$ and $B$ represent the number of giant and baby steps, respectively, with $G \times B = R$. The total number of rotations is reduced from $R$ to $B + G - 2 \approx 2\sqrt{R}$. It can be extended to compute $\sum_{i=0}^R A_i\times\operatorname{Rot}^i(B)$ when $A_i$ is a plaintext.

The BSGS optimization can be applied to step 1 and step 2 of our $\operatorname{matmul_{cc}}$ protocol to reduce rotations. In Step 1, which involves plaintext multiplication and accumulation, BSGS can be directly applied. Step 2, involving ciphertext rotation and multiplication, does not immediately benefit from BSGS since there is no accumulation to reduce rotations during the giant step. However, we observe that the giant-step rotations can be fused with step 3. As shown in Figure~\ref{fig:bsgs}, we postpone the giant-step rotations until step 3, where we complete the partial sum accumulation by adding one additional mask multiplication and rotation to each ciphertext generated by the giant step. This integration allows us to maintain a multiplication depth of 4 while reducing the number of rotations from $B \times G$ to $B + 2G$. Although an additional $G$ is introduced, the complexity remains reduced from $R$ to $2\sqrt{2R}$.

% \vspace{-1.0em}
\subsection{Diagonal Ciphertext-Plaintext MatMul} \label{app: diagcp}
% \xts{concat can be overlooked}
When using \method's \( \operatorname{ct-ct} \) MatMul protocol to compute \( \operatorname{Att}_h = \operatorname{Softmax}(\cdot) \times V_h \), the output \( \operatorname{Att}_h \) is in diagonal packing. However, the current \( \operatorname{ct-pt} \) MatMul protocol in BOLT only supports inputs with spatial-first packing and cannot directly handle \( \operatorname{Concat}(\operatorname{Att}_h) W_O \). Moreover, the final result must be in spatial-first packing to ease subsequent computations. Thus, we develop a new \( \operatorname{ct-pt} \) MatMul protocol that supports diagonal-packed inputs and outputs in spatial-first packing.

For clarity, consider the example \( E = C \times W \), where \( C \in \mathbb{R}^{L \times D} \) is a ciphertext and \( W \in \mathbb{R}^{D \times D} \) is a plaintext. Figure~\ref{fig:diag} illustrates this with \( L = 4 \) and \( D = 2 \). Notably, \( C \) is produced by a \( \mmcc \) in a diagonal packing. If zero padding is used in $\mmcc$ to ensure the input and output dimensions are the same, the dimension of \( W \) must be adjusted accordingly.
However, we find that if zero padding is exactly doubled, we can add adjacent ciphertexts of $C$ to obtain the output in a dense diagonal packing, as shown in Figure \ref{fig:diag} (b). Therefore, zero padding at most doubles the dimension $D$. 
Figure \ref{fig:diag} (c) illustrates our optimized protocol. This protocol is based on the observation that the $i$-th column of $E$ can be represented as a combination of diagonals in $C$, multiplied by the elements of the $i$-th column in $W$. With $C$ in diagonal packing, we adjust the weight matrix $W$ to a duplicated spatial-first packing. We then rotate $C$ and multiply it with $W$ to produce the output $E$ in spatial-first packing. This step can also utilize BSGS to reduce the number of rotations. Consequently, our modified $\mmcp$ protocol maintains the same computational complexity as BOLT. Additionally, the $\operatorname{Concat}$ is inherently handled in the ciphertext domain, as all ciphertexts are tightly packed. Therefore, no additional protocol is needed.

% As shown in Figure \ref{fig:diag}, we expand the computation of the first column in $C$. If we rotate the partial sum of $y_{10}$ to the left, we can observe that computation is restructured as summations of diagonals in $A$ multiplied by elements in the first column in $B$, which is shown in Figure \ref{fig:bsgs} (c). This computation form is similar to BOLT's \( \operatorname{ct-pt} \) MatMul protocol except that in BOLT, the same elements is multiplied with a whole column whereas, in \method's protocol, elements multiplied with one diagonal is different. As shown in Figure \ref{fig:diag} (d), when multiplied with a large plaintext matrix, we could pack elements in different columns into one plaintext. With this packing, the multiplication will yield partial sums for different columns to eliminated the need to aggregate partial sums in one ciphertext and also enable the use of the BSGS strategy.

% \vspace{-1.5em}
\subsection{Ring and Field Conversion}\label{app:ring-field}
We now describe the protocols for conversions between ring and field, as proposed in BOLT~\cite{pang2023bolt} and SiRNN~\cite{rathee2021sirnn}.For field-to-ring conversion, we employ the signed extension protocol $\Pi_{\mathrm{Ext}}^{2^{l_1}, 2^{l_2}}$ in~\cite{rathee2021sirnn}, which extends shares from $\mathbb{Z}{2^{l_1}}$ to $\mathbb{Z}{2^{l_2}}$ ($l_1 < l_2$). Following BOLT~\cite{pang2023bolt}, we adapt it to the field-to-ring setting as $\Pi_{\mathrm{Ext}}^{q, 2^{\lceil \log_2 q \rceil}}$.
% We briefly introduce the underlying principle. For conversion from $\mathbb{Z}_{2^{l_1}}$ to $\mathbb{Z}_{2^{l_2}}$, the core idea is:
% $m \bmod 2^{l_2}= \ashare{m}_0^{2^{l_1}} + \ashare{m}_1^{2^{l_1}} - \mbm{1}\{\ashare{m}_0^{2^{l_1}}+\ashare{m}_1^{2^{l_1}}\ge 2^{l_1}\} * 2^{l_1} \bmod 2^{l_2}$,
% where $\mbm{1}\{\ashare{m}_0^{2^{l_1}}+\ashare{m}_1^{2^{l_1}}\ge 2^{l_1}\}$ is obtained using a single call to the comparison protocol $\operatorname{cmp}$.
% This naturally extends to field-to-ring conversion:
% $m \bmod 2^{\lceil \log_2 q \rceil} = \ashare{m}_0^{q} + \ashare{m}_1^{q} - \bold{1}\{\ashare{m}_0^{q}+\ashare{m}_1^{q}\ge q\} * q \bmod 2^{\lceil \log_2 q \rceil}$,
% where $\mbm{1}\{\ashare{m}_0^{q}+\ashare{m}_1^{q}\ge q\}$ is similarly obtained via one invocation of $\operatorname{cmp}$.
Then, we perform a local modulus reduction to $\mathbb{Z}_{2^l}$. We can avoid local modulus reduction errors by setting a proper $l$ such that the output range is within $\mathbb{Z}_{2^l}$.

For $\operatorname{ring-to-field}$ conversion, we use $\Pi_{\mathrm{Ext}}$ to extend the shares $\ashare{m}$ to a larger ring, e.g., from $\mathbb{Z}_{2^l}$ to $\mathbb{Z}_{2^{l+40}}$. In this case, the probability that $\ashare{m}_0 + \ashare{m}_1 > 2^{l+40}$ is greater than $1 - 2^{-40}$. This enables us to locally change the modulus to $q$ with an error rate below $2^{-40} \approx 10^{-12}$:

\vspace{-1em}
\begin{equation*}\label{eq:ring-to-field}
    \begin{aligned}
        [\![ m ]\!]^q_0 &= [\![ m ]\!]^{2^{l+40}}_0 \bmod q,\quad
        [\![ m ]\!]^q_1 = [\![ m ]\!]^{2^{l+40}}_1 - 2^{l+40} \bmod q \\
        [\![ m ]\!]^q_0 &+ [\![ m ]\!]^q_1 = [\![ m ]\!]^{2^{l+40}}_0 + [\![ m ]\!]^{2^{l+40}}_1 -2^{l+40} \bmod q \\ 
        &= m + 2^{l+40} -2^{l+40} \bmod q 
        = m \bmod q
    \end{aligned}
\end{equation*}
\subsection{Probabilistic Local Truncation Method}\label{app:accurate_encoding}
\delete{
Local truncations can introduce MSB errors during the encoding and decoding process~\cite{rathee2021sirnn}. To mitigate this issue, we propose a probabilistic local truncation method akin to the approach in Equation~\ref{eq:ring-to-field}. Given a secret share $\ashare{m}^{2^l}$, we extend it to a larger ring using an extension protocol, resulting in $\ashare{m}^{2^{l+40}}$.

In the larger ring, the probability that $\ashare{m}_0 + \ashare{m}_1 > 2^{l+40}$ exceeds $1 - \frac{1}{2^{40}}$. This enables us to perform local truncation with an error rate below $\frac{1}{2^{40}} \approx 10^{-12}$, as detailed in Equation~\ref{eq:local_trunc}.
% Notably, Least Significant Bit (LSB) errors are not critical since CKKS HE is an approximate scheme.
}
\delete{
After encoding or decoding, a local modulus reduction is performed to return to the original ring $\mathbb{Z}_{2^l}$.

The additional extension protocol can be fused in the $\operatorname{field-to-ring}$ and $\operatorname{ring-to-field}$ protocols, eliminating additional communication costs. For $\operatorname{field-to-ring}$, we can directly change the modulus to $2^{l+40}$ instead of $2^l$, and then conduct decoding. For $\operatorname{ring-to-field}$, since the ring is already extended to $2^{l+40}$ after encoding, we can directly change the modulus to $q$ without additional communication costs.
}

\revise{
We leverage the (non-interactive) probabilistic truncation protocol introduced in SecureML~\cite{mohassel2017secureml}. 
The protocol takes as input a secret-shared value $[\![ m ]\!]^{2^{l}}$ and outputs $[\![ m' ]\!]^{2^{l}}$, where $m' = \lfloor m / 2^s \rfloor + u $, with $\Pr(u = 0) =1-\frac{m\mod 2^s}{2^s}, \Pr(u = 1) = \frac{m\mod 2^s}{2^s}$. 
% and $\Pr(u' = 2^{l-1}) = 1 - \Pr(u' = 0) = m/2^l$.
% That is, the rounding error $u$ is independent of the message $m$. 
Li et al.~\cite{li2023efficient} point out that some probabilistic truncation protocols are not secure in the simulation-based sense. 
To prove the security of our truncation protocol in the simulation-based paradigm, we change our truncation functionality in Figure~\ref{fig:truncpr}, which gives the cutoff point $k$ to the adversary. Note that this extra element reveals only as much information as the definition of probabilistic truncation itself, where a rounding error $u = 1$ is introduced with probability $\frac{m \mod 2^s}{2^s}$.
The detailed proof under the simulation-based paradigm can be found in Section 3 in~\cite{santos2024curl}.

However, the protocol~\cite{mohassel2017secureml} may fail, potentially causing a max significant bit (MSB) error, with a probability of $|m|/2^{l-1}$.
To overcome this, we take an extend-then-truncate approach.
We first use the \cite{rathee2021sirnn} protocol to extend the modulus size: 
$\ashr{m}^{2^{l+40}} \leftarrow {\tt Extend}(\ashr{m}^{2^l}).$
% to increase the modulus size to $l + 40$ bits.
Then we perform SecureML's truncation over the larger ring using arithmetic right-shift ($\gg_a$):
$
\ashr{m'}^{2^{l+40}}_b = \ashr{m}^{2^{l+40}}_b \gg_a s
$.
The failure probability is now smaller than $2^{-40}$.
With one more modulo operation, i.e., $\ashr{m'}^{2^l}_b = \ashr{m'}^{2^{l+40}}_b\bmod 2^l$ for $b = 0, 1$, we obtain the truncated result over the ring $\mathbb{Z}_{2^l}$.

% By assuming the truncation shift $s < 40$, we can obtain a uniform distributed 

% might not realize the standard ideal functionality for probabilistic truncation 
% as their rounding error is a deterministic function of the input $m / 2^s$, and certain information revealed during those protocols.
% However, as formally proved in~\cite{santos2024curl} (Appendix A), the protocol from~\cite{mohassel2017secureml,escudero2020improved} realizes a modified probabilistic truncation functionality, which we define in Figure~\ref{fig:truncpr}. This functionality explicitly reveals the cutoff point $s_{(s)}$. The leakage of this cutoff point conveys no more information than what is already implied by the standard semantics of probabilistic truncation—i.e., that the least significant bit error occurs with probability $m / 2^s$. This formulation has been widely accepted in the literature~\cite{mohassel2017secureml,computation2010improved,escudero2020improved,santos2024curl,knott2021crypten,dalskov2019secure,damgaard2019new} and is considered practically sound for most use cases.
}
% \begin{figure}[htbp]
% \centering
% \begin{tcolorbox}[title={$\mathcal{F}^{\text{TruncPr}}_{l,s}\left( \llbracket m \rrbracket \right)(s \le 40)$}]
% % \textit{Input:} Assume both parties hold their shares of $\llbracket m \rrbracket^{2^l}$.
% \begin{enumerate}
%     \item Reconstruct $m$ from sharing $\llbracket m \rrbracket^{2^l}$.
%     \item Sample a bit $u \leftarrow \{0, 1\}$  uniformly at random.
%     % \item With a chance of $m/2^l$, set a value $u' = 2^{l-s}$ or $u' = 0$ otherwise.
%     \item Set $m^{(s)} = \lfloor m / 2^s \rfloor$.
%     \item Set $m' = m^{(s)} - u\bmod 2^l$ .
%     \item Output a random sharing $\llbracket m' \rrbracket^{2^l}$ of $m'$.
% \end{enumerate}
% \end{tcolorbox}
% \caption{\revise{The modified probabilistic truncation functionality}}
% \label{fig:truncpr}
% \end{figure}

\begin{figure}[t]
\revise{
\fbox{
\centering
\begin{minipage}{0.46\textwidth}
  \begin{center}
  \underline{\centering $\mathcal{F}^{\text{TruncPr}}_{l,s}\left( \llbracket m \rrbracket \right)(s \le 40)$ }
\end{center}
  \begin{enumerate}
    \setlength{\itemsep}{0pt}
    \setlength{\topsep}{-1pt}
    \setlength{\parsep}{0pt}
    \setlength{\parskip}{0pt}
    \item Reconstruct $m$ from sharing $\llbracket m \rrbracket$.
    \item Set $m^{(s)} = \lfloor m / 2^s \rfloor$ and $m_{(s)} = (m \bmod 2^s)$.
    \item Pick cutoff point $s_{(s)}$ at random in $\mathbb{Z}_{2^s}$.
    \item Set $m' = m^{(s)} + u$ where
    $u = \mbm{1}\{m_{(s)} > s_{(s)}\}$
    \item Generate a random sharing $\llbracket m' \rrbracket$ of $m'$.
    \item Output $\llbracket m' \rrbracket$ and leak $s_{(s)}$ to the adversary.
  \end{enumerate}
\end{minipage}
}
\caption{\revise{The modified probabilistic truncation functionality}} \label{fig:truncpr}
}
\vspace{-5pt}
\end{figure}

\section{\revise{Additional Experiments}}\label{app:exp}
\begin{table}[!tb]
    \centering
    % \huge
    \begingroup
    \color{black}
    \caption{\revise{Amortized latency comparison on BERT-base.}}
    \label{tab:exp_end2end}
    \resizebox{0.8\linewidth}{!}{
    % \begin{threeparttable}
    \begin{tabular}{cSSS}
    \toprule
    \multicolumn{1}{c}{\multirow{2}{*}{Framework}} & \multicolumn{3}{c}{{Amortized Latency (min)}}  \\
    \cmidrule{2-4}
    & LAN & {WAN$_2$} & {WAN$_3$} \\
    \midrule
    % \cmidrule{3-4}
    \method & 2.5 & 6.6 &13.2\\
    \midrule
    \method~+batch size 32 & 0.79 & 2.1 & 4.1\\
    \midrule
    NEXUS+batch size 32 & 0.63 & 0.64 & 1.0 \\
    \bottomrule
    \end{tabular}
    }
    \endgroup
\end{table}
\paragraph{\revise{Comparison with NEXUS under Batched Inputs.}}
\revise{
Although BLB defaults to a batch size of 1, it also supports batched inputs. Batching enables BLB to pack more data along the spatial dimension, reduce rotations, improve GPU utilization, and amortize communication rounds, resulting in lower amortized latency. In Table~\ref{tab:exp_end2end}, BLB shows a $3\times$ speedup with batch. Compared to NEXUS, BLB performs similarly under LAN but is slower under WAN due to higher communication costs.}

% \begin{figure}[htbp]
% \centering
% \begin{tcolorbox}[title={$\mathcal{F}^{\text{TruncPr}}_{l,s}\left( \llbracket m \rrbracket \right)$}]
% \textit{Input:} Assume both parties hold $\llbracket m \rrbracket^{2^l}$.
% \begin{enumerate}
%     \item Reconstruct $m$ from sharing $\llbracket m \rrbracket$.
%     \item Set $m^{(s)} = \lfloor m / 2^s \rfloor$ and $m_{(s)} = (m \bmod 2^s)$.
%     \item Pick cutoff point $s_{(s)}$ at random in $\mathbb{Z}_{2^s}$.
%     \item Set $m' = m^{(s)} + u$ where
%     \[
%     u = 
%     \begin{cases}
%         0 & \text{if } m_{(s)} \leq s_{(s)}, \\
%         1 & \text{otherwise}.
%     \end{cases}
%     \]
%     \item Generate a random sharing $\llbracket m' \rrbracket$ of $m'$.
%     \item Output $\llbracket m' \rrbracket$ and leak $s_{(s)}$ to the adversary.
% \end{enumerate}
% \end{tcolorbox}
% \caption{\revise{Ideal modified probabilistic truncation functionality}}
% \label{fig:truncpr}
% \end{figure}

%%%%%%%%%%%%%%%%%%%%%%%%%%%%%%%%%%%%%%%%%%%%%%%%%%%%%%%%%%%%%%%%%%%%%%%%%%%%%%%%
\end{document}